\documentclass[12pt, hyper]{article}
\usepackage{amsmath,amsopn,amssymb,amsfonts}
\numberwithin{equation}{section}
\usepackage{amsthm}
\usepackage{hyperref}
\usepackage[pdftex]{graphicx}
\usepackage{epsfig}
\usepackage{cleveref}
\usepackage{subfigure}
\usepackage{mathrsfs}
\DeclareMathAlphabet{\mathpzc}{OT1}{pzc}{m}{it}
\usepackage{verbatim}
\usepackage{multirow}
\usepackage{tikz}
\usepackage{pgfplots}
\usepackage{makecell}
\usepackage{arydshln}
\usepackage{enumitem}
\setlist{leftmargin=2mm}
\usepackage{cite}
\usepackage{amsthm}
\usepackage[margin=1cm]{caption}
\usepackage[font={small}]{caption}

\crefname{equation}{}{}

\def\IZ{\mathbb{Z}}

\def\CD{{\cal D}}

\def\CH{{\cal H}}\def\CI{{\cal I}}

\def\CN{{\cal N}}\def\CO{{\cal O}}

\def\CV{{\cal V}}

\def\CZ{{\cal Z}}

\def\a{\alpha}\def\b{\beta}
\def\d{\delta}\def\e{\epsilon}
\def\th{\theta}

\def\m{\mu}\def\n{\nu}
\def\r{\rho}\def\s{\sigma}
\def\t{\tau}
\def\G{\Gamma}
\def\D{\Delta}
\def\O{\Omega}

\def\half{\frac{1}{2}}

\def\p{\partial}

\def\Tr{{\rm Tr}}

\usepackage{cleveref}

\begin{document}

\begin{titlepage}
\vfill
\begin{flushright}
{\tt\normalsize KIAS-P18096}\\
%{\tt\normalsize hep-th/yymmnnn}\\
\end{flushright}
\vfill
\begin{center}
{\LARGE \bf Modular Constraints on \\ \vskip 0.5cm Superconformal Field Theories}

\vfill

Jin-Beom Bae, Sungjay Lee, and Jaewon Song

\vskip 5mm
{\it Korea Institute for Advanced Study \\
85 Hoegiro, Dongdaemun-Gu, Seoul 02455, Korea}

\end{center}
\vfill

\begin{abstract}
\noindent
We constrain the spectrum of $\CN=(1, 1)$ and $\CN=(2, 2)$ superconformal field theories in two-dimensions by requiring the NS-NS sector partition function to be invariant under the $\Gamma_\theta$ congruence subgroup of the full modular group $SL(2, \mathbb{Z})$. We employ semi-definite programming to find constraints on the allowed spectrum of operators with or without $U(1)$ charges. Especially, the upper bounds on the twist gap for the non-current primaries exhibit interesting peaks, kinks, and plateau.  We identify a number of candidate rational (S)CFTs realized at the numerical boundaries and find that they are realized as the solutions to modular differential equations associated to $\Gamma_\theta$. Some of the candidate theories have been discussed by H\"ohn in the context of self-dual extremal vertex operator (super)algebra.
We also obtain bounds for the charged operators and study their implications to the weak gravity conjecture in AdS$_3$.
\end{abstract}

\vfill
\end{titlepage}

\parskip 0.1 cm
\tableofcontents
\renewcommand{\thefootnote}{\#\arabic{footnote}}
\setcounter{footnote}{0}

\parskip 0.2 cm

%%%%%%%%%%%%%%%%%%%%%%%%%%%%%%%%%%%%%%%%%%%%%%%%%%%%%%%%%%%%%%%%%%%%%
%%%%%%%%%%%%%%%%%%%%%%%%%%%%%%%%%%%%%%%%%%%%%%%%%%%%%%%%%%%%%%%%%%%%%
\section{Introduction}
%%%%%%%%%%%%%%%%%%%%%%%%%%%%%%%%%%%%%%%%%%%%%%%%%%%%%%%%%%%%%%%%%%%%%
%%%%%%%%%%%%%%%%%%%%%%%%%%%%%%%%%%%%%%%%%%%%%%%%%%%%%%%%%%%%%%%%%%%%%

Modular invariance is a fundamental constraint for any consistent two-dimensional conformal field theory (CFT). It means that the partition function should not depend on the large diffeomorphisms of the torus, which is $SL(2, \IZ)$. Combined with underlying Virasoro symmetry or extended chiral algebra, modular invariance constraints the spectrum of operators for a consistent CFT.
Via AdS/CFT correspondence \cite{Maldacena1999f, Gubser:1998bc,Witten:1998qj}, modular constraints on two-dimensional CFT can be translated to universal features on arbitrary quantum gravity in three-dimensional Anti-de Sitter (AdS) space.

In a previous paper of the current authors \cite{BLS}, we have investigated the consequences of modular invariance for a non-supersymmetric two-dimensional CFT with (higher-spin) conserved currents along the line of \cite{Hellerman2011c, Friedan2013, Collier2016}. See also \cite{Keller2012,Qualls:2013eha,Qualls:2015bta,Benjamin2016,Apolo:2017xip, Afkhami-Jeddi:2017idc,Dyer2017,Anous:2018hjh}. In this paper, we further investigate the modular constraints for $\CN=(1, 1)$ and $\CN=(2, 2)$ superconformal field theories. We also consider (S)CFTs with conserved currents in addition to the (super)conformal symmetry. Here, we follow the method of \cite{Benjamin2016, Dyer2017} to obtain bounds that depend on the charges. Since $\CN=2$ superconformal algebra contains $U(1)_R$ global symmetry, we also investigate the difference between R-symmetry and (non-R) flavor symmetry.

The modular bootstrap equation for the superconformal theories differs from the non-supersymmetric case in the following way. For any CFT, modular invariance demands the partition function on a torus to be invariant
\begin{align}
 Z (\tau, \bar{\tau} ) = Z \left( \frac{a \tau + b}{c \tau + d},  \frac{a \bar\tau + b}{c \bar\tau + d} \right) \ ,
\end{align}
for any element 
\begin{align}
  \left( \begin{array}{cc} a & b \\ c & d \end{array} \right)\in \G \ ,
\end{align}
in the modular group $\G = SL(2, \IZ)$.
It is an elementary fact that the modular group is generated by two elements commonly denoted as $S$ and $T$, which acts on $\tau$ as
\begin{align}
 S: \tau \mapsto -\frac{1}{\tau} \ , \qquad T: \tau \mapsto \tau + 1 \ .
\end{align}
For a supersymmetric theory or any theory with fermions, we need to specify the spin structure on the torus. In order to obtain a modular invariant partition function, we need to sum over all the spins as is suggested by the path-integral description:
\begin{align}
  Z (\tau, \bar{\tau} ) = Z_{\textrm{NS-NS}}(\tau, \bar{\tau} ) + Z_{\textrm{NS-R}}(\tau, \bar{\tau} ) + Z_{\text{R-NS}}(\tau, \bar{\tau} ) \pm Z_{\text{R-R}}(\tau, \bar{\tau} ) \ ,
\end{align}
where NS (Neveu-Schwarz) and R (Ramond) refer to the anti-periodic and periodic boundary conditions for the two circle directions.\footnote{Throughout this paper, we only consider  parity invariant CFTs so that left-moving and right-moving fermions have the same boundary condition.} Even though the full partition function is modular-invariant, partition functions for each sector is not since four different spin structures get swapped under the $T$ and $S$ transformations.
In the Hamiltonian picture, the NS sector partition function is defined as
\begin{align}
 Z_{\textrm{NS}} (\tau, \bar\tau) \equiv \Tr_{\CH_{\textrm{NS}}} e^{2\pi i \tau (L_0 - \frac{c}{24}) } e^{-2\pi i \bar{\tau} (\bar{L}_0 - \frac{c}{24}) } \ ,
\end{align}
which is identical to $Z_{\textrm{NS-NS}}$ given in the path-integral description. The $Z_{\textrm{R-NS}}$ can be defined in a similar way as the $Z_{\textrm{NS}}$ except that we insert $(-1)^F$ in the trace. The NS sector partition function is invariant only under $T^2$ and $S$, which generates $\Gamma_\theta$ congruence subgroup of the modular group.

In this paper, we consider the NS partition function and the constraints coming from the $\G_\theta$, instead of the full partition function under the full modular group $SL(2, \IZ)$. 
This is because the $\G_\theta$-invariant NS partition function is enough to obtain NS-R, R-NS sectors by acting $T$ and $S$. Additional constraint can in principle come from the RR sector, which is modular invariant by itself. But the RR partition function is either trivial for the $\CN=1$ case or rather well understood for $\CN\ge2$, thanks to the powerful constraint coming from the weak Jacobi form. See, for instance \cite{Gaberdiel:2008xb}.
One crucial point for us is that it is possible to decompose the partition function for each sector in terms of superconformal characters. But the full partition function, once decomposed in terms of superconformal characters, can be plagued with the minus sign coming from the insertion of $(-1)^F$ for the fermions. This hampers us from imposing the positivity constraints for the degeneracies that is necessary to utilize the semi-definite programming.
In addition, in the context of AdS$_3$/CFT$_2$ correspondence, it is the NS sector in the CFT$_2$ that is mapped to the AdS$_3$ gravity. Therefore, we focus on the NS sector only. 

The NS partition function admits a character decomposition under the superconformal algebra (or extended chiral algebra) so that we can write schematically
\begin{align} \label{eq:NSpartfunc}
 Z_{\textrm{NS}} (\tau, \bar{\tau}) = \chi_0(\tau) \bar{\chi}_0(\bar{\tau}) + \sum_{h, \bar h} d_{h,\bar h} \chi_h (\tau) \chi_{\bar h} (\bar\tau) \ ,
\end{align}
where the sum is over various representations other than the vacuum of the superconformal algebra and $\chi_h$ refers to the corresponding character. Here we split the vacuum and non-vacuum contributions. For the $\CN=(2, 2)$ SCFT, there exist short multiplets that we need to take care separately. This turns out to be rather involved, and we discuss in detail in section \ref{subsec:N2SCA}. Note that what we do here is different from $\CN=(2, 2)$ analysis done in \cite{Keller2012, Friedan2013} where they used the extended $\CN=(2, 2)$ supersymmetry \cite{Odake:1989dm, Odake:1989ev} that includes the spectral flow symmetry.
Our bootstrap problem now boils down to finding constraints for the NS partition function \eqref{eq:NSpartfunc} under $\G_\theta$. The invariance under $T^2: \tau \mapsto \tau + 2$ forces the spin in each state to be a half-integer. The non-trivial condition is under the $S$ transformation:
\begin{align}
 Z_{\textrm{NS}} (\tau, \bar{\tau}) = \textrm{(phase factor)} \times Z_{\textrm{NS}} \left(-\frac{1}{\tau}, -\frac{1}{\bar{\tau}} \right)
\end{align}
We may have an extra phase factor under the $S$-transformation once we turn on the chemical potential for the $U(1)_R$ or the flavor symmetry \cite{Dyer2017}. This is the bootstrap equation we would like to exploit using the semi-definite programming.

\paragraph{Summary of Results}
\begin{figure}[h]
\begin{center}\includegraphics[width=.85\textwidth]{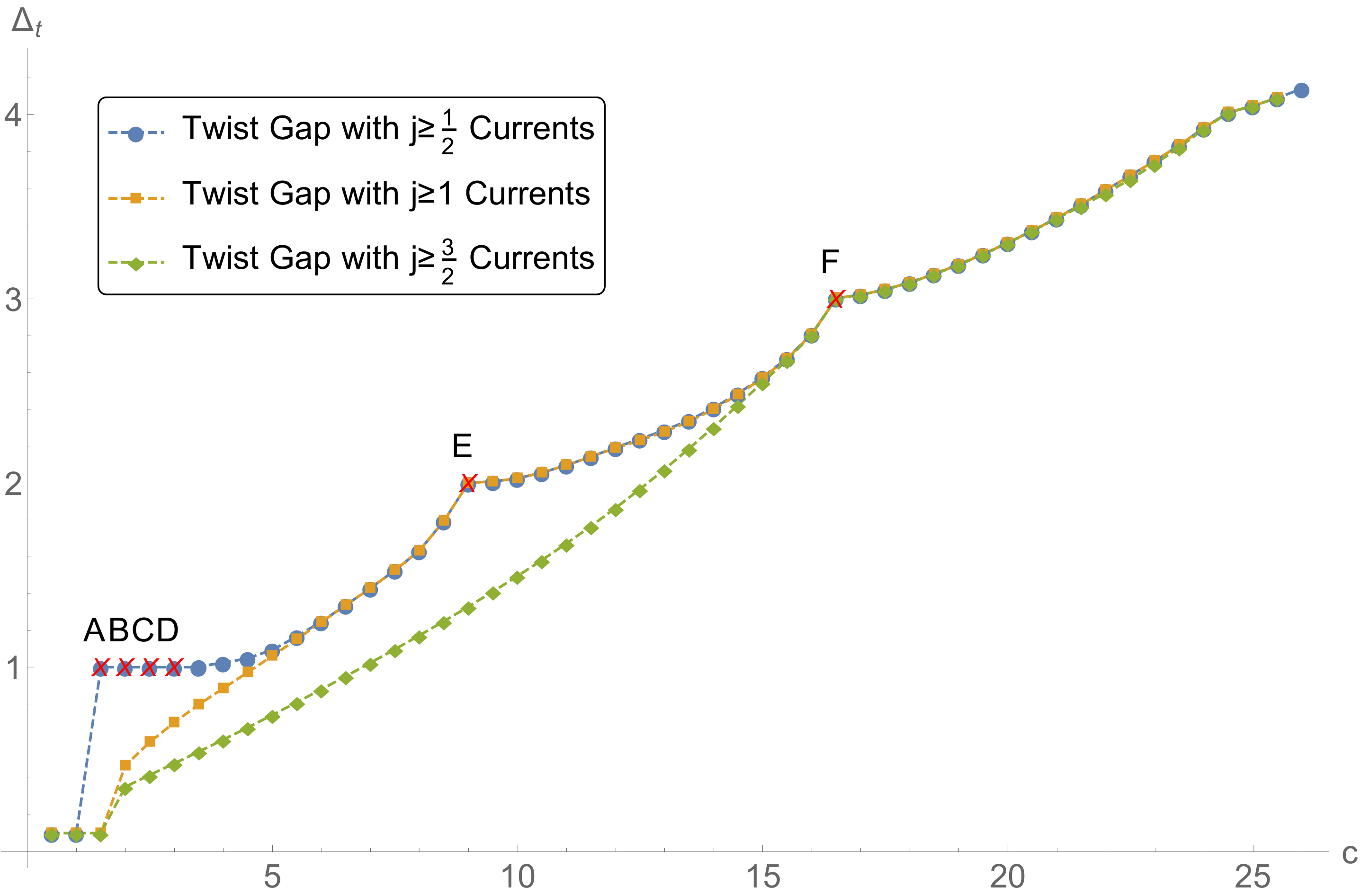}
\end{center}
\caption{Numerical upper bounds on the twist gap for the $\mathcal{N}=1$ SCFTs with imposing the
conserved currents of $j \ge \frac{1}{2}$, $j \ge 1$ and $j \ge \frac{3}{2}$.}
\label{fig:N1TwistGap}
\end{figure}
Let us summarize some of the highlights of we obtain.  As was shown in the previous paper \cite{BLS}, it is interesting to consider the twist gap for the operators with assuming the existence of holomorphic currents with spin $j\ge j_0$. It has been found that the numerical bounds exhibit dramatic peaks and kinks. It was also noticed that at these special points live interesting rational CFTs. We repeat this analysis for the NS partition function of superconformal theory and find similar features as in the case of non-SUSY analysis.

For the $\CN=(1, 1)$ theory, we obtain the numerical bounds as depicted in figure \ref{fig:N1TwistGap}.
As we can see, there are numbers of kinks that we name as points A, E and F. One of the salient feature is that there exists a plateau in the region $\frac{3}{2} \le c \le 3$. We denote the points with central charge $c=\frac{3}{2}, 2, \frac{5}{2}, 4$ as A, B, C, D respectively. By using the extremal functional method (EFM) \cite{El-Showk2012}, we obtain the candidate partition functions at these points. The partition functions at these points agree with that of the free fermions.

We also identify the partition functions at the kinks $c=9$ (E) and $c=\frac{33}{2}$ (F) that appear even when we remove spin $j=\half$ currents. We were able to write an ansatz for the partition functions at each point and find that the bounds on the degeneracies indeed approach to the ansatz as we increase the numerical precision. For example at $c=9$ with the twist gap $\Delta_t=2$, the partition function can be holomorphically factorized into
\begin{align}
 Z_{\textrm{NS}}^{c=9} (\tau, \bar{\tau}) = f^{c=9}(\tau) \bar{f}^{c=9}(\bar\tau) \ ,
\end{align}
and the holomorphic part $f^{c=9}(\tau)$ turns out to be a solution to a modular differential equation associated to $\G_\theta$
\begin{align}
 \left[ q \frac{\p}{\p q} + N_2 (q) \right] f^{c=9}(\tau) = 0 \ ,
\end{align}
where $q=e^{2\pi i \tau}$ and $N_2(q)$ is defined in \eqref{N2}. The modular differential equation has been used 30 years ago by \cite{Mathur1988} in an attempts to classify rational conformal field theories (RCFT). It has been revived rather recently in many contexts \cite{Gaberdiel:2008pr,Bantay:2010uy, Hampapura:2015cea,Hampapura:2016mmz,BLS, Mukhi:2017ugw,Chandra:2018pjq,Harvey:2018rdc,Arakawa:2016hkg,Beem:2017ooy,Buican:2017rya}.
The kink at $c=\frac{33}{2}$ (F) is the one that has twist gap $\Delta_t=3$. We find that the putative partition function at this point also factorizes. Our result suggests that it may be possible to find a rational (S)CFT at each points with the single character given by $f^{c=9, \frac{33}{2}} (\tau)$. The same partition function has been obtained in \cite{Hoehn2007}.

\begin{figure}[h]
\begin{center}\includegraphics[width=0.85\textwidth]{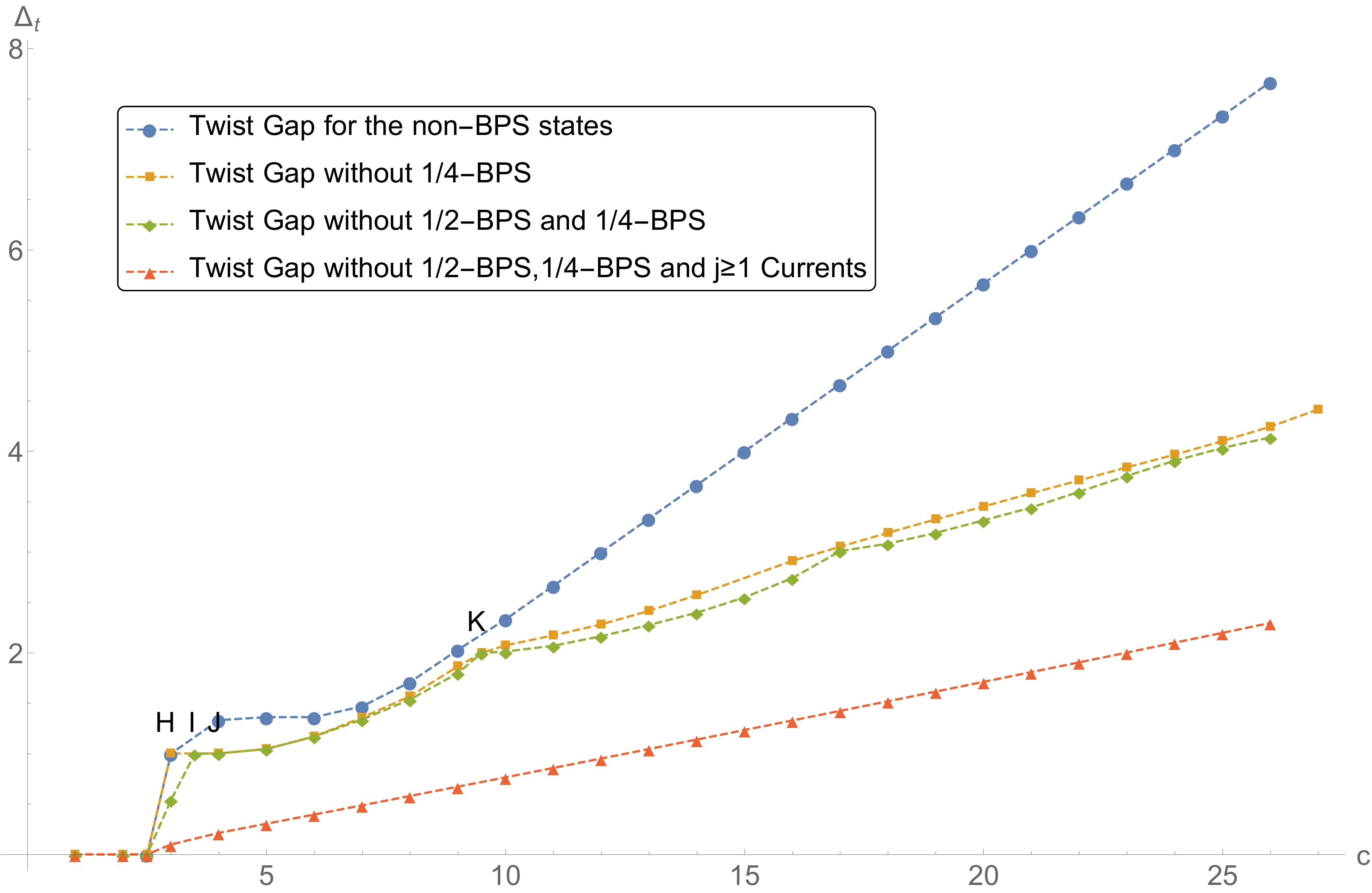}
\end{center}
\caption{Numerical upper bounds on the twist gap for the $\CN=2$ SCFTs under various assumptions. The blue line represents the most generic bounds for the $\CN=(2, 2)$ SCFT.}
\label{fig:N2TwistGap}
\end{figure}

We perform a similar analysis with $\CN=(2, 2)$ superconformal symmetry. The result is given in figure \ref{fig:N2TwistGap}.
We find the twist gap in the NS sector for the non-BPS sector is bounded above as the blue curve in the figure. It asymptotes to a linear line as
\begin{align}
\D_t = \frac{1}{3}(c-3) \ ,
\end{align}
for $c \gtrsim 6$.
There is a kink (H) at $c=3$, $\Delta_t = 1$, at which we observe the partition function given by that of 6 free fermions. The theory of 6 free fermions indeed preserve $\CN=2$ supersymmetry since upon bosonization, it can be mapped to the theory of a free $\CN=(2, 2)$ chiral multiplet.
Quite interestingly, more kinks appear at the numerical boundary of the twist gap $\Delta_t$ once we drop some of the BPS sectors.
At the kink (K) with $c=19/2, \Delta_t=2$, we find the putative partition function holomorphically factorizes and again the holomorphic part solves a $\G_\theta$ modular differential equation. Again, this suggests that there may exist an $\CN=2$ superconformal vertex operator algebra (VOA) at $c=19/2$ with the partition function given as the holomorphic part.

Finally, we study constraints on the charged states coming from modular invariance. Especially, we test weak gravity conjecture (WGC) \cite{Arkani-Hamed2006} in AdS$_3$ \cite{Nakayama2015, Montero2016}. Weak gravity conjecture asserts that any consistent quantum theory of gravity with $U(1)$ gauge field should contain a light charged particle.
Using the modular bootstrap, we find further supporting evidence for the WGC conjecture.
We find that there must exist at least one state satisfying the relation 
\begin{align}
\D \le \frac{7 c}{48} + \frac{3Q^2}{2c} \ ,
\end{align}
where $Q$ and $\Delta$ are its $U(1)$ charge and scaling dimension respectively. 
This is consistent with the version of the WGC proposed by \cite{Montero2016}. Here we obtained the result without assuming quantization of the $U(1)$ charge.

\paragraph{Outline of the paper}
The outline of this paper is as follows. In section \ref{sec:Prelim} we review aspects of representation theory of superconformal algebra and their characters. Then we formulate the modular bootstrap problem in the form of semi-definite programming. In section \ref{sec:QIndepBounds}, we present the result of numerical analysis on various bounds that does not depend on charges. We also identify CFTs at some special points on the numerical boundary. In section \ref{sec:QBounds}, we refine the analysis by including the effect of charges to study constraints on the charged states. We, in particular, investigate the implication to the weak gravity conjecture. In appendix \ref{app:MDE}, we formulate the modular differential equations for $\G_\theta$ which will be of some interest. We use some part of this result in identifying rational SCFTs living at the boundary in section \ref{sec:QIndepBounds}. Finally, in appendix \ref{app:PMP}, we summarize the polynomial matrix problem that is relevant in the study of charge dependent bounds.

%%%%%%%%%%%%%%%%%%%%%%%%%%%%%%%%%%%%%%%%%%%%%%%%%%%%%%%%%%%%%%%%%%%%%
%%%%%%%%%%%%%%%%%%%%%%%%%%%%%%%%%%%%%%%%%%%%%%%%%%%%%%%%%%%%%%%%%%%%%
\section{Preliminaries} \label{sec:Prelim}
%%%%%%%%%%%%%%%%%%%%%%%%%%%%%%%%%%%%%%%%%%%%%%%%%%%%%%%%%%%%%%%%%%%%%
%%%%%%%%%%%%%%%%%%%%%%%%%%%%%%%%%%%%%%%%%%%%%%%%%%%%%%%%%%%%%%%%%%%%%

%%%%%%%%%%%%%%%%%%%%%%%%%%%%%%%%%%%%%%%%%%%%%%%%%%%%%%%%%%%%%%%%%%%%%
\subsection{Representations of $\mathcal{N}=1$ Superconformal Algebra}
%%%%%%%%%%%%%%%%%%%%%%%%%%%%%%%%%%%%%%%%%%%%%%%%%%%%%%%%%%%%%%%%%%%%%
The two-dimensional $\mathcal{N}=1$ super-Virasoro algebra is generated by the Fourier coefficients of the stress-energy tensor ($L_m$) and the superconformal currents ($G_r$). Their commutation relations are given by
\begin{align}
\begin{split}
[L_m, L_n] &= (m-n) L_{m+n} + \frac{c}{12}(m^3-m) \delta_{m+n,0} \ , \\
[L_m, G_r] &= (\frac{m}{2}-r) G_{m+r} \ ,  \\
\{G_r, G_s\} &= 2 L_{r+s} +\frac{c}{3} (r^2-\frac{1}{4}) \delta_{r+s,0},
\end{split}
\end{align}
where $m, n \in \mathbb{Z}$ and $r,s$ are half-integer valued for the Neveu-Schwarz (NS) sector and integer valued for the Ramond (R) sector. When we have a flavor symmetry, we have additional generators $\mathcal{J}_m$ from the conserved current, that commutes with other super-Virasoro generators. In this article, we will mainly consider SCFT with $U(1)$ flavor symmetry, for simplicity.

An irreducible highest-weight representation of the $\mathcal{N}=1$ super-Virasoro algebra is labeled by the conformal weight $h$. If there is a $U(1)$ symmetry in the theory, the highest-weight state is also labelled by the  $U(1)$ charge $Q$.
When $c \ge \frac{3}{2}$, a generic highest-weight representation (module) is non-degenerate. Therefore the states in the module $\CV_h$ of the NS sector are simply generated by acting super-Virasoro raising operators $L_m (m \ge 1)$ and $G_r (r \ge \frac{1}{2})$ on the highest-weight state $| h \rangle$. The highest-weight state $|h\rangle$ is annihilated by $L_n$ and $G_{n-\half}$ for $n > 0$.
The vacuum module $\CV_0$ always contains null states generated by $L_{-1}$ and $G_{-\half}$ that have to be removed. Then, the super-Virasoro characters for the vacuum and non-vacuum modules can be written as
\begin{align}
\chi_{0}(\tau) &= \Tr_{\CV_0} q^{L_0 - \frac{c}{24}} = q^{-\frac{c}{24}} \prod_{n=2}^{\infty} \frac{1+q^{n-\frac{1}{2}}}{1-q^{n}},\\
\chi_{h}(\tau) &= \Tr_{\CV_{h}} q^{L_0 - \frac{c}{24}} = q^{h-\frac{c}{24}} \prod_{n=1}^{\infty} \frac{1+q^{n-\frac{1}{2}}}{1-q^{n}},
\end{align}
where  $q \equiv e^{2 \pi i \tau}$.
% and $y \equiv e^{2 \pi i z}$.
%(Add characters with z dependence for CFTs with global symmetry!!!)
Using the identity
\begin{align}
\vartheta(0;\tau) = \prod_{n=1}^{\infty} (1-q^n)(1+ q^{n-\frac{1}{2}})^2 = \frac{(\eta(\tau))^5}{(\eta(\frac{\tau}{2}))^2 (\eta(2\tau))^2} ,
\end{align}
the vacuum and non-degenerate super-Virasoro characters can be expressed as follows:
\begin{align}
\chi_{0}(\tau) = q^{-\frac{c}{24}} q^{\frac{1}{16}} (1-q^{\frac{1}{2}}) \frac{\eta(\tau)}{\eta(\frac{\tau}{2}) \eta(2\tau)}, \qquad \chi_{h}(\tau) = q^{h-\frac{c}{24}} q^{\frac{1}{16}}  \frac{\eta(\tau)}{\eta(\frac{\tau}{2}) \eta(2\tau)}.
\end{align}
Here $\eta(\tau)$ is the Dedekind eta-function. This latter expression will turn out to be useful in the semi-definite programming.

The presence of superconformal currents naturally defines a CFT with fermions, thus we should specify the spin structure on the torus.
The fermions can have either periodic or anti-periodic boundary conditions along the two circles of the torus, giving four possible spin structures.
In this article, we focus on the spin structure in which fermions are anti-periodic (NS boundary condition) for both cycles. This is the choice of the spin structure that is dual to the AdS$_3$ gravity. The torus partition function we will be studying is of the form
\begin{align}
  Z_\text{NS}(\t,\bar \t,z, \bar{z}) =
  \text{Tr}_{\CH_{\text{NS}}}\Big[ q^{L_0-\frac{c}{24}}
  \bar q^{\bar L_0 -\frac{c}{24} } y^Q  {\bar y}^{\bar Q} \Big]\ ,
  \label{NSpartition}
\end{align}
where $Q$ and $\bar Q$ denote the left- and right-moving conserved charges of a $U(1)$ flavor symmetry and $y=e^{2\pi i z}$.
This is often referred to as the NS partition function graded by the $U(1)$ charge.
(\ref{NSpartition}) can be decomposed into a sum over $\CN=1$ superconformal families as
\begin{align}
\label{N1_partition}
\begin{split}
  Z(\t,\bar{\t},z,\bar z)  =  Z_\text{vac}(\t,\bar{\t}) + \sum_{h,\bar h}
  d_{h,\bar h}^{Q,\bar{Q}} Z_{h,\bar{h}}^{Q,\bar{Q}}(\t,\bar{\t},z,\bar z)
\end{split}
\end{align}
with
\begin{align}
\begin{split}
  Z_\text{vac}(\t,\bar \t) & = \chi_0(\t) \bar{\chi}_0(\bar \t) ,
  \\
  Z_{h,\bar h}^{Q,\bar Q}(\t,\bar \t,z,\bar z) & =
  \chi_h^Q(\t,z) \bar{\chi}_{\bar h}^{\bar Q} (\bar \t,\bar z) \qquad
  (\chi_h^Q(\t,z)=y^Q \chi_Q(\t)).
\end{split}
\end{align}
Here $d_{h,{\bar{h}}}^{Q,\bar Q}$ is a positive integer counting
the degeneracy of primary state of conformal weight $(h,\bar{h})$ and charge $(Q,\bar Q)$.

%%%%%%%%%%%%%%%%%%%%%%%%%%%%%%%%%%%%%%%%%%%%%%%%%%%%%%%%%%%
\subsection{Representations of $\mathcal{N}=2$ Superconformal Algebra} \label{subsec:N2SCA}
%%%%%%%%%%%%%%%%%%%%%%%%%%%%%%%%%%%%%%%%%%%%%%%%%%%%%%%%%%%
In this section, we briefly review necessary aspects of $\CN=2$ superconformal algebra.
The two-dimensional $\CN=2$ superconformal algebra reads
\begin{align}
\begin{split} \label{N=2 SCA}
  [L_m, L_n] &= (m-n) L_{m+n} + \frac{c}{12} (m^3-m) \delta_{m+n, 0}, \\
  [L_m, G_r^{\pm}] &= \left( \frac{m}{2} - r \right) G_{m+r}^{\pm}, \\
  [L_m, J_n] &= -n J_{m+n}, \\
  [J_m, J_n] &= \frac{c}{3} m \delta_{m+n, 0}, \\
  [J_n, G_r^{\pm}] &= \pm G_{r+n}^{\pm}, \\
  \{G_r^+, G_s^-\} &= 2 L_{r+s} + (r-s) J_{r+s} + \frac{c}{3} \left( r^2 - \frac{1}{4} \right) \delta_{r+s,0},
\end{split}
\end{align}
where $L_m$, $J_m$ and $G^\pm_r$ are the Fourier modes of the stress-energy tensor $T(z)=\sum_m L_m z^{-m-2}$, the $U(1)$ R-current $J(z)=\sum_m J_m z^{-m-1}$ and the supercurrents $G^\pm(z) = \sum_r G^\pm_r z^{-r-3/2}$ respectively. Here $m,n \in \mathbb{Z}$ while $r,s \in \IZ+\half$ for the Neveu-Schwarz (NS) sector or $r, s \in \IZ$ for the Ramond (R) sector.

The $\CN=2$ superconformal algebra has an automorphism, commonly referred to as the spectral flow isomorphism. Namely, the superconformal algebra is invariant under the following transformation:
\begin{align}
\begin{split} \label{Spectral Flow}
  L_n & \rightarrow L_n' = L_n + \eta J_n + \frac{c}{6} \eta^2 \delta_{n,0}\ ,  \\
  J_n & \rightarrow J_n' = J_n + \frac{c}{3} \eta \delta_{n,0},  \\
  G_r^{\pm} & \rightarrow {G_r^\pm}' = G_{r \pm \eta}^{\pm},
\end{split}
\end{align}
where $\eta$ is a continuous parameter. The spectral flow with $\eta=\half$ allows us to relate the NS and R-sector spectrum, which plays a role of spacetime supersymmetry for the string world-sheet theory. Quite often, the spectral flow by an integer $\eta$ can be realized
as a symmetry in an $\CN=2$ SCFT. For a holographic SCFT, the spectral flow by an integer $\eta$ can be understood as the large $U(1)$ gauge transformation in the bulk, hence it becomes a symmetry of the SCFT. For a Calabi-Yau sigma-model, the spectral flow by one unit is also a part of the symmetry algebra of the SCFT. In this case, the holomorphic $d$-form $\O$ of a given CY$_d$ leads to an operator for one unit of the spectral flow in the corresponding SCFT.
In the present work, we \emph{do not} restrict our attention to theories invariant under the spectral flow by one unit and search for the constraints for a generic $\CN=(2,2)$ SCFT.

The unitary highest-weight representations of $\CN=2$ superconformal algebra have been classified in \cite{Boucher1986}. The highest-weight states of $\CN=2$ algebra can be labeled by two quantum numbers $h$ and $Q$ where $h$ is the conformal weight and $Q$ is the $U(1)$ R-charge. The unitarity of the representation constrains the allowed values of $(h,Q)$: Let us assume $c\geq 3$. Then, for the NS sector, there exist two classes of unitary representations.
\begin{figure}[h!]
\begin{center}
\includegraphics[width=.7\textwidth]{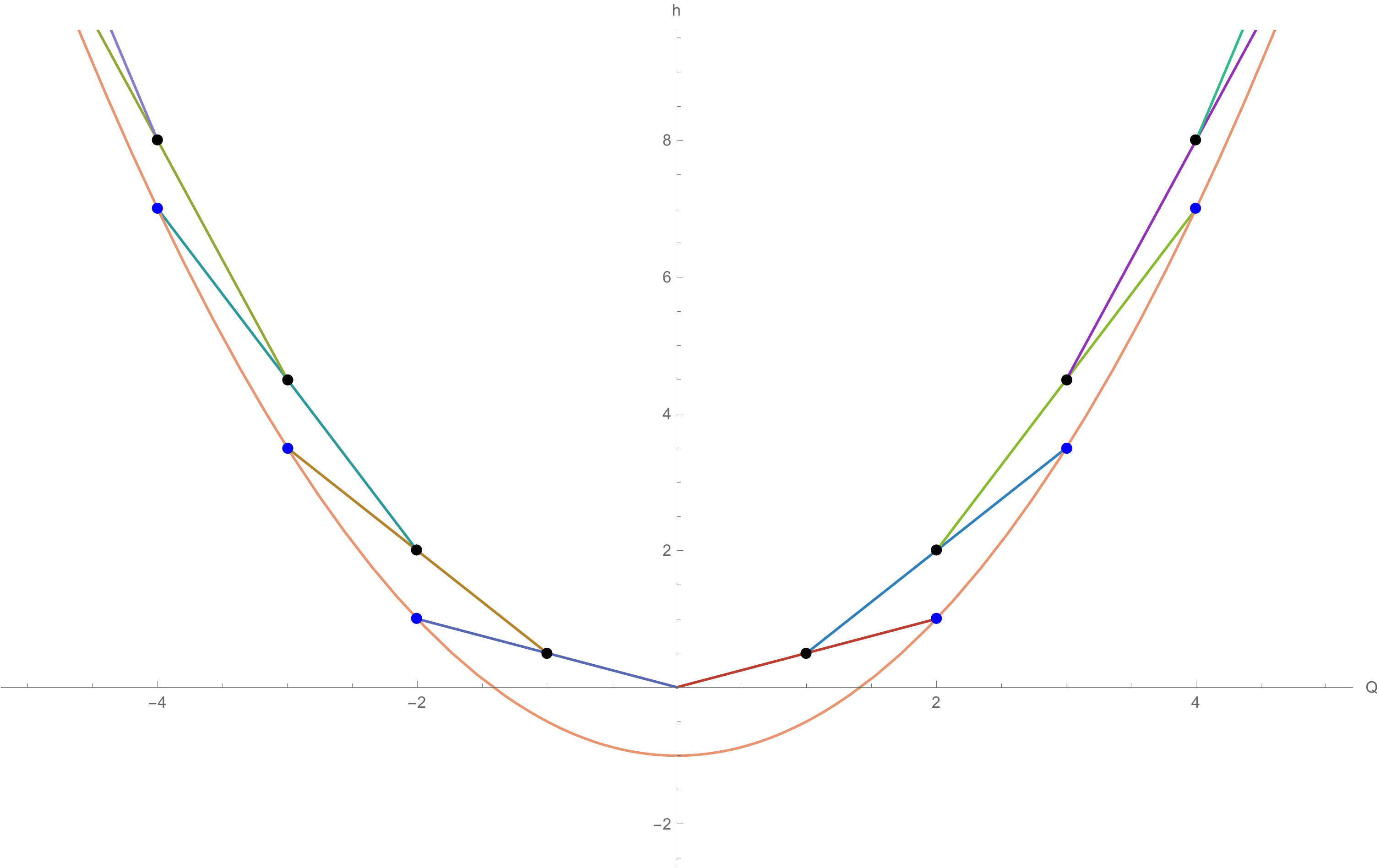}
\end{center}
\caption{Domain of the unitary irreducible representation of $c=6$, $\mathcal{N}=2$ super-Virasoro algebra. The orange curve passes through the points that satisfy a relation $ 2 h - Q^2 + 2 = 0$.}
\label{N2Rep}
\end{figure}
\begin{itemize}[leftmargin=0.8cm]
\item[(1)] {\it Massive representation}: The massive representations must lie inside the closed convex region defined by the chords joining the successive points
\begin{align}
  \big(h,Q\big) = \big(  (\frac c3 - 1) \frac{s^2}{2} , (\frac c3-1)s \big) \qquad
  \big(s\in \mathbb{Z}\big) \ ,
\end{align}
as depicted in figure \ref{N2Rep}. The $\CN=2$ character of the massive representation is
\begin{align}
  \text{Ch}_h^Q(\t,z) = q^{h - \frac{c}{24}} y^Q
  \prod_{n=1}^\infty \frac{(1+q^{n-\frac{1}{2}} y)(1 + q^{n-\frac{1}{2}} y^{-1})}{(1-q^n)^2}\ .
\end{align}

\item[(2)] {\it Massless representation}:
The massless representations lie on the line segments
\begin{align}
  2 h - 2 r Q +  \big(\frac c3 - 1\big) \big( r^2 - \frac14\big) = 0 \qquad
  \big(r\in \frac12 + \mathbb{Z}\big)  ,
  \label{massless}
\end{align}
where
\begin{align}
  \big(\frac c3 - 1 \big) \big(|r|-\frac12\big) \leq \text{sign}(r) Q \leq
  \big(\frac c3-1\big)\big(|r|+\frac12\big) + 1  \ ,
\end{align}
Note that the highest weight state of the massless representation on a line segment with $r$ can be annihilated by $G_{-|r|}^{\text{sign}(r)}$. The $\CN=2$ character of the massless representations is given by
\begin{align}
  \chi_h^r(\t,z) = q^{h-\frac{c}{24}} y^{Q(h,r)} \frac{1}{1+ q^{|r|} y^{\text{sign}(r)}}
  \prod_{n=1}^{\infty} \frac{(1+q^{n-\frac12} y)( 1 + q^{n-\frac12} y^{-1})}{(1-q^n)^2}
\end{align}
with $Q(h,r)$ satisfying (\ref{massless}) unless $\big(h,Q(h,r)\big)$ additionally satisfies
\begin{align}
  2 \big( \frac c3 - 1\big) h - Q(h,r)^2 - \frac14\big( \frac c3 - 1\big)^2 + \frac14 \Big(
  \big( \frac c3 - 1\big) + 2\Big)^2 = 0\ .
\end{align}
For such special values of $\big(h,Q\big)$, i.e.,
\begin{align}
  h(r) & =\frac{c}{6} \eta(r)^2 - \frac12\big(|r|-\frac12\big)^2, \nonumber \\
  Q(r) & = \text{sign}(r)\Big[ \big(\frac c3-1\big)\big(|r|+\frac12\big) + 1 \Big],
  \label{specialvalues}
\end{align}
the corresponding characters can be obtained by the spectral flow images of the $\CN=2$ vacuum character $\chi_\text{vac}(\t,z)$ given by
\begin{align}
  \chi_\text{vac}(\t,z) = q^{-\frac{c}{24}} \frac{(1-q)}{(1+q^{\frac12}y )(1+q^{\frac12}y^{-1})}
  \prod_{n=1}^{\infty} \frac{(1+q^{n-\frac12} y)( 1 + q^{n-\frac12} y^{-1})}{(1-q^n)^2} \ .
\end{align}
More precisely, the $\CN=2$ characters with (\ref{specialvalues}) are
\begin{align}
  \chi^r_h(\t,z) = q^{\frac{c}{6}\eta(r)^2} y^{\frac{c}{3}\eta(r)}
  \chi_\text{vac}\big(\t,z+\eta(r) \t\big)
\end{align}
with
\begin{align}
  \text{sign}(r) \eta(r) = |r| +  \frac{1}{2}\ .
\end{align}
\end{itemize}
Since the $\CN=2$ superconformal algebra has the $U(1)$ R-current with level $k=c/3$,
it is natural to consider the NS partition function graded by the $U(1)$ charge,
\begin{align}
  Z(\t,\bar \t,z,\bar{z}) = \text{Tr}_{\CH_\text{NS}}\Big[ q^{L_0-\frac{c}{24}}
  {\bar q}^{{\bar L}_0 - \frac{c}{24}} y^{J_0} {\bar y}^{{\bar J}_0}\Big]\ ,
\label{tp02}
\end{align}
where $J_0$ and ${\bar J}_0$ are the left and right-moving $U(1)$ charges.
The partition function (\ref{tp02})
can be expressed as a sum of products of the $\CN=2$
superconformal characters. It will be useful to decompose (\ref{tp02})
into the four different sectors
\begin{align}
\label{N2_partition}
  \hspace*{-0.1cm}
  Z(\tau, \bar{\tau}, z,\bar z) &= Z_0(\tau, \bar{\tau}, z, \bar z) +
  Z_{\frac 12} (\tau, \bar{\tau}, z, \bar z) +
  Z_{\frac 14}(\tau, \bar{\tau}, z, \bar z)  +
  Z_\text{non-BPS} (\tau, \bar{\tau}, z, \bar z).
\end{align}
Let us discuss each term separately: $Z_0$ contains the $\CN=2$ descendants of the vacuum
\begin{align}
  Z_0(\t,\bar \t,z,\bar z) =  \chi_\text{vac}(\t,z) {\bar \chi}_\text{vac}(\bar \t,z) \ ,
\end{align}
while $Z_{1/2}$ has the $1/2$-BPS contributions where both left and right-movers
are in massless representations
\begin{align}
  Z_{\frac12}(\t,\bar \t,z,\bar z) = \sum_{h,\bar h} \sum_{r,\bar r} d_{h,\bar h}^{r,\bar r}
  \cdot \chi_h^r(\t,z) {\bar \chi}_{\bar h}^{\bar r}(\bar \t,\bar z)\ .
\end{align}
The $1/4$-BPS states of which either left or right-mover is in a massless
representation are contained in $Z_{1/4}$,
\begin{align}
  \hspace*{-0.1cm}
   Z_{\frac14}(\t,\bar \t,z,\bar z) & = \sum_{h,\bar h}
   \bigg[
   \sum_{r, \bar Q} d_{h,\bar h}^{r,\bar Q}
   \cdot \chi_h^r(\t,z) \overline{\text{Ch}}_{\bar h}^{{\bar Q}} (\bar \t,\bar z) +
   \sum_{Q, \bar r} d_{h,\bar h}^{Q,\bar r}
   \cdot \text{Ch}_h^Q(\t,z) \overline{\chi}_{\bar h}^{{\bar r}} (\bar \t,\bar z) \bigg],
\end{align}
and finally $Z_\text{non-BPS}$ contains the massive representations for
both left and right-movers,
\begin{align}
  Z_\text{non-BPS}(\t,\bar \t,z,\bar z) =
  \sum_{h,\bar h} \sum_{Q,\bar Q} d_{h,\bar h}^{Q,\bar Q}
  \cdot
  \text{Ch}_h^Q(\t,z) \overline{\text{Ch}}_{\bar h}^{\bar Q}(\bar \t, \bar z) \ .
\end{align}
For the parity-preserving $\CN=(2,2)$ SCFTs, degeneracies are required to satisfy
\begin{align}
  d_{h,\bar h}^{r,\bar r}  = d_{\bar h, h}^{\bar r, r}, \quad
  d_{h,\bar h}^{r, \bar Q}  = d_{\bar h,h}^{Q,\bar r}, \quad
  d_{h,\bar h}^{Q,\bar Q}  = d_{\bar h, h}^{\bar Q, Q}.
\end{align}
%

%%%%%%%%%%%%%%%%%%%%%%%%%%%%%%%%%%%%%%%%%%%%%%%%%%%%%%%%%%%
\subsection{Semi-Definite Programming}
%%%%%%%%%%%%%%%%%%%%%%%%%%%%%%%%%%%%%%%%%%%%%%%%%%%%%%%%%%%

For SCFTs having no large diffeomorphism anomaly,
the NS partition function graded by the $U(1)$ charge
has to be invariant under $T^2$ ($T: \t \to \t+1$)
transformation while it transforms under the $S$ modular
transformation as
\begin{align}
  Z\left(-\frac{1}{\t}, - \frac{1}{\bar \t}, \frac{z}{\t}, \frac{\bar z}{\bar \t} \right)
  = e^{ i \pi k  (\frac{z^2}{\tau} - \frac{{\bar z}^2}{\bar \t})} Z\big(\t,\bar \t,z,\bar z \big),
  \label{S01}
\end{align}
where $k$ denotes the level of the $U(1)$ current.
The invariance under $T^2$ simply requires that each spin
has to be either half-integral or integral.
On the other hand, one can make use of the character decomposition
of the partition function (\ref{N1_partition}) or (\ref{N2_partition}) to reformulate
(\ref{S01}) as the so-called modular bootstrap equation,
\begin{align}
  0 = \CZ_\text{vac}(\t,\bar\t,z,\bar z) + \sum_{h,\bar h} \sum_{Q,\bar Q}
  d_{h,\bar h}^{Q,\bar Q} \CZ_{h,\bar h}^{Q,\bar Q}(\t,\bar \t,z, \bar z) \ ,
  \label{N1bootstrap}
\end{align}
with
\begin{align}
\begin{split}
  \CZ_\text{vac}(\t,\bar\t,z,\bar z) & = \chi_0\big(-\frac{1}{\t}\big)
  {\bar \chi}_0\big( -\frac{1}{\bar \t}\big)
  - e^{ i \pi k  ( \frac{z^2}{\tau} - \frac{{\bar z}^2}{\bar \t} )}
  \chi_0\big(\t\big) {\bar \chi}_0\big(\bar \t\big) \ ,
  \\
  \CZ_{h,\bar h}^{Q,\bar Q}(\t,\bar \t,z, \bar z) & =
  \chi_h^Q \big(-\frac{1}{\t},\frac{z}{\t}\big)
  {\bar \chi}_{\bar h}^{\bar Q} \big(-\frac{1}{\bar \t},\frac{\bar z}{\bar \t}\big)
  - e^{ i \pi k  ( \frac{z^2}{\tau} - \frac{{\bar z}^2}{\bar \t} )}
  \chi_h^Q \big(\t,z\big) {\bar \chi}_{\bar h}^{\bar Q} \big(\bar \t,\bar z\big) \ ,
\end{split}
\end{align}
for $\CN=(1,1)$ superconformal theories, or
\begin{align}
\begin{split}
   \CZ_\text{vac}(\t,\bar\t,z,\bar z) & +
  \sum_{h,\bar h} \Bigg[
   \sum_{r,\bar r} d_{h,\bar h}^{r,\bar r}
  \CZ_{h,\bar h}^{r,\bar r}(\t,\bar \t,z, \bar z) +
  \sum_{r,\bar Q} d_{h,\bar h}^{r,\bar Q}
  \CZ_{h,\bar h}^{r,\bar Q}(\t,\bar \t,z, \bar z)
   \\   & ~~~ +
  \sum_{Q,\bar r} d_{h,\bar h}^{Q,\bar r}
  \CZ_{h,\bar h}^{Q,\bar r}(\t,\bar \t,z, \bar z) +
  \sum_{Q,\bar Q} d_{h,\bar h}^{Q,\bar Q}
  \CZ_{h,\bar h}^{Q,\bar Q}(\t,\bar \t,z, \bar z) \Bigg] = 0,
  \label{N2bootstrap}
\end{split}
\end{align}
with
\begin{align}
\begin{split}
  \CZ_{h,\bar h}^{r,\bar r}(\t,\bar \t,z, \bar z) & =
  \chi_h^r \big(-\frac{1}{\t},\frac{z}{\t}\big)
  {\bar \chi}_{\bar h}^{\bar r} \big(-\frac{1}{\bar \t},\frac{\bar z}{\bar \t}\big)
  - e^{ i \pi k  ( \frac{z^2}{\tau} - \frac{{\bar z}^2}{\bar \t} )}
  \chi_h^r \big(\t,z\big) {\bar \chi}_{\bar h}^{\bar r} \big(\bar \t,\bar z\big),
   \\
  \CZ_{h,\bar h}^{r,\bar Q}(\t,\bar \t,z, \bar z) & =
  \chi_h^r \big(-\frac{1}{\t},\frac{z}{\t}\big)
  \overline{\text{Ch}}_{\bar h}^{\bar Q} \big(-\frac{1}{\bar \t},\frac{\bar z}{\bar \t}\big)
  - e^{ i \pi k  ( \frac{z^2}{\tau} - \frac{{\bar z}^2}{\bar \t} )}
  \chi_h^r \big(\t,z\big) \overline{\text{Ch}}_{\bar h}^{\bar Q} \big(\bar \t,\bar z\big),
   \\
  \CZ_{h,\bar h}^{Q,\bar r}(\t,\bar \t,z, \bar z) & =
  \text{Ch}_h^Q \big(-\frac{1}{\t},\frac{z}{\t}\big)
  \bar{\chi}_{\bar h}^{\bar r} \big(-\frac{1}{\bar \t},\frac{\bar z}{\bar \t}\big)
  - e^{ i \pi k  ( \frac{z^2}{\tau} - \frac{{\bar z}^2}{\bar \t} )}
  \text{Ch}_h^Q\big(\t,z\big)\bar{\chi}_{\bar h}^{\bar r}  \big(\bar \t,\bar z\big),
  \\
  \CZ_{h,\bar h}^{Q,\bar Q}(\t,\bar \t,z, \bar z) & =
  \text{Ch}_h^Q \big(-\frac{1}{\t},\frac{z}{\t}\big)
  \overline{\text{Ch}}_{\bar h}^{\bar Q} \big(-\frac{1}{\bar \t},\frac{\bar z}{\bar \t}\big)
  - e^{ i \pi k  ( \frac{z^2}{\tau} - \frac{{\bar z}^2}{\bar \t} )}
  \text{Ch}_h^Q\big(\t,z\big) \overline{\text{Ch}}_{\bar h}^{\bar Q} \big(\bar \t,\bar z\big) ,
\end{split}
\end{align}
for $\CN=(2,2)$ superconformal theories.

One can utilize the numerical method of semi-definite programming (SDP)
to explore the consequence of modular bootstrap equation. To see this,
let us describe how to apply the SDP method to (\ref{N1bootstrap})
for the sake of conciseness.
It is straightforward to apply the method
to (\ref{N2bootstrap}). One proceeds by making a hypothesis
on a CFT spectrum, and search for a linear functional $\a$
such that the conditions below are satisfied
\begin{align}
  \a\Big[ \CZ_\text{vac}(\t,\bar \t,z,\bar z) \Big]  = 0, \qquad
  \a\Big[ \CZ_{h,\bar h}^{Q,\bar Q} (\t,\bar \t,z,\bar z)\Big] \geq 0.
\end{align}
If such an $\a$ exists, the non-negative integers
$d_{h,\bar h}^{Q,\bar Q}$ then lead to a contradiction that
(\ref{N1bootstrap}) cannot be satisfied. In other words,
the hypothetical spectrum is ruled out. 
We solve the semi-definite programming problem using the \verb"SDPB" package \cite{Simmons-Duffin:2015qma}. For more
detailed reviews, see \cite{Poland:2018epd}.

An explicit form of the linear functional $\a$
will be presented in the next sections where
various hypotheses under investigation are specified.

%%%%%%%%%%%%%%%%%%%%%%%%%%%%%%%%%%%%%%%%%%%%%%%%%%%%%%%%%%%%%%%%%%%%%
%%%%%%%%%%%%%%%%%%%%%%%%%%%%%%%%%%%%%%%%%%%%%%%%%%%%%%%%%%%%%%%%%%%%%
\section{Charge Independent Bounds} \label{sec:QIndepBounds}
%%%%%%%%%%%%%%%%%%%%%%%%%%%%%%%%%%%%%%%%%%%%%%%%%%%%%%%%%%%%%%%%%%%%%
%%%%%%%%%%%%%%%%%%%%%%%%%%%%%%%%%%%%%%%%%%%%%%%%%%%%%%%%%%%%%%%%%%%%%

In this section, we investigate the modular constraint using the semi-definite programming, without assuming flavor symmetry in the two-dimensional SCFTs. In other words, we turn off the chemical potential for the flavor symmetry. We will also turn off the chemical potential for the R-symmetry in the $\CN=2$ super-Virasoro characters.

In order to make the numerical analysis efficient, we solve a
little different but essentially the same SDP problem with
the NS partition function and characters multiplied
certain $\G_\th$-invariant factors. For examples,
we work with the reduced partition function and characters
given by
\begin{align}
\begin{split}
\widehat{\chi}_{h}(\t,\bar\t) & =
\tau^{\frac{1}{4}} \frac{\eta(\frac{\tau}{2}) \eta(2\tau)}{\eta(\tau)}
\chi_{h}(\t,\bar \t)
\\
\widehat{Z}(\tau, \bar{\tau}) &= |\tau|^{\frac{1}{2}} \frac{|\eta(\frac{\tau}{2})|^2 |\eta(2\tau)|^2}{|\eta(\tau)|^2}   Z(\tau, \bar{\tau}),
\end{split}
\end{align}
for $\mathcal{N}=(1, 1)$ SCFTs and
\begin{align}
\widehat{\chi}_{h}(\t,\bar\t) & =
\tau \frac{\eta^3(\tau)}{\vartheta(0;\tau)}
\chi_{h}(\t,\bar \t)
\\
\widehat{Z}(\tau, \bar{\tau}) & = |\tau|^2 \frac{ |\eta(\tau)|^6}{|\vartheta(0;\tau)|^2} Z(\tau, \bar{\tau}) \ ,
\end{align}
for $\mathcal{N}=(2, 2)$ SCFTs. Here $\eta(\tau)$ is the usual Dedekind eta function, and
the Jacobi theta function is defined as
\begin{align}
\vartheta(z; \tau) = \sum_{n \in \IZ} \exp \left( \pi i n^2 \tau + 2\pi i n z \right) = \sum_{n \in \IZ} q^{n^2} y^n \ .
\end{align}

A universal bound on the lowest primary operator in two-dimensional CFTs is obtained by the spin-independent linear functional, as pioneered by Hellerman and Friedan-Keller \cite{Hellerman2011c, Friedan2013}. A refined Hellerman-Friedan-Keller (HFK) bounds of SCFTs will be presented in this section. In addition, we find the upper bounds on the lowest primaries using the spin-dependent linear functional, assuming the scalar gap or twist gap in the spectrum.

%%%%%%%%%%%%%%%%%%%%%%%%%%%%%%%%%%%%%%%%%%%%%%%%%%%%%%%%%%%%
\subsection{Spin-independent bounds on the gap}
%%%%%%%%%%%%%%%%%%%%%%%%%%%%%%%%%%%%%%%%%%%%%%%%%%%%%%%%%%%%

The modular invariance requires that
any unitary two-dimensional CFT with $c>1$ to
contain an infinite number of Virasoro primary operators. In the holographic dual,
it implies the existence of massive states in addition to
the boundary gravitons in AdS$_3$.

It was not however obvious whether there is a universal energy scale
at which such massive states begin to appear.
In a beautiful paper by Hellerman \cite{Hellerman2011c}, it was shown that such an energy scale indeed exists
in quantum gravity with negative cosmological constant.
The derivation relies only on the unitarity and modular invariance of the boundary CFT.
More precisely, it was shown that
the modular invariance leads to a universal upper bound,
insensitive to spin, on the scaling dimension of
the lowest-lying primary state. The bound is often referred to as
the Hellerman-Friedan-Keller (HFK) bound, initially given by
$\D_\text{HFK} \leq \frac c6 + \CO(1)$ \cite{Hellerman2011c}
and further improved to
\begin{align}
 \D_\text{HFK} \leq \frac{c}{9} + \CO(1)
 \label{HFK}
\end{align}
in the large central charge limit \cite{Collier2016}.
Using the AdS$_3$/CFT$_2$ correspondence (more specifically the Brown-Henneaux relation \cite{Brown:1986nw}),
the HFK bound (\ref{HFK}) proves that any theory of
quantum gravity must contain a massive state
no heavier than $1/(6G_\text{N})$. Given that the
(semi-classical) mass of the lightest BTZ black hole
is $1/(8G_\text{N})$, the result suggests that the
lightest massive state can be identified as
the black hole state near the threshold.

We would like to examine the constraints
on the lightest massive state in
supersymmetric quantum gravity in AdS$_3$.
To this end, let us consider the NS partition function (\ref{NSpartition})
in the spin and charge blind limit, i.e., $\t=-\bar \t=i\b$ and $z=\bar z=0$,
\begin{align}
  Z_\text{NS}(\b) = \text{Tr}_{\CH_\text{NS}}\Big[
  e^{-2\pi \b (L_0+\bar L_0 -\frac{c}{12})} \Big],
\end{align}
and take the linear functional $\a$ of the form
\begin{align}
  \a_1= \left. \sum_{l=0}^N \alpha_{l} \left( \b \frac{\partial}{\partial \b} \right)^{2l+1} \right|_{\b=1}\ .
\end{align}
We then proceed by assuming a gap $\D_\ast$ in the spectrum.
If one finds a linear functional $\a$ satisfying
\begin{align}
  \a_1\Big[  \CZ_\text{vac} (\b) \Big]  =1 \ ,
  \qquad
  \a_1 \Big[ \CZ_\D(\b)\Big]  \geq 0 \
  \text{ for } \ \D \geq \D_\ast,
  \label{linear functional N1}
\end{align}
then such a spectrum is ruled-out so that $\Delta_*$ provides an upper limit on the (spin-independent) gap.
Here $\CZ_{\D}(\b)$ denotes
\begin{align}
  \CZ_{\D}(\b) = \CZ_{\frac{\D+s}{2},\frac{\D-s}{2}}^{Q,\bar Q}(\t=-\bar \t =i\b,z=\bar z=0)
\end{align}
with any spin $s$ and any charge $Q$ for the $\CN=(1,1)$ SCFTs. In the case of $\CN=(2, 2)$ SCFTs, $\CZ_{\D}(\b)$ denotes any of the non-vacuum contributions
\begin{align}
\begin{split}
  \CZ_{\D}(\b) & =
  \CZ_{\frac{\D+s}{2},\frac{\D-s}{2}}^{r,\bar r}(\t=-\bar \t =i\b,z=\bar z=0),
  \\
  & \mbox{or} \ = \CZ_{\frac{\D+s}{2},\frac{\D-s}{2}}^{r,\bar Q}(\t=-\bar \t =i\b,z=\bar z=0),
   \\
  & \mbox{or} \ = \CZ_{\frac{\D+s}{2},\frac{\D-s}{2}}^{Q,\bar r}(\t=-\bar \t =i\b,z=\bar z=0),
  \\
  & \mbox{or} \ = \CZ_{\frac{\D+s}{2},\frac{\D-s}{2}}^{Q,\bar Q}(\t=-\bar \t =i\b,z=\bar z=0) ,
\end{split}
\end{align}
with arbitrary $s$, $r$ and $Q$ for the $\CN=(2,2)$ SCFTs. We set all four sectors to be the same because we are using the linear functional that is insensitive to the spin and charges.

\begin{figure}[h!]
\begin{center}\includegraphics[width=.74\textwidth]{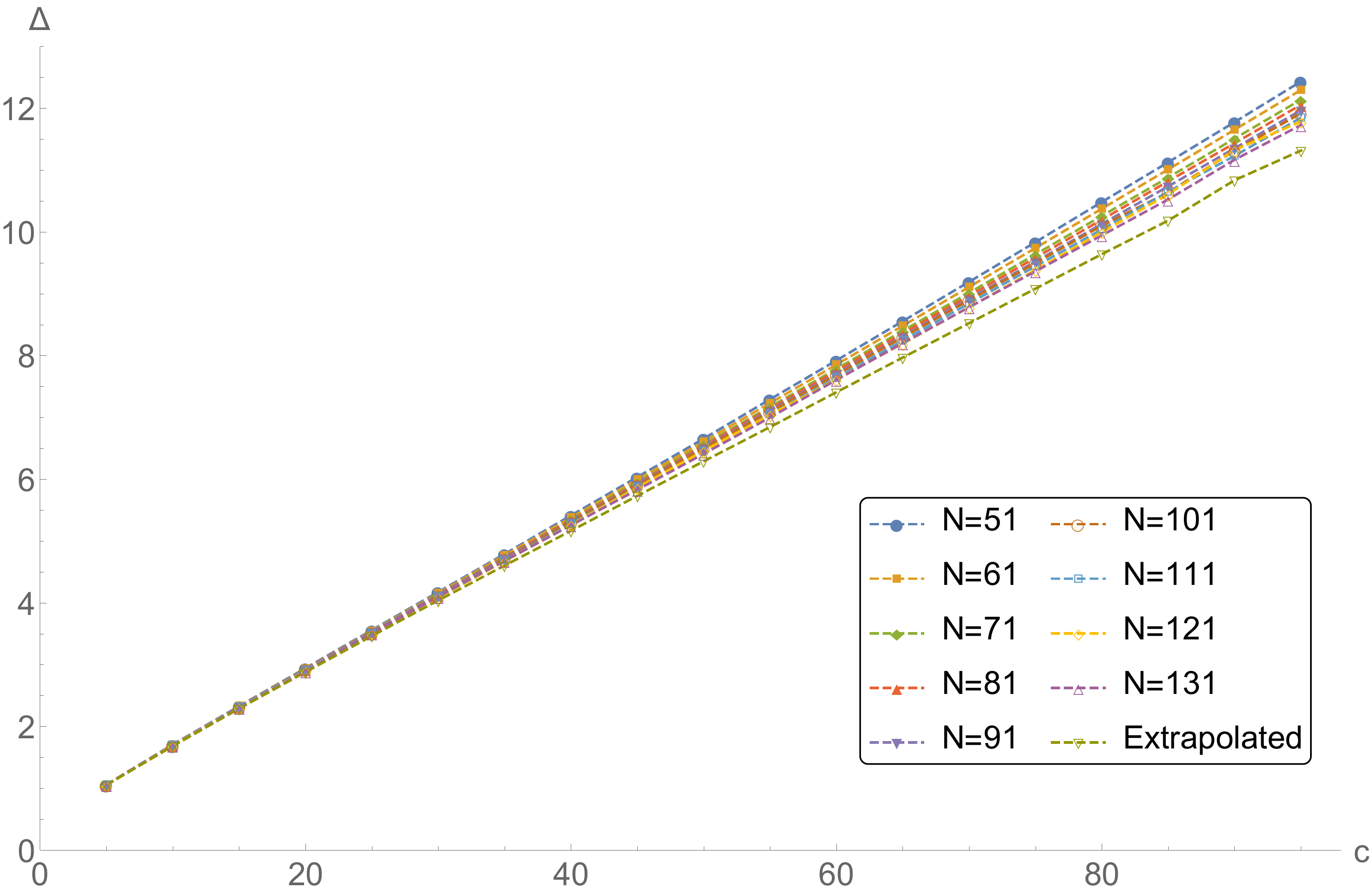} \\
\includegraphics[width=.74\textwidth]{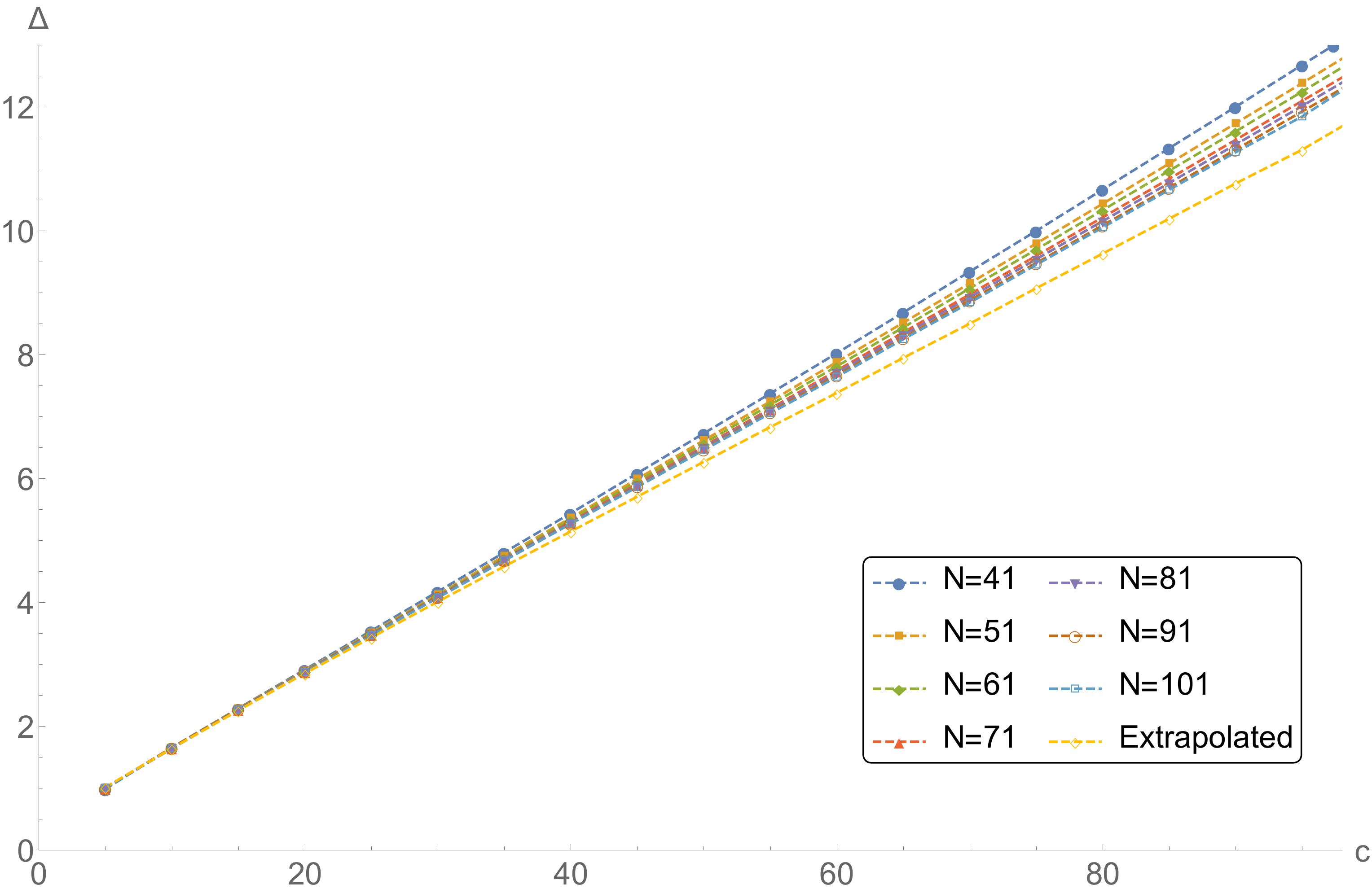}
\end{center}
\caption{The spin-independent bound on the lowest primary for the $\CN=(1,1)$ (upper) and $\CN=(2,2)$ (lower) SCFTs. The upper bounds decrease as we increase the parameter $N$ for the linear functional $\a$ from 51 to 131.
The stringiest bounds represent the extrapolated result at each $c$, to $N \rightarrow \infty$.
}
\label{N=1HFK}
\end{figure}
The HFK-type bounds for the $\CN=(1,1)$ and $\CN=(2,2)$ SCFTs we obtain are depicted in figure \ref{N=1HFK}. We find the numerical upper bound on the $\D_\ast$ as a function of the central charge, where the dimension of the linear functional $N$ ranging from $51$ to $131$. We also extrapolated them to $N\to \infty$ at each value of $c$ using the method employed in \cite{Collier2016}. In the large central charge limit, we observed that the numerical value of the asymptotic slopes $d\D_\text{HFK}/dc$ approach $0.113366$ and $0.113656$ for the $\mathcal{N}=(1,1)$ and $\mathcal{N}=(2,2)$ SCFTs respectively. Based on the numerical results, we propose that the supersymmetric HFK-bounds are again given by
\begin{align}
 \Delta_\text{HFK} \le \frac{c}{9}  + \CO(1) \ .
\end{align}
We do not find any improvements on the numerical bounds for the gap compared to the result of \cite{Collier2016} without supersymmetry. This is not surprising, since the threshold mass for the BTZ black hole is still $1/(8G_N)$ in AdS$_3$ supergravity.

%%%%%%%%%%%%%%%%%%%%%%%%%%%%%%%%%%%%%%%%%%%%%%%%%%%%%%%%%%%
\subsection{Spin-dependent Bounds for $\mathcal{N}=(1,1)$ SCFT}
%%%%%%%%%%%%%%%%%%%%%%%%%%%%%%%%%%%%%%%%%%%%%%%%%%%%%%%%%%%

In order to refine various bounds on the spectrum discriminating
primaries of different spins, let us consider the partition (\ref{NSpartition})
function blind to charge but aware of spin, i.e., $z=\bar z=0$,
\begin{align}
  Z_\text{NS}(\t,\bar \t) = \text{Tr}_{\CH_\text{NS}}\Big[
  e^{2\pi i\t (L_0 -\frac{c}{24})}e^{-2\pi i \bar \t (\bar L_0 -\frac{c}{24})} \Big],
\end{align}
and take the linear functional $\a_2$ of the form
\begin{align}
\label{LF_type2}
  \a_2 = \sum_{p=0}^N \sum_{m+n=2p+1} \tilde \a_{m,n}
  \left. \left(  \t \frac{\partial}{\partial \t} \right)^m
  \left( \bar \t \frac{\partial}{\partial \bar \t} \right)^n
  \right|_{\t=i,\bar \t=-i}.
\end{align}
One can prove that there is a gap $\Delta_*(j)$ in each spin-$j$ spectrum other than conserved currents ($j \neq 1$)
when a linear functional $\a_2$ satisfying the following conditions can be found \cite{BLS}:
\begin{align}
  \a_2\Big[  \CZ_\text{vac} (\t,\bar \t) \Big]  & =1,
  \nonumber \\
  \a_2 \Big[ \CZ_{\frac{\D\pm j}{2},\frac{\D\mp j}{2}}(\t,\bar \t)\Big]  & \geq 0 \
  \text{ for } \ \D \geq \D_\ast(j),
  \label{linear functional N1_2}
\end{align}
for a half-integer spin $j$.
For $\CN=(1,1)$ SCFTs, we simply take
\begin{align}
  \CZ_{\frac{\D+j}{2},\frac{\D-j}{2}}(\t,\bar \t) =
  \CZ_{\frac{\D+j}{2},\frac{\D-j}{2}}^{Q,\bar Q}(\t,\bar \t,z=\bar z=0)\ ,
\end{align}
and ignore any charge dependence in this section.
In the present work, we examine two kinds of the spin-dependent
gap $\D_\ast(j)$, namely the scalar gap and the twist gap.

\paragraph{Twist Gap}
\begin{figure}[h]
\begin{center}\includegraphics[width=.85\textwidth]{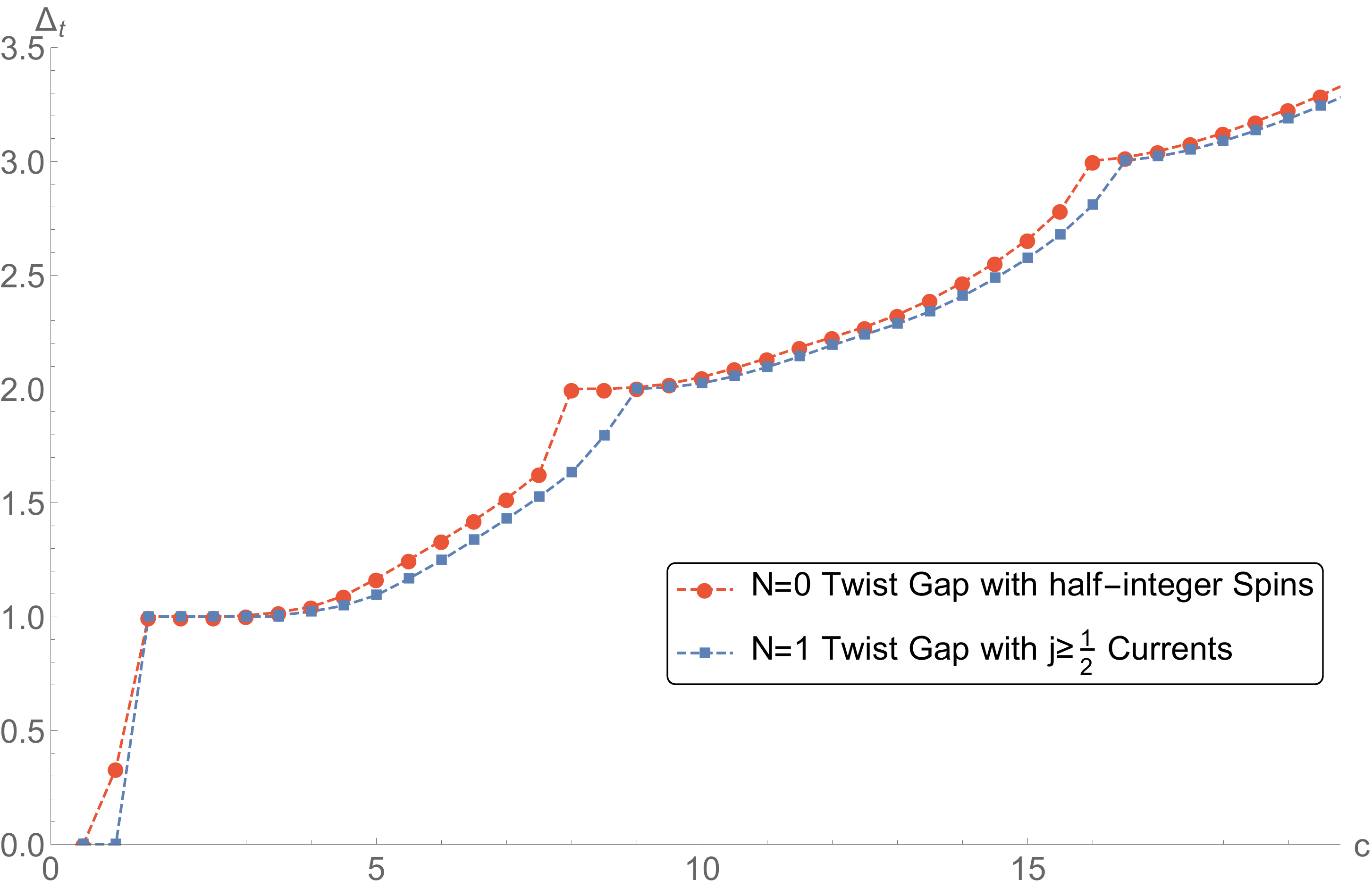}
\end{center}
\caption{Numerical upper bounds on the twist gap for the $\mathcal{N}=0, 1$ SCFTs with the
conserved currents of $j \ge \frac{1}{2}$.}
\label{fig:N0vsN1}
\end{figure}
Let us assume that the spectrum is constrained
to have primaries with conformal dimension with $\D\geq j + \D_t$, i.e.,
\begin{align}
  \D_\ast(j) = j + \D_t.
\end{align}
$\D_t$ is often referred to as the twist gap. Here we do not impose the gap condition to the conserved currents (that always satisfy $\Delta = j$).

\begin{figure}[h]
\begin{center}\includegraphics[width=.85\textwidth]{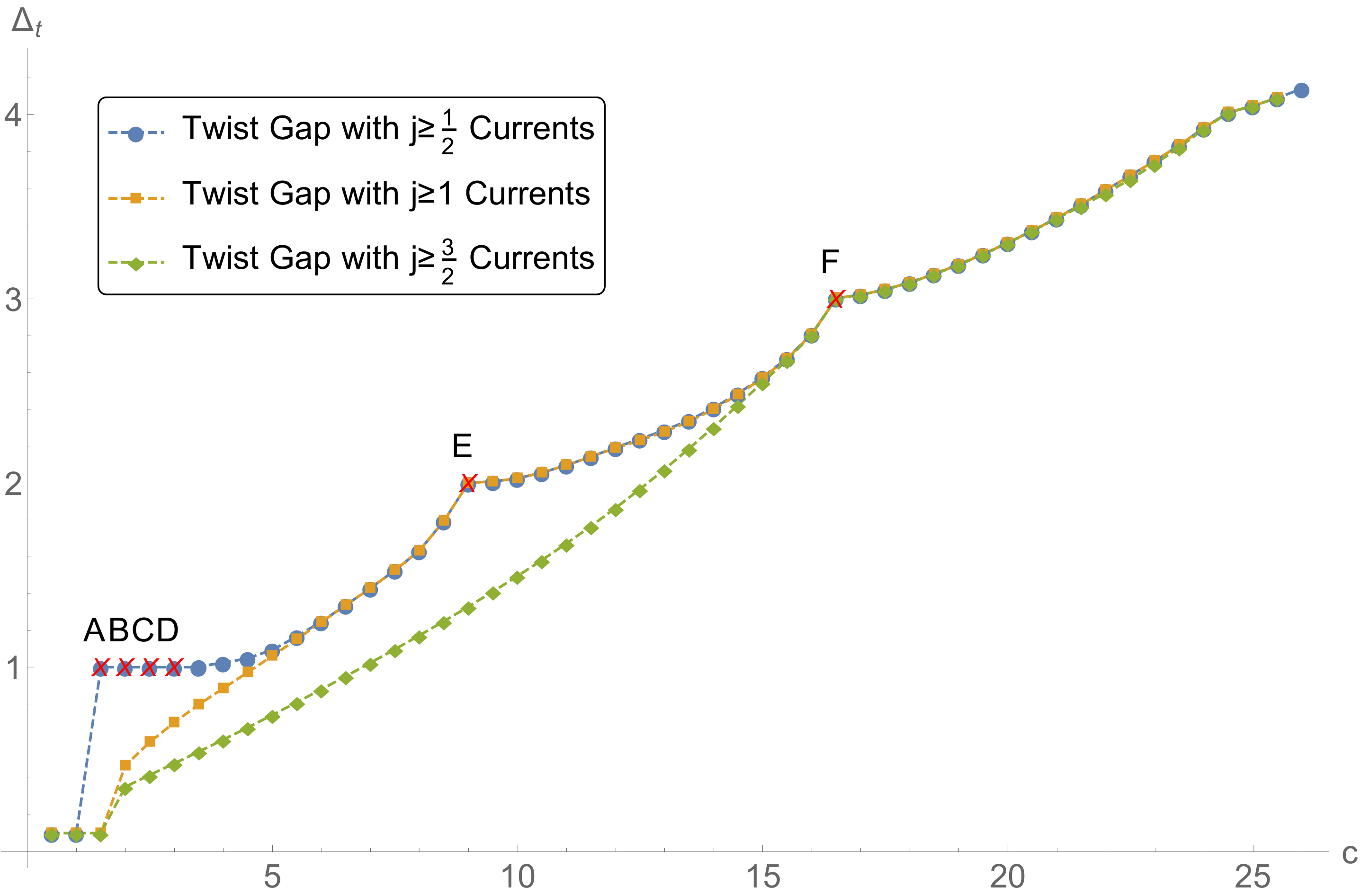}
\end{center}
\caption{Numerical upper bounds on the twist gap for the $\mathcal{N}=1$ SCFTs with imposing the
conserved currents of $j \ge \frac{1}{2}$, $j \ge 1$ and $j \ge \frac{3}{2}$.}
\label{Super_Virasoro_N1}
\end{figure}
We solve the semi-definite programming problem under the above twist gap assumptions.
Figure \ref{Super_Virasoro_N1} summarizes the numerical upper bounds on
$\D_t$ we obtain. We also present the bounds without assuming supersymmetry in Figure \ref{fig:N0vsN1} for comparison.
As we can see, the bounds become stronger for the supersymmetric case.

Notice that there are several kinks, peaks, and even plateau.
It is possible to utilize the extremal functional method (EFM) to read off
the entire spectrum of a hypothetical CFT saturating the numerical
upper bound on the twist gap. We refer the details of the EFM
method to \cite{El-Showk2012}. Now let us give an interpretation of the special points we obtain via numerical computation.

\begin{figure}[h]
\begin{center}
\includegraphics[width=.45\textwidth]{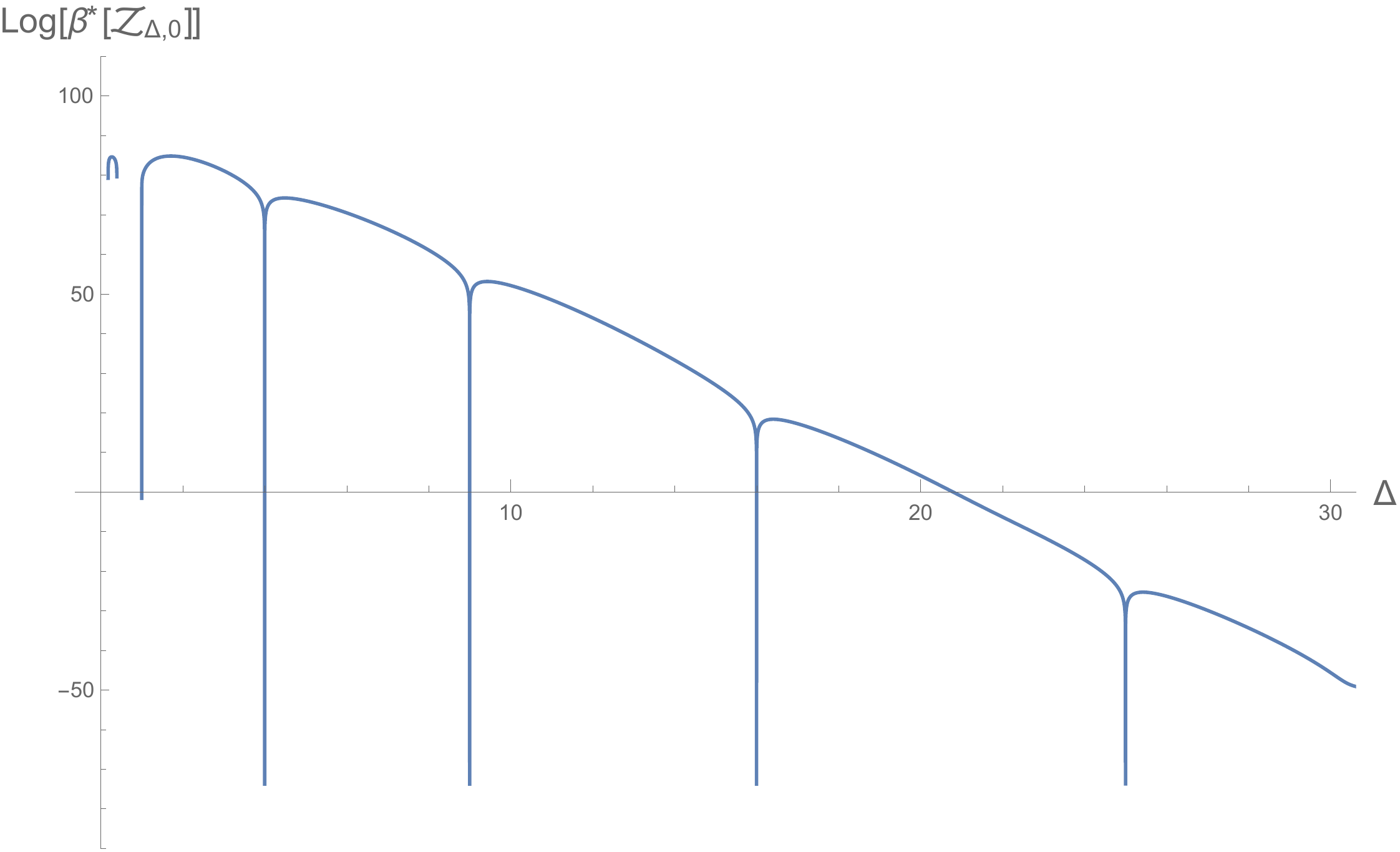} \quad
\includegraphics[width=.45\textwidth]{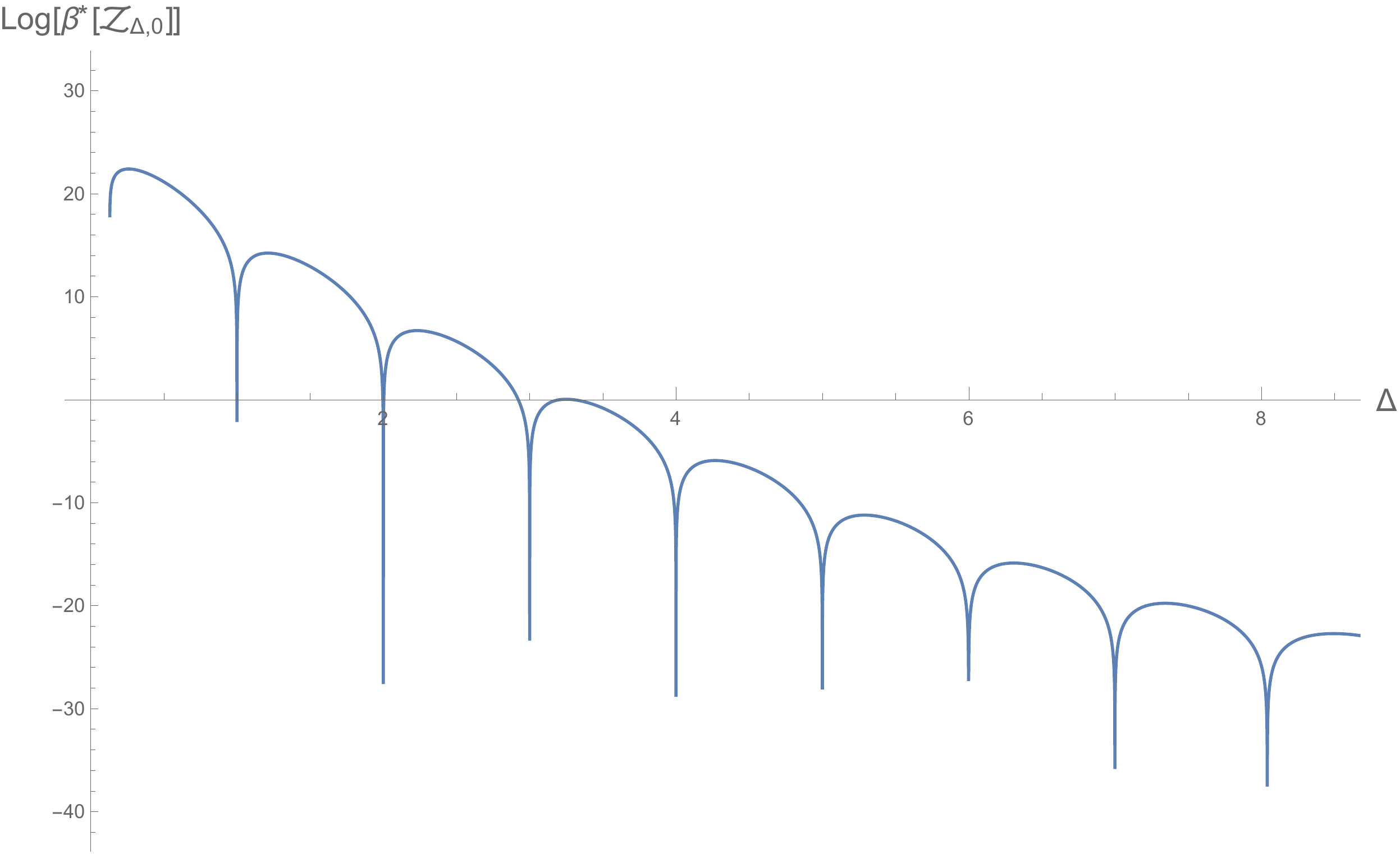}\\
\includegraphics[width=.45\textwidth]{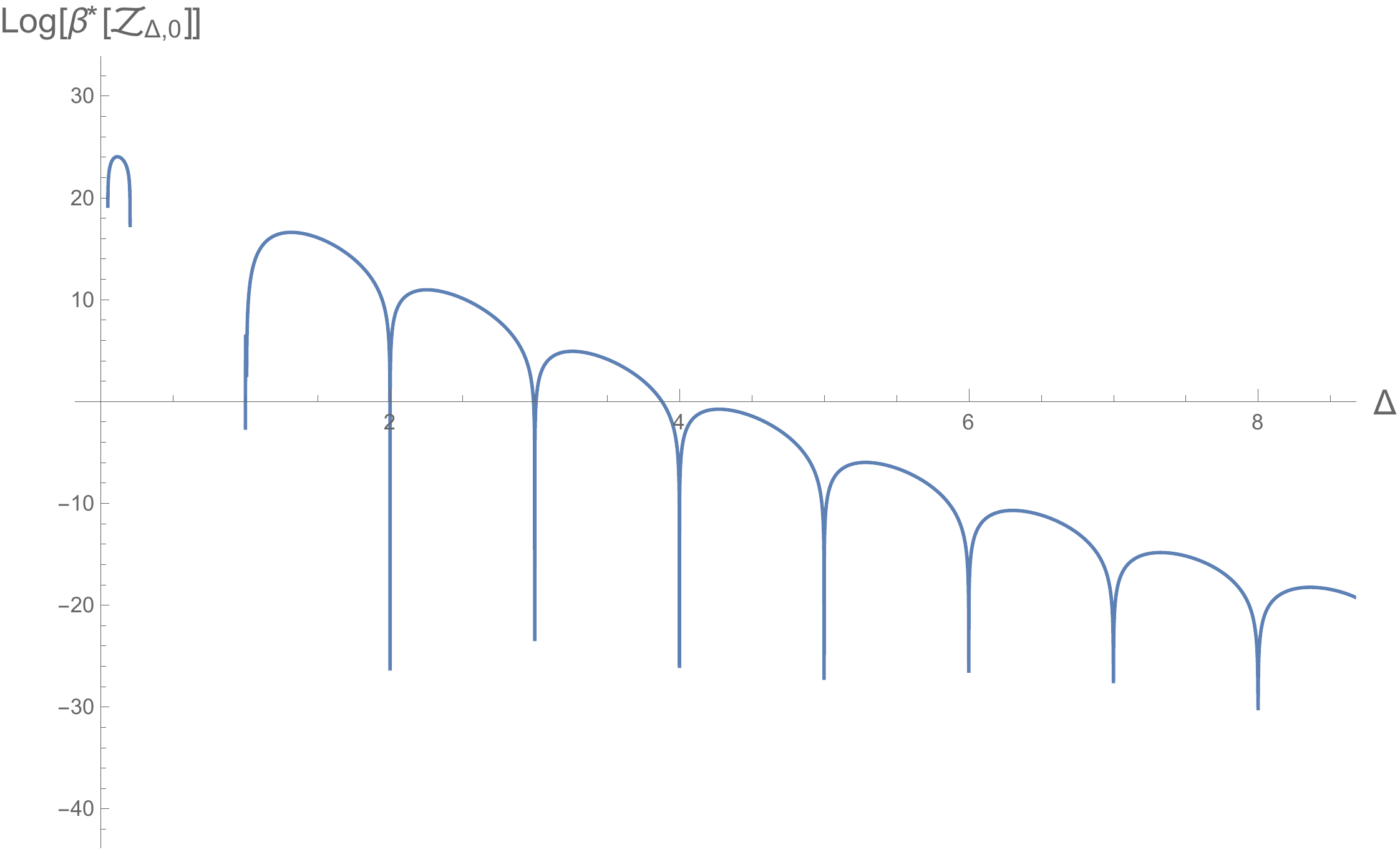} \quad
\includegraphics[width=.45\textwidth]{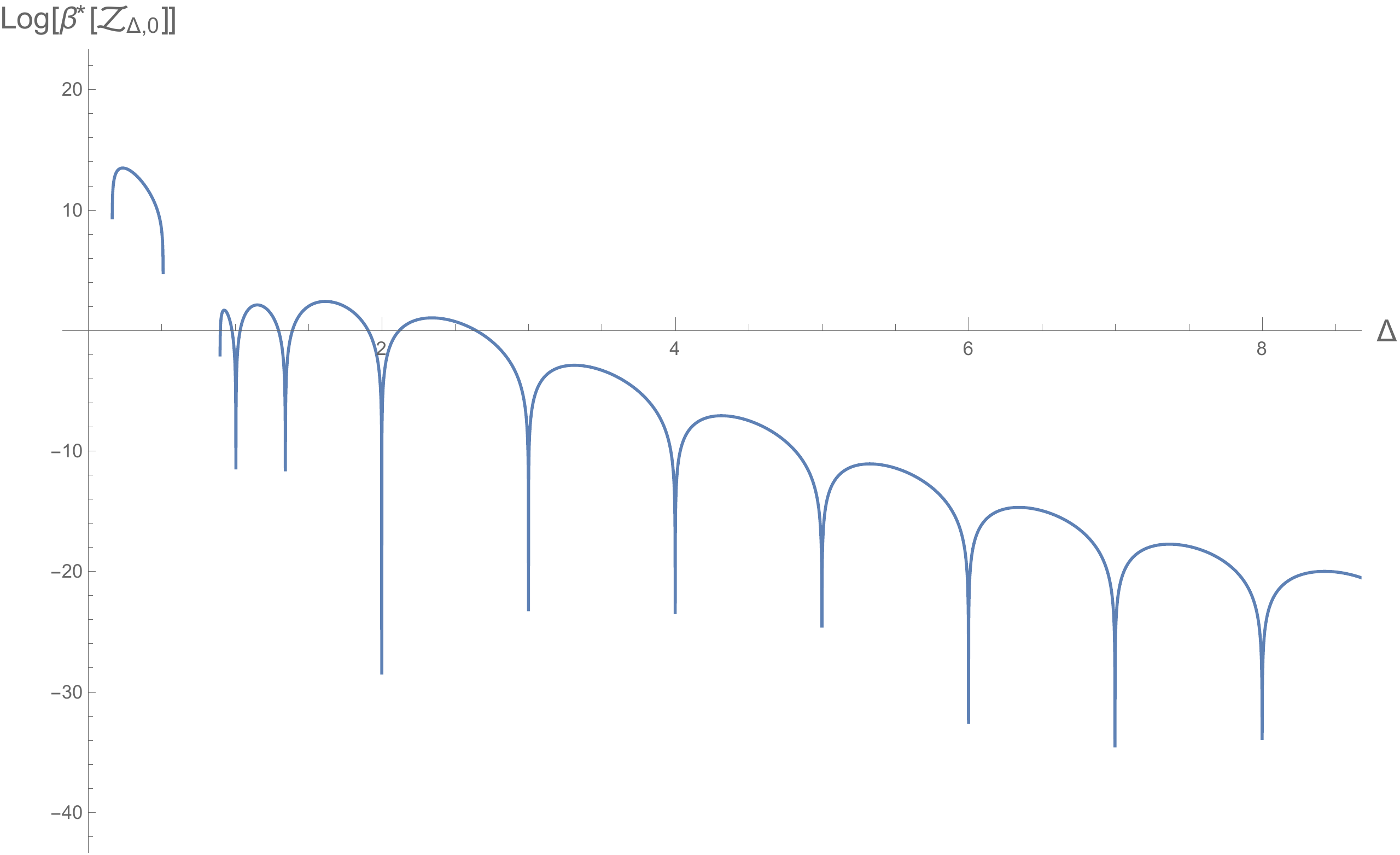}
\end{center}
\caption{The extremal functional method applied to the putative CFTs with $c=3/2$(upper left), $2$(upper right), $5/2$(lower left) and $3$(lower right). Four figures are showing the expected spectrum of the scalar states.}
\label{N=1 EFM}
\end{figure}

\paragraph{CFTs on the plateau}
It turns out that the spectra of putative CFTs saturating the gap
on the plateau, labeled by A,B,C and D ($c=\frac{3}{2}, 2, \frac{5}{2}, 3$) in figure \ref{Super_Virasoro_N1},
agree with those of theories of free fermions. Using the EFM method, we find that for
a hypothetical CFT that lives on the numerical boundary $c=\frac32, 2, \frac52, 3$ and $\D_t=1$ has the spectrum
as given in figure \ref{N=1 EFM} and table \ref{N1pointA}.

\begin{table}[t]
\centering
{
\begin{tabular}{|c |c || c| c || c | c|| c | c |}
\hline
      \multicolumn{2}{|c||}{$c=\frac{3}{2}$} &
      \multicolumn{2}{|c|}{$c=2$} &
      \multicolumn{2}{|c||}{$c=\frac{5}{2}$} &
      \multicolumn{2}{|c|}{$c=3$} \\
\hline
\hline
 \rule{0in}{3ex}     $(h,\bar{h})$  &  Max. Deg & $(h,\bar{h})$ & Max. Deg & $(h,\bar{h})$  &  Max. Deg & $(h,\bar{h})$ & Max. Deg\\
\hline
\rule{0in}{3ex} $(\frac{1}{2}, \frac{1}{2})$ & 9.00000  & $(\frac{1}{2}, \frac{1}{2})$ &  16.0000 & $(\frac{1}{2}, \frac{1}{2})$ & 25.0001 & $(\frac{1}{2}, \frac{1}{2})$ &  36.0082  \\
\hline
\rule{0in}{3ex} $(2, \frac{1}{2})$ & 6.00000& $(1, \frac{1}{2})$ & $8.0001$ &  $(1, \frac{1}{2})$ & 25.0009 &  $(1, \frac{1}{2})$ & 54.0426 \\
\hline
\rule{0in}{3ex} $(2, 2) $  & 4.00000  & $(\frac{3}{2}, \frac{1}{2})$ & $4.0004$  &  (1,1) & 25.0035  & $(1, 1)$ & 81.1453 \\
 \hline
\rule{0in}{3ex} $(\frac{9}{2}, 2)$  & 4.00000 & $(\frac{3}{2}, 1)$ & $2.0014$ & $(\frac{3}{2},\frac{1}{2})$   & 20.0029 & $(\frac{3}{2}, \frac{1}{2})$ & 60.1393  \\
 \hline
\rule{0in}{3ex} $ (\frac{9}{2}, \frac{9}{2}) $  &  4.00000   & $(\frac{3}{2}, \frac{3}{2})$ & $1.0093$  & $(2,\frac{1}{2})$  & 50.0111 & $(2, 1)$ & 172.428\\
 \hline
\end{tabular}
\caption{ The maximum value of degeneracies for the low-lying states in a putative CFTs on the plateau of figure \ref{Super_Virasoro_N1}.}
\label{N1pointA}
}
\end{table}

We find that the partition function for the $c=\frac{3}{2}, \Delta_t = 1$ agrees with
\begin{align}
  Z_\text{NS}(\t,\bar \t) =
  \left| q^{-\frac{1}{16}} \prod_{n=1} \big(1+q^{n-\frac12} \big)^3
  \right|^2 \ .
  \label{3freefermions}
\end{align}
This is nothing but the partition function of three free Majorana fermions.
Using the identity
\begin{align}
  \sum_{k\in\mathbb{Z}} q^{k^2/2} \prod_{n=1} \frac{1}{1-q^n} =
  \prod_{n=1} (1+q^{n-1/2})^2 \ ,
\end{align}
the partition function (\ref{3freefermions}) can be decomposed in terms of $\CN=1$ super-Virasoro characters
as
\begin{align}
  Z_\text{NS}(\t,\bar \t) =
  \left| \chi_\text{vac}(q)+ 3 \chi_{h=\frac12}(q) + 2 \sum_{k=2}^\infty
  \chi_{h=\frac{k^2}{2}}(q) \right|^2.
\end{align}
Similarly, our numerical results also suggest that the
putative partition functions for theories at $c=2,\frac52,3$
on the plateau are given as
\begin{align}
  Z_\text{NS}(\t,\bar \t) =
  \left| q^{-\frac{c}{24}} \prod_{n=1} \big(1+q^{n-\frac12} \big)^{2c}
  \right|^2,
\end{align}
which again admit $\CN=1$ character decompositions.

It is in fact well-known that the supersymmetry can be non-linearly realized in
the theory of free massless fermions \cite{Antoniadis:1985az, Windey1986}. To see this,
let us start with the (Euclidean) action of $N$ free Weyl-Majorana fermions
\begin{align}
  \CI = \frac{1}{4\pi} \int d^2z ~\sum_{i=1}^N \psi^i \bar \partial \psi^i.
  \label{actionfreefermion}
\end{align}
The action (\ref{actionfreefermion}) is invariant under not only the two-dimensional
conformal transformation but also a fermionic transformation
\begin{align}
  \d\psi_i = f_{ijk} \psi_j \psi_k \e \ ,
  \label{fermiontransf}
\end{align}
where $f_{ijk}$ is totally anti-symmetric and $\e$ is a Grassmannian parameter.
The current $G(z)$ generating (\ref{fermiontransf}) can be written
\begin{align}
  G(z) = \frac13 f_{ijk} :\psi_i(z) \psi_j(z) \psi_k(z):.
  \label{current}
\end{align}
One can show that (\ref{current}) becomes the supersymmetry current
if and only if $f_{ijk}$ are proportional
to the structure constants of a semi-simple Lie
group $G$. More precisely, the OPE between $G(z)$ agrees
with that of $\CN=1$ superconformal algebra
\begin{align}
  G(z) G(0) \sim \frac{2c}{3z^3} + \frac{2 T(z)}{z},
\end{align}
if and only if two conditions below are satisfied,
\begin{align}
  f_{ijk} f_{l m k} + f_{ilk} f_{mjk} + f_{imk}f_{jlk} = 0,
  \label{former}
\end{align}
and
\begin{align}
  f_{ikl} f_{jkl} \propto \d_{ij}.
  \label{latter}
\end{align}
The former (\ref{former}) is the Jacobi identity
which can be solved when $f_{ijk}$ are the structure constants of a Lie group $G$.
On the other hand, the latter (\ref{latter}) implies that $G$ is semi-simple.

The classification of supersymmetric theories of free fermions given above
immediately implies that the hypothetical SCFTs on the plateau at $c=3/2$ and $c=3$
are indeed theories of three and six free fermions respectively, associated to the
Lie algebras $SO(3)$ and $SO(4)$. However, the theories of four and five free fermions
cannot be associated with the appropriate Lie algebra,
although their partition functions allows the decomposition in terms of the $\CN=1$ super-Virasoro characters.
Therefore, the $c=2, 5/2$ theory on the boundary does not exist as a consistent superconformal theory. Nevertheless, the theory of four and five fermions are consistent non-supersymmetric CFTs. Here we see that the modularity is not enough to show the existence of SCFT.

When we consider the upper bounds on $\D_t$, only allowing the conserved currents
with $j \ge 1$ or $j \ge \frac{3}{2}$, the plateau at $\frac{3}{2} \le c \le 3$
disappears. This is because the theory of free fermions always has the conserved
spin-$\half$ current, where the conservation simply comes from the equation of motion.

\begin{table}[t]
\centering
{
\begin{tabular}{|c|c |c |c| c| }
\hline
  Label &  $c$     & $\Delta_{t}$  & Expected CFT \\
\hline\hline
 \rule{0in}{2.5ex}A & $3/2$  & $1$  & Three free fermions \\ [0.5ex]
 \hline
 \rule{0in}{2.5ex}B & $2$  & $1$  &  Four free fermions (Non-SUSY) \\ [0.5ex]
 \hline
 \rule{0in}{2.5ex}C & $5/2$  & $1$ & Five free fermions (Non-SUSY) \\[0.5ex]
 \hline
\rule{0in}{2.5ex}D&  $3$  & $1  $  & Six free fermions \\[0.5ex]
\hline
\hline
\rule{0in}{2.5ex}E& $9$  & $2$  &  Single character rational SCFT \\[0.5ex]
\hline
 \rule{0in}{2.5ex}F& $33/2$  & $3$ & Single character rational SCFT \\[0.5ex]
\hline
\end{tabular}
\caption{List of $\mathcal{N}=(1,1)$ theories on the numerical boundary of  figure \ref{Super_Virasoro_N1}, when the conserved currents with $j \ge \frac{1}{2}$ included in the spectrum.}
\label{List_j1}
}
\end{table}

\begin{figure}[h]
\begin{center}
\includegraphics[width=.45\textwidth]{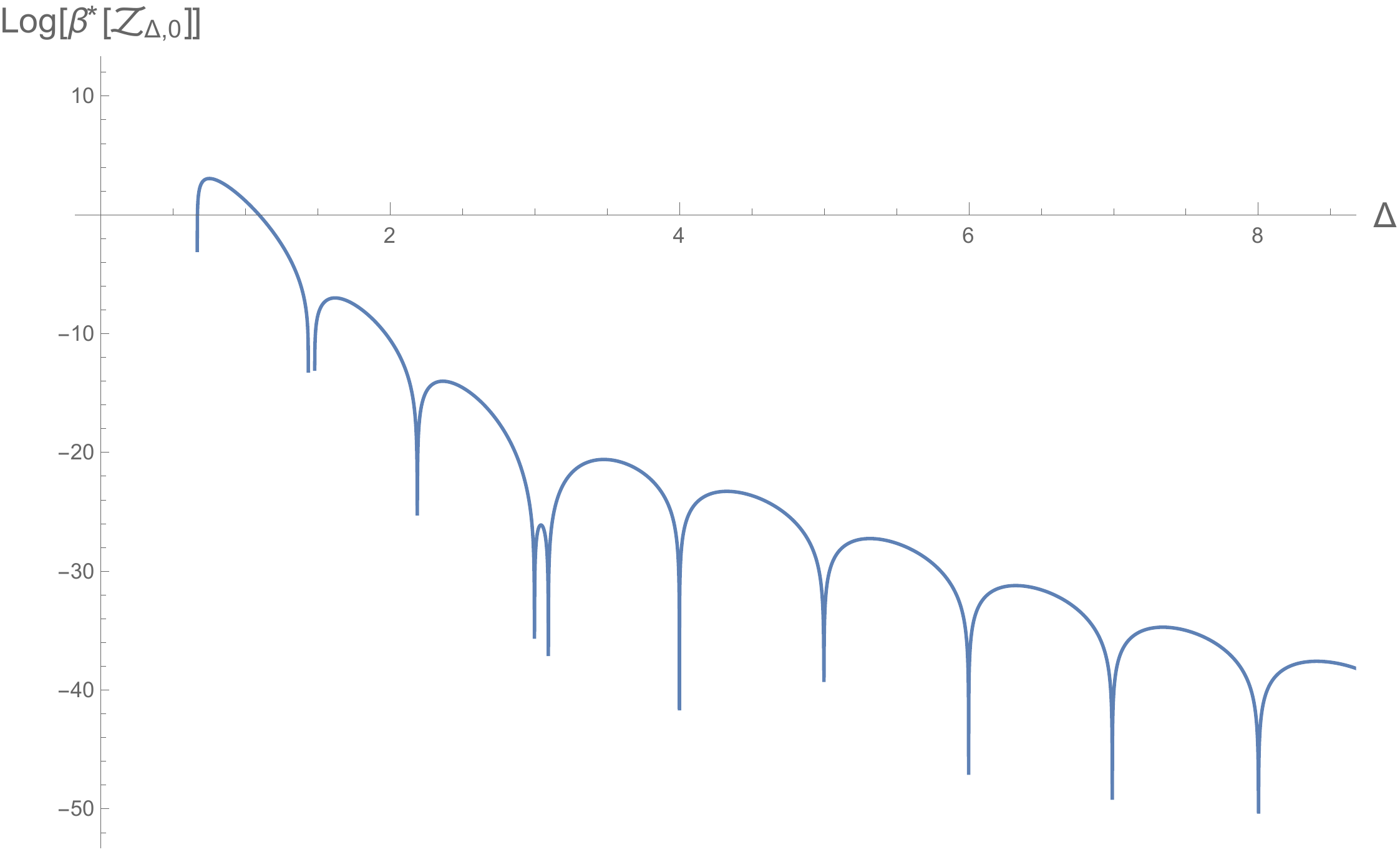} \quad
\includegraphics[width=.45\textwidth]{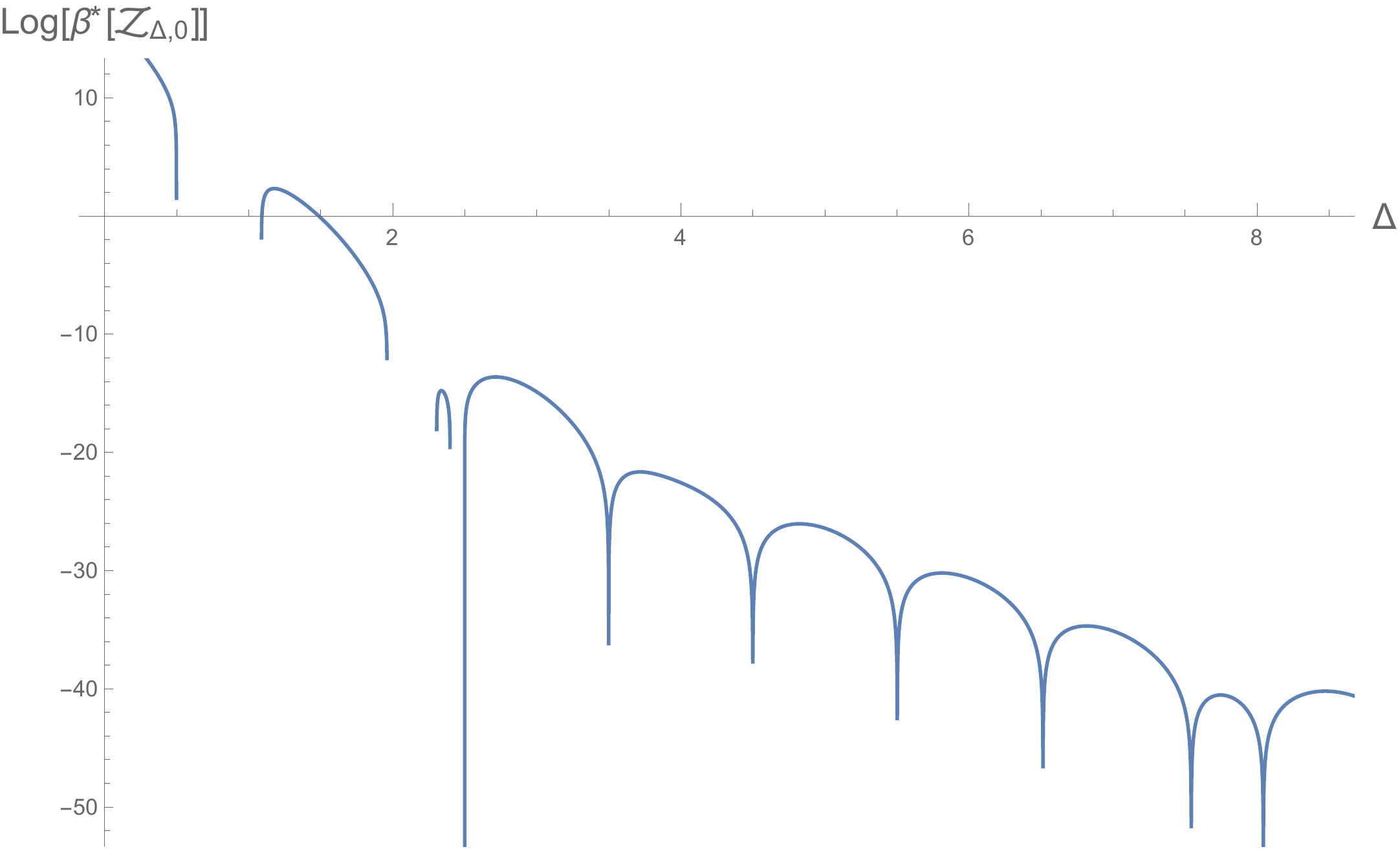}
\end{center}
\caption{The extremal functional method applied to the $c=9$ putative CFT. The scalar spectrum (left) and spin-one spectrum (right) are presented.}
\label{N=1 EFM c9}
\end{figure}

\paragraph{CFT at the kinks}
In addition to the plateau, we also find kinks at $c=9$
and $c=\frac{33}{2}$, labelled as E and F in figure \ref{Super_Virasoro_N1}.
We conjecture that at the kink $c=9$ lies a certain single-character
rational CFT that we are going to specify. To see this, we apply the EFM method to the putative CFT at
$c=9$ with $\D_t=2$ to determine the entire (extremal) spectrum.
As illustrated in figure \ref{N=1 EFM c9} and table \ref{N1pointEandF}, we observe that
the superconformal primaries are integer-spaced in $\D$ and
their degeneracies get closer to
integer values as $N$ in (\ref{LF_type2}) becomes greater.

\begin{table}[ht]
\centering
{
\begin{tabular}{|c ||c | c| c | c|| c |}
\hline
 \rule{0in}{3ex}    & $N=31$  &  $N=41$ & $N=51$ & $N=61$ & Expected \\
\hline
\hline
\rule{0in}{3ex} $(1, 1)$ & 68347.367 & 68183.991 & 68134.348  & 68125.588 & 68121 \\
\hline
\rule{0in}{3ex} $(\frac{3}{2}, 1)$ & 52626.077 & 51047.268 & 50771.242  & 50667.153 & 50634\\
\hline
\rule{0in}{3ex} $(\frac{3}{2}, \frac{3}{2})$ & 59814.440 & 43901.832 & 38955.363 & 38105.517 & 37636\\
\hline
\rule{0in}{3ex} $(2, \frac{3}{2})$ & 933518.17  & 820426.78 & 796158.81 & 787307.63 & 784536 \\
\hline
\rule{0in}{3ex} $(\frac{5}{2}, 1)$ & 1062307.93 & 980071.44 & 963664.49  & 957864.26 & 956043 \\
\hline
\rule{0in}{3ex} $(2, 2)$ & 18091641.87 & 16924088.86 & 16477154.75  & 16398071.96 & 16353936 \\
\hline
\end{tabular}
\caption{ The maximum value of degeneracies for low-lying states in a putative $\mathcal{N}=(1,1)$ CFT at $c=9$. }
\label{N1pointEandF}
}
\end{table}

Our numerical results imply that the partition function of the putative CFT is
\begin{align}
  Z_{c=9}(\t,\bar{\t}) & =
  \chi_0(\tau) \bar{\chi}_0(\bar{\tau})+ 68121 \chi_{1}(\tau) \bar{\chi}_{1}(\bar{\tau})
  + 37636  \chi_{\frac{3}{2}}(\tau) \bar{\chi}_{\frac{3}{2}}(\bar{\tau})
  \nonumber \\ & +
  50634\left(\chi_{\frac{3}{2}}(\tau) \bar{\chi}_{1}(\bar{\tau}) +
  \chi_{1}(\tau) {\bar \chi}_{\frac{3}{2}}(\bar \tau)  \right)+ \cdots.
  \label{c=9_N=1_decomposition}
\end{align}
We notice that (\ref{c=9_N=1_decomposition}) can be holomorphically
factorized,
\begin{align}
\label{factorization01}
  Z_{c=9}(\t,\bar{\t}) &= f^{c=9}(\t) \bar{f}^{c=9}(\bar{\t}),
\end{align}
where $f^{c=9}(\t)$ is a solution to a differential
equation given as,
\begin{align}
  \left[ q\frac{\partial}{\partial q}   +  N_2(q) \right] f(\t) = 0 \ ,
  \label{MDE01}
\end{align}
with
\begin{align}
\label{N2}
  N_2(q) = \frac{3}{16} \Big[ \vartheta_{01}^4(q) -\vartheta_{10}^4(q) \Big]
  \left[ \frac{K(q) + 30}{K(q) + 6} \right].
\end{align}
Here $N_2(q)$ is a modular form of weight two under the congruence subgroup $\G_\th$
of the full modular group $SL(2,\mathbb{Z})$, and (\ref{MDE01}) is invariant under $\G_\th$.
The function $K(q)$ is the well-known NS partition function of the $\CN=1$ super-extremal
CFT with $c=12$ studied in \cite{Witten2007} given by,
\begin{align}
  K(q) & = \frac{q^{-\frac12}}{2} \left[\prod_{n=1}^\infty ( 1+ q^{n-\frac 12} )^{24}
  - \prod_{n=1}^\infty (1 - q^{n -\frac12} )^{24} \right] +
  2048 q \prod_{n=1}^\infty (1 + q^n )^{24}
  \nonumber \\ & =
  \left[ q^{- \frac{1}{48}} \prod_{n=1}^\infty(  1+ q^{n-\frac12})^{24}\right]^{24} - 24.
\end{align}
The form of the partition function (\ref{factorization01}) then guarantees that the putative SCFT
is rational with a single character. See appendix \ref{app:MDE} for the normalization of the theta functions we use and also for a more detailed discussion on the differential equation invariant under $\G_\th$.

As a side remark, we find a bilinear relation,
\begin{align}
  K(q) + 6 = f^{c=9}(\t) \cdot f^{c=3}(\t),
  \label{bilinear01}
\end{align}
where $f^{c=3}(\t)$ denotes the contribution from left-movers to the NS partition
function of the theory of six free fermions,
\begin{align}
  f^{c=3}(\t) =  q^{-\frac{1}{4}} \prod_{n=1} \big(1+q^{n-\frac12} \big)^{6}
  = \left( \frac{\eta(q)^2}{\eta(q^2) \eta(\sqrt{q}) } \right)^6.
\end{align}
The above expression (\ref{bilinear01}) is analogous to the bilinear relation between the
Ising model and the Baby Monster CFT\cite{Hampapura:2016mmz, Harvey:2018rdc}, or
$c=8$ and $c=16$ CFT without Kac-Moody symmetry\cite{BLS,MukhiPrivateComm}.
It is therefore natural to expect that there is
a group-theoretic reason behind (\ref{bilinear01}), i.e.,
the superconformal primaries of the hypothetical SCFT at $c=9$
could be in certain representations under a finite group, often sporadic.

Let us comment on the hypothetical SCFT of $c=\frac{33}{2}$, that lie at the point F in figure \ref{Super_Virasoro_N1}. We conjecture that the partition function of  $c=\frac{33}{2}$ SCFT is holomorphically factorized into
\begin{align}
\label{factorization02}
  Z_{c=\frac{33}{2}}(\t,\bar{\t}) &= f^{c=\frac{33}{2}}(\t) \bar{f}^{c=\frac{33}{2}}(\bar{\t}),
\end{align}
where $f^{c=\frac{33}{2}}(\t)$ is given by
\begin{align}
\label{Ch33h}
\begin{split}
f^{c=\frac{33}{2}}(\t) = q^{-\frac{33}{48}} \left( 1 + 7766 q^{\frac{3}{2}} + 11220 q^2 + 408507 q^{\frac{5}{2}} + 515251 q^3 + \cdots \right) .
\end{split}
\end{align}
The character \eqref{Ch33h} is derived in \cite{Hoehn2007}, and equivalently obtained by a third order modular differential equation in appendix \ref{app:MDE}. Now, the partition function \eqref{factorization02} is decomposed into the $\mathcal{N}=1$ super-Virasoro characters as below.
\begin{align}
\begin{split}
  Z_{c=\frac{33}{2}}(\t,\bar{\t}) & =
  \chi_0(\tau) \bar{\chi}_0(\bar{\tau})+ 60295225 \chi_{\frac{3}{2}}(\tau) \bar{\chi}_{\frac{3}{2}}(\bar{\tau})
  + 11930116  \chi_{2}(\tau) \bar{\chi}_{2}(\bar{\tau})
   \\ & +
  26820310 \left(\chi_{\frac{3}{2}}(\tau) \bar{\chi}_{2}(\bar{\tau}) +
  \chi_{2}(\tau) {\bar \chi}_{\frac{3}{2}}(\bar \tau)  \right)+ \cdots.
  \label{c=33h_N=1_decomposition}
\end{split}
\end{align}

\begin{table}[ht]
\centering
{
\begin{tabular}{|c ||c | c| c || c|}
\hline
 \rule{0in}{3ex}     &  $N=41$ & $N=51$ & $N=61$ & Expected \\
\hline
\hline
\rule{0in}{3ex} $(\frac{3}{2}, \frac{3}{2})$ & 60471375.088  &  60342184.510  &  60313636.206 & 60295225 \\
\hline
\rule{0in}{3ex} $(2, \frac{3}{2})$ & 27685500.543  &  27059008.764  &  26895957.918 & 26820310 \\
\hline
\rule{0in}{3ex} $(2, 2)$ & 43743886.2618  &  20862868.0446  &  15440948.5135  & 11930116 \\
\hline
\rule{0in}{3ex} $(\frac{5}{2}, \frac{3}{2})$ & 3127027597.42  &  3098009417.73  &  3090529976.19 & 3084933555 \\
\hline
\rule{0in}{3ex} $(\frac{5}{2}, \frac{5}{2})$ & 164728420016.60  & 160027206364.24  &  158841761761.11 & 157836960369 \\
\hline
\end{tabular}
\caption{ The maximum value of degeneracies for low-lying states in a putative $\mathcal{N}=(1,1)$ CFT at c=$\frac{33}{2}$. }
\label{N1pointEandF}
}
\end{table}

We investigated the maximal degeneracies of  $c=\frac{33}{2}$ hypothetical CFT in table \ref{N1pointEandF}. We expect that the numerical values of maximal degeneracies will be closed to the degeneracies in \eqref{c=33h_N=1_decomposition}, as increasing the number of derivatives $N$.

\begin{figure}[h!]
\begin{center}\includegraphics[width=.85\textwidth]{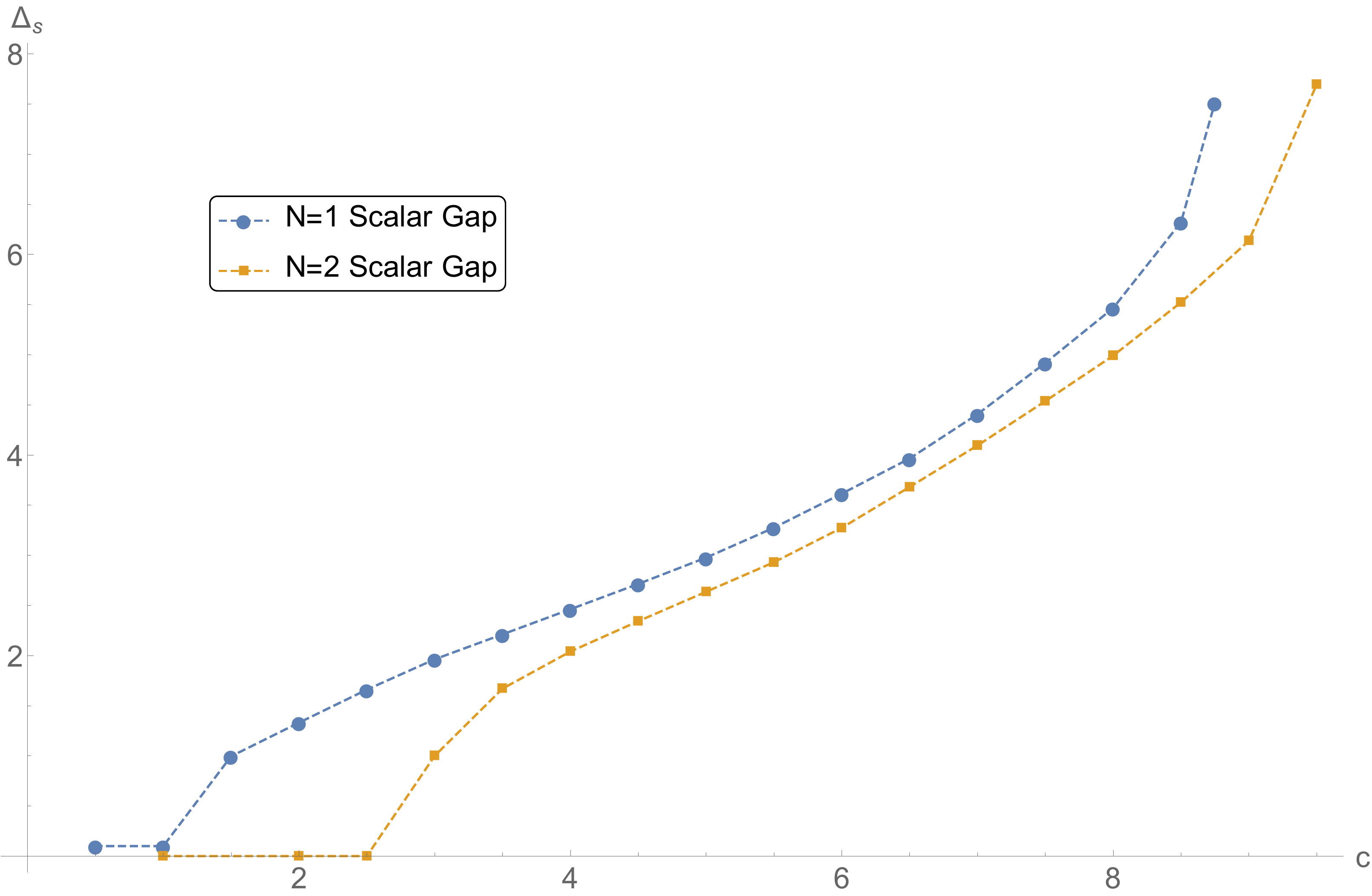}
\end{center}
\caption{Numerical upper bounds on the scalar gap for the $\mathcal{N}=(1,1)$ and $\mathcal{N}=(2,2)$ SCFTs.}
\label{fig:ScalarGap}
\end{figure}

\paragraph{Scalar Gap}
Here a gap $\D_s$ is imposed in the spectrum of scalar primaries while
other primaries with non-zero spin are subject to unitary bound, i.e.,
\begin{align}
  \D_\ast(j=0) = \D_s , \quad \D_\ast(j) = j \text{ for  } j\neq 0.
\end{align}
We find the numerical upper bounds as in figure \ref{fig:ScalarGap}.
From this result, we find that it is not possible to have a theory without relevant operators in the NS sector for $c \lesssim 3.1$ and $c \lesssim 4.0$ for the $\CN=(1, 1)$ and $\CN=(2, 2)$ SCFT respectively.
Notice that the numerical upper bound for the scalar gap seems to disappear for $c \gtrsim 9$ for the $\CN=(1, 1)$ case and $c \gtrsim 10$ for the $\CN=(2, 2)$ case. This is reminiscent of the non-SUSY analysis where the scalar gap bound blows up near $c=25$ \cite{Collier2016}, which is identical to the central charge of the Liouville theory at $b=1$.

%%%%%%%%%%%%%%%%%%%%%%%%%%%%%%%%%%%%%%%%%%%%%%%%%%%%%%%%%%%
\subsection{Spin-dependent Bounds for $\mathcal{N}=(2,2)$ SCFT}
%%%%%%%%%%%%%%%%%%%%%%%%%%%%%%%%%%%%%%%%%%%%%%%%%%%%%%%%%%%

%
\begin{figure}[h]
\begin{center}\includegraphics[width=0.85\textwidth]{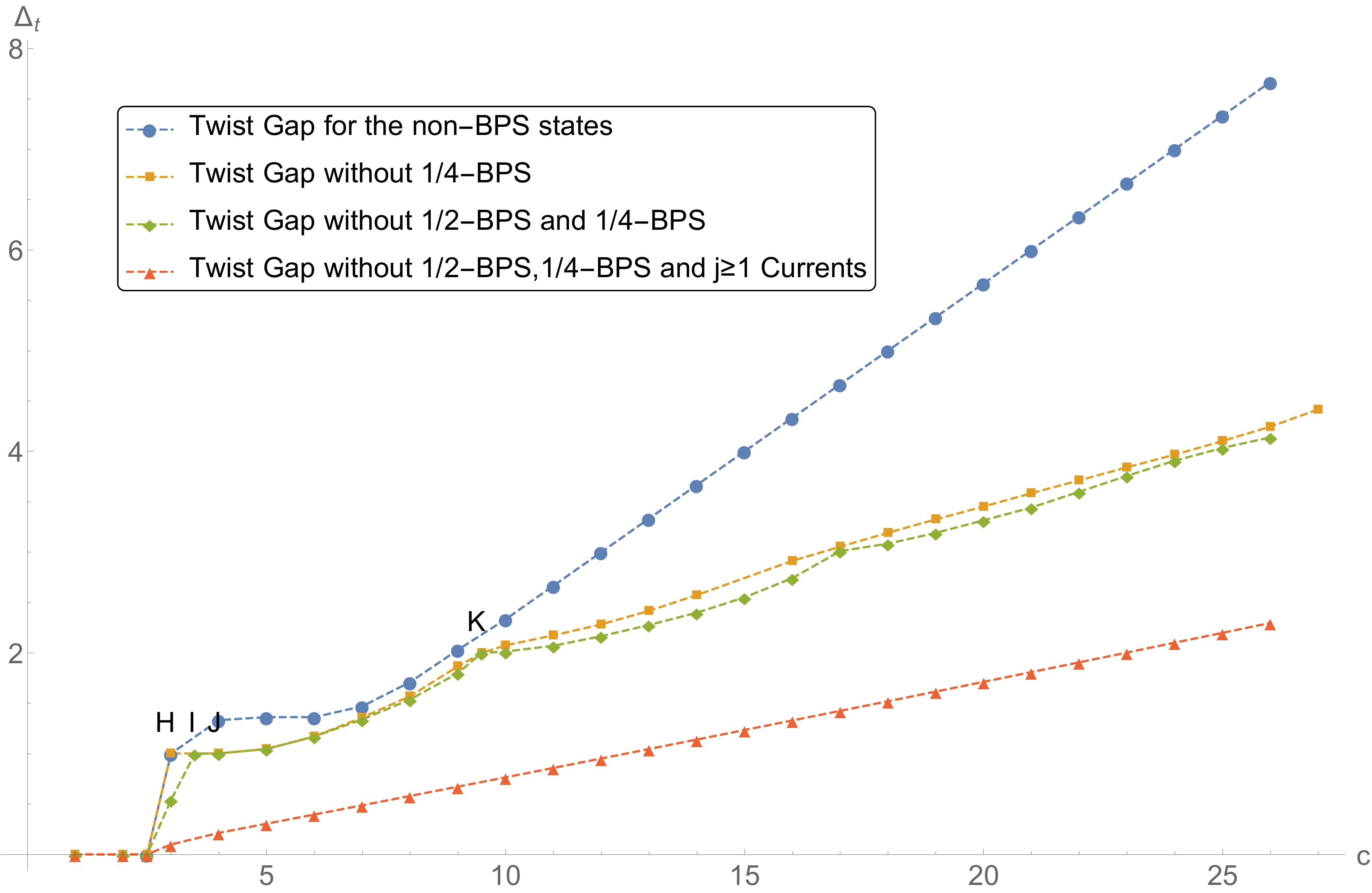}
\end{center}
\caption{Numerical upper bounds on the twist gap for the $\CN=2$ SCFTs under various assumptions. The blue line represents the most generic bounds for the $\CN=(2, 2)$ SCFT.}
\label{N2TwistGap}
\end{figure}

\begin{table}[t]
\centering
{
\begin{tabular}{|c|c |c |c| c| }
\hline
  Label &  $c$     & $\Delta_{t}$  & Expected CFT \\
\hline\hline
 \rule{0in}{2.5ex}H & $3$  & $1$  & Six free fermions\\ [0.5ex]
 \hline
 \rule{0in}{2.5ex}I & $7/2$  & $1$  &  Seven free fermions (Non-SUSY) \\ [0.5ex]
 \hline
 \rule{0in}{2.5ex}J & $4$  & $1$ & Eight free fermions (Non-SUSY) \\[0.5ex]
 \hline
\hline
\rule{0in}{2.5ex}K&  $19/2$  & $2 $  & Single character rational SCFT \\[0.5ex]
\hline
\end{tabular}
\caption{List of theories on the numerical boundary of figure \ref{N2TwistGap}, when the conserved currents with $j \ge \frac{1}{2}$ included in the spectrum.}
\label{N=2 List_j1}
}
\end{table}

In this subsection, we investigate the numerical bounds on the twist
gap $\D_t$ for $\mathcal{N}=(2,2)$ SCFTs. In order to place bounds on
the twist gap, we solve the SDP problem (\ref{linear functional N1_2})
with
\begin{align}
  \CZ_{\frac{\D+j}{2},\frac{\D-j}{2}}(\b) & =
  \CZ_{\frac{\D+j}{2},\frac{\D-j}{2}}^{r,\bar r}(\t,\bar \t,z=\bar z=0)
  \hspace*{0.78cm} \text{ for 1/2-BPS states},
  \nonumber \\ \text{or, }  & =
  \begin{cases}
  \CZ_{\frac{\D+j}{2},\frac{\D-j}{2}}^{r,\bar Q}(\t,\bar \t,z=\bar z=0)
  \\
  \CZ_{\frac{\D+j}{2},\frac{\D-j}{2}}^{Q,\bar r}(\t,\bar \t,z=\bar z=0)
  \end{cases} \text{for 1/4-BPS states},
  \\ \text{or, } & =
  \CZ_{\frac{\D+j}{2},\frac{\D-j}{2}}^{Q,\bar Q}(\t,\bar \t,z=\bar z=0)
  \hspace*{0.78cm} \text{ for non-BPS states}
  \nonumber
\end{align}
for $\D \geq j+\D_t$. figure \ref{N2TwistGap} shows the numerical upper bounds on
$\D_t$ for the $\mathcal{N}=(2,2)$ SCFTs with and without BPS representations.

The blue curve in figure \ref{N2TwistGap} denotes the upper bounds on the twist gap for the non-BPS states in the NS sector. (Holomorphic currents belong to BPS sector, which has zero twist gap)
We find the upper bounds asymptotes to a linear line with slope $\frac{1}{3}$ as $\D_t = \frac{1}{3}(c-3)$
for $c \gtrsim 6$.

Interestingly, as far as the $\G_\th$ invariance of the partition function is concerned,
we cannot rule out the possibility of $\CN=(2,2)$ SCFTs without any BPS states except for the vacuum.
The asymptotic slope of the upper bounds (the red curve in figure \ref{N2TwistGap}) is close to $c/12$.

Once BPS states are included, the numerical upper bounds become relaxed.
It shows a sudden jump at $c=3$ labelled by H. 
%We summarize in table some of the
%spectrum of this hypothetical SCFT H obtained by the EFM. It implies that
%its NS partition function would be
%
%\begin{align}
%  Z_\text{NS}^{c=3}(\t,\bar \t) = \left| q^{-\frac 18} \prod_{n=1}^\infty (1+ q^{n-\frac12} )^6 \right|^2,
%\end{align}
%
%which agrees with the partition function of six free fermions.
Note that the partition function of six free fermion theory cannot be decomposed 
in terms of $\CN=2$ massive characters only. To decompose it, one should consider the
$\CN=2$ massless characters, as well as the massive characters.
As we discussed in the previous subsection, the theory already has $\CN=1$
supersymmetry non-linearly realized. It is easy to understand the
supersymmetry enhancement from $\CN=1$ to $\CN=2$ via
bosonizing the theory to the theory of a free $\CN=2$ chiral
multiplet.

We also found that the spectra of putative CFTs saturating the
twist gap bounds at $c=7/2$ and $c=4$ match with those of
theories of seven and eight free fermions.
Although their partition functions admit $\CN=2$ character
decomposition, we do not expect them to have even $\CN=1$ supersymmetry
due to the non-existence of semi-simple Lie groups of dimensions seven and eight
 that satisfy \eqref{former}, \eqref{latter}.

\begin{figure}[h!]
\begin{center}
\includegraphics[width=.45\textwidth]{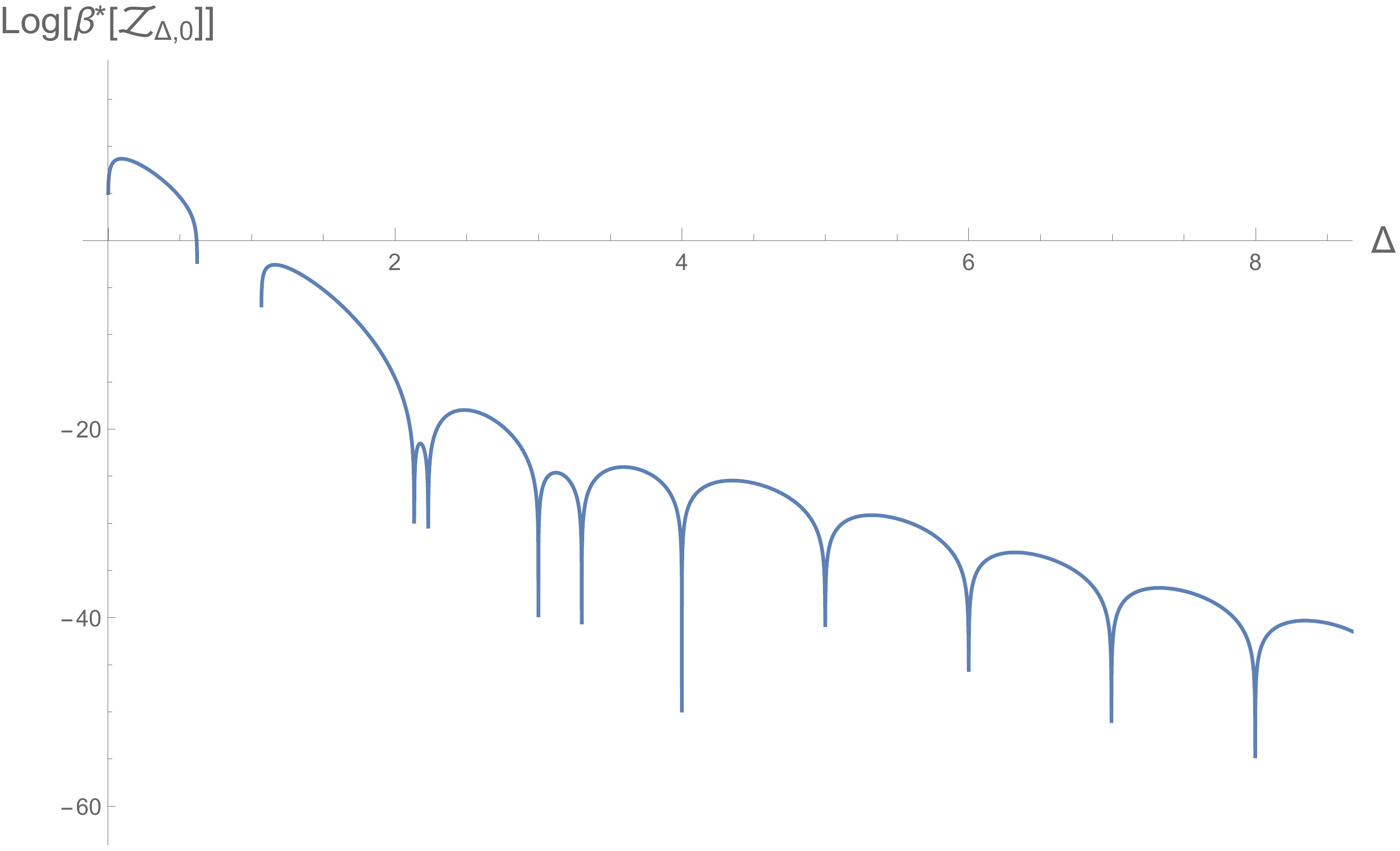} \quad
\includegraphics[width=.45\textwidth]{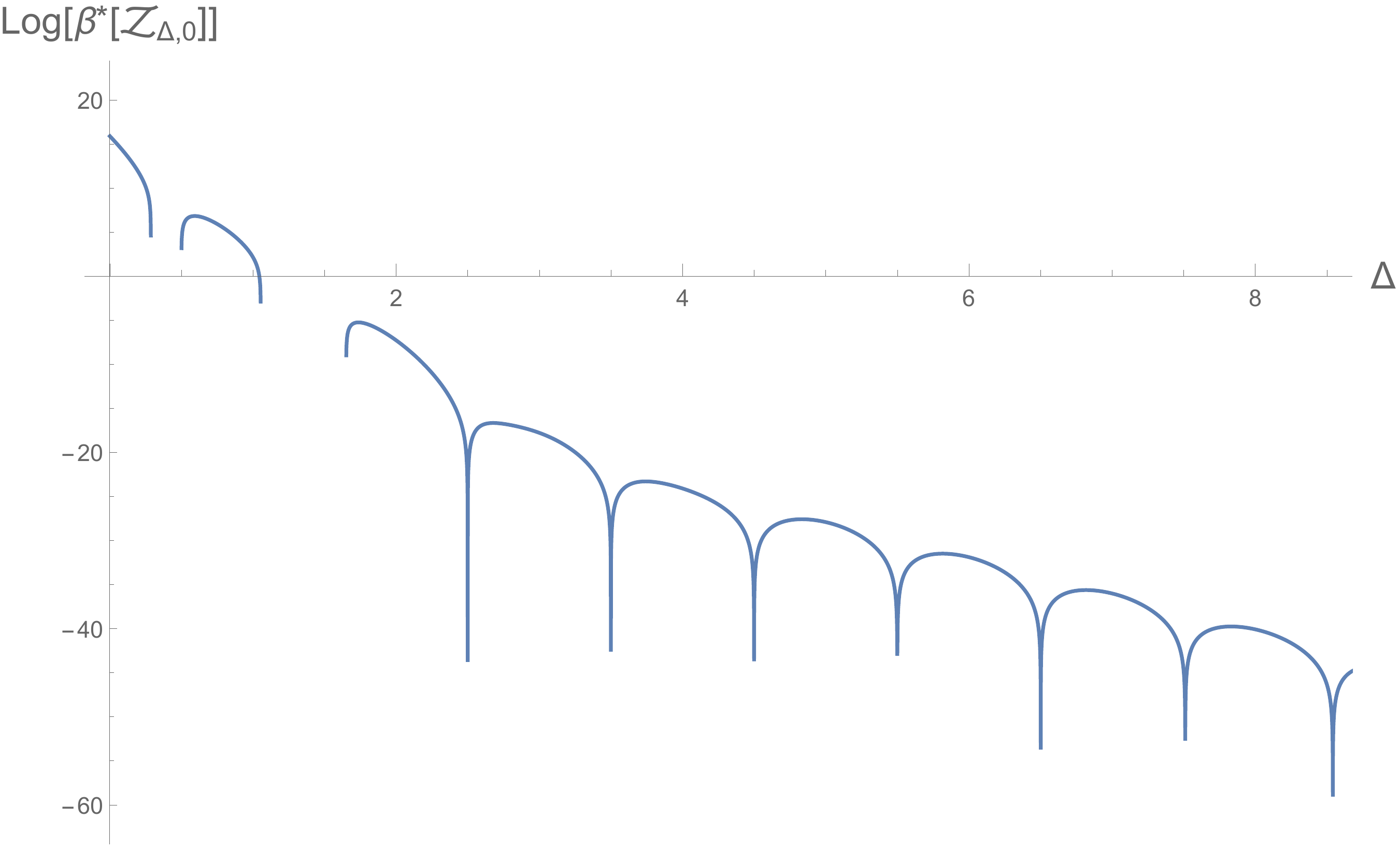}
\end{center}
\caption{EFM method applied to the putative CFT with $c=\frac{19}{2}$. The results include scalar spectrum of $c=\frac{19}{2}$ SCFT(left) and spin-half spectrum of $c=\frac{19}{2}$ SCFT(right). }
\label{N=2 EFM}
\end{figure}

\begin{table}[ht]
\centering
{
\begin{tabular}{|c ||c | c| c || c|}
\hline
 \rule{0in}{3ex}     &  $N=41$ & $N=51$ & $N=61$ & Expected \\
\hline
\hline
\rule{0in}{3ex} $(1, 1)$ & 70268.354  &  70233.498  &  70227.854 & 70225 \\
\hline
\rule{0in}{3ex} $(\frac{3}{2}, \frac{3}{2})$ & 33425.565  &  30049.592  &  29521.540 & 29241 \\
\hline
\rule{0in}{3ex} $(\frac{3}{2}, 1)$ & 45599.452  &  45404.124  &  45335.672  & 45315 \\
\hline
\rule{0in}{3ex} $(\frac{5}{2}, \frac{3}{2})$ & 817352.378  & 608772.507  &  576606.246 & 559170 \\
\hline
\rule{0in}{3ex} $(2, 1)$ & 1027480.82  & 1023105.62  &  1022437.70 & 1022105 \\
\hline
\rule{0in}{3ex} $(2, 2)$ & 15257461.76  &  14950965.34  &  14902363.14 & 14876449 \\
\hline
\end{tabular}
\caption{ The maximum value of degeneracies for low-lying states in a putative $\mathcal{N}=(2,2)$ CFT at $c=\frac{19}{2}$. }
\label{N1pointEandF}
}
\end{table}

Let us now in turn consider the kink of $c=19/2$ and $\D_t=2$ labelled by
$J$ in figure \ref{N2TwistGap}. It turns out that
the hypothetical SCFT at the kink has
no BPS states other than the vacuum states.
Furthermore, we see from figure \ref{N=2 EFM} that
the conformal dimensions of scalar and spin-half extremal spectrum
are all integer spaced. This implies the holomorphic factorization
of the NS partition proposed below,
\begin{align}
\begin{split}
  Z_{c=\frac{19}{2}} &=
  \chi_0(\tau) \bar{\chi}_0(\bar{\tau}) + 70225 \text{Ch}_{h=1}(\tau) \bar{\text{Ch}}_{\bar h=1}(\bar{\tau})
  + 29241 \text{Ch}_{h=\frac{3}{2}}(\tau) \bar{\text{Ch}}_{\bar h= \frac{3}{2}}(\bar{\tau})
   \\
  & + 45315\left(\text{Ch}_{h=1}(\tau) \bar{\text{Ch}}_{\bar h=\frac{3}{2}}(\bar{\tau})
  + \mbox{c.c.}\right) + \cdots.
  \label{N=2_J}
\end{split}
\end{align}
Note here that one can not determine the $U(1)$ R-charge of each
massive character since the chemical potential is turned off.
After non-trivial numerical experimentation,
indeed one can show
that (\ref{N=2_J}) can be expressed simply as
\begin{align}
  Z_{c=\frac{19}{2}} = f_{c=\frac{19}{2}}(\t) \cdot
  {\bar f}_{c=\frac{19}{2}}(\bar \t),
\end{align}
where $f_{c=\frac{19}{2}}(\t)$ is the solution
to a $\G_\th$-invariant differential equation
\begin{align}
   \left[ q \frac{\partial}{\partial q} + N_2(q) \right] f(\t) = 0,
   \label{MDE02}
\end{align}
with
\begin{align}
  L_2(q) = \frac{19}{48} \Big[ \vartheta_{01}^4(q) -\vartheta_{10}^4(q) \Big]
  \left[ \frac{K(q) + 29}{K(q) + 5} \right].
\end{align}
The above factorization suggests that a certain
rational (super)conformal field theory with a single character
can saturate the twist gap bound and degeneracy bound as well.

%%%%%%%%%%%%%%%%%%%%%%%%%%%%%%%%%%%%%%%%%%%%%%%%%%%%%%%%%%%%%%%%%%%%%
%%%%%%%%%%%%%%%%%%%%%%%%%%%%%%%%%%%%%%%%%%%%%%%%%%%%%%%%%%%%%%%%%%%%%
\section{Charge Dependent Bounds} \label{sec:QBounds}
%%%%%%%%%%%%%%%%%%%%%%%%%%%%%%%%%%%%%%%%%%%%%%%%%%%%%%%%%%%%%%%%%%%%%
%%%%%%%%%%%%%%%%%%%%%%%%%%%%%%%%%%%%%%%%%%%%%%%%%%%%%%%%%%%%%%%%%%%%%

In this section, we explore the modular constraints on the charged states of SCFT. This requires us to turn on the chemical potential associated to the global symmetry. Let us assume there is an $U(1)$ global symmetry and write the partition function as
\begin{align}
 Z_{\textrm{NS}} (\tau, \bar{\tau}, z) = \Tr_{\mathcal{H}_{\textrm{NS}}} \left[ e^{2\pi i \tau (L_0 - \frac{c}{24})} e^{-2\pi i \bar{\tau} (\bar{L}_0 - \frac{c}{24})} e^{2\pi i z Q} \right] \ ,
\end{align}
where $z$ is the fugacity for the flavor symmetry. (We take $\bar{z}=0$ for simplicity.)
We also discuss the implication of modular bootstrap on the weak gravity conjecture in AdS$_3$.

One can construct the reduced partition function of the below form,
\begin{align}
\begin{split}
\widehat{Z}(\tau, \bar{\tau},z) = |\tau|^{\frac{1}{2}} \frac{|\eta(\frac{\tau}{2})|^2 |\eta(2\tau)|^2}{|\eta(\tau)|^2}  e^{\frac{\pi i k}{2} \frac{z^2}{\tau}} Z(\tau, \bar{\tau},z),
\end{split}
\end{align}
for the partition function of $\mathcal{N}=(1, 1)$ SCFT and
\begin{align}
\widehat{Z}(\tau, \bar{\tau},z) = \frac{|\tau|^2 \cdot |\eta(\tau)|^6}{|e^{\frac{i \pi z^2}{2\tau}}|^2 \cdot \vartheta(z;\tau) \cdot \vartheta(0;\tau)} e^{\frac{\pi i c}{6} \frac{z^2}{\tau}}  Z(\tau, \bar{\tau},z) \ ,
\end{align}
for the partition function of $\mathcal{N}=(2, 2)$ SCFT.
Meanwhile, we reparametrize $\tau \equiv i e^{t}$ and $y \equiv w e^{\frac{t}{2}}$, so that the modular crossing point corresponds to $t=0, \bar{t}=0$ and $\omega=0$.
We use the reduced characters and partition function to express
the modular bootstrap equations \eqref{N1bootstrap}, \eqref{N2bootstrap} without the phase factor,
which makes the conversion to the SDP problems simple and efficient.

%%%%%%%%%%%%%%%%%%%%%%%%%%%%%%%%%%%%%%%%%%%%%%%%%%%%%%
\subsection{HFK-type Bound for the Charged States}
%%%%%%%%%%%%%%%%%%%%%%%%%%%%%%%%%%%%%%%%%%%%%%%%%%%%%%

Let us consider a quantum theory of gravity in AdS$_3$
coupled to a gauge field. It is natural to ask if one can prove
the universal energy scale beyond which a new massive charged state has to appear.
If not, there may exist a consistent theory of gravity
with a gauge field having only neutral states. Although
various arguments have been made on the existence of light charged
states under some assumptions, a generic and rigorous proof is still
missing. The above question can be translated into a more concrete
question in the boundary CFT with global symmetry, searching
for an upper bound below which light charged primaries must be
present.

To place a bound on the gap to the lightest charged primary,
let us consider the NS partition function (\ref{NSpartition})
blind to spin but aware of the charge,
\begin{align}
  Z (\tau = -\bar{\tau}=i\b, z, \bar z)_{\textrm{NS}} = \Tr_{\mathcal{H}_{\textrm{NS}}}
  \left[ e^{- 2\pi \b (L_0+\bar L_0 - \frac{c}{12})} y^Q
  \bar y^{\bar Q} \right].
\end{align}
where $y$ and $\bar y$ are the fugacities for the left and right-moving flavor symmetry.
For simplicity, let us restrict our attention to a theory with an abelian global symmetry,
and turn off the fugacity $\bar y$ for the right-moving charge $\bar Q$.
It is straightforward to generalize the analysis given below
to theories with non-abelian global symmetry and with the $\bar y$ dependence restored.
An explicit expression of the linear functional used in this and the next
subsection is
\begin{align}
\label{HFK linear functional2}
  \b_1 = \sum_{\ell=0}^N \sum_{m+2n=2\ell+1} \b_{m,n}
  \left. \left( \b\frac{\partial}{\partial \b} \right)^m
  \left( \frac{\partial}{\partial \omega} \right)^{2n}
  \right|_{\b=i, \omega=0},
\end{align}
where $y = \sqrt{\b}\omega$.

In \cite{Benjamin2016}, the authors proposed an upper bound on the dimension of $U(1)$ charged primaries in the large central charge limit. They showed that the scaling dimension of a charged primary is bounded above as
\begin{align}
\label{Charged HFK}
\D \le \frac{c}{6} + \mathcal{O}(1) \ .
\end{align}
It was shown by constructing an explicit linear functional of type \eqref{HFK linear functional2} with dimension $N=4$.
Our goal in this subsection is to improve this upper bound \eqref{Charged HFK}
using the semi-definite programming and a higher dimensional linear functional.
Specifically, we investigate the bound for the $U(1)$ R-charge of non-BPS states in
$\CN=(2,2)$ theories. To this end, we seek a linear functional $\b_1$ that satisfy,
\begin{align}
\label{N=2 CHFK}
\begin{split}
  \b_1 \Big[\CZ_\text{vac} (i\b, -i\b, z,0) \Big] & = 1,
  \\
  \b_1 \Big[ \CZ_{h=\frac{\D \pm j}{2},\bar h=\frac{\D\mp j}{2}}^{Q=0,\bar Q} (i\b, -i\b, z,0)  \Big] & \geq 0 \quad \mbox{for} \quad \D \ge 0 \\
  \b_1 \Big[\CZ_{h=\frac{\D \pm j}{2},\bar h= \frac{\D\mp j}{2}}^{Q\neq0,\bar Q} (i\b, -i\b, z,0)   \Big] & \geq 0 \quad \mbox{for} \quad \D \ge \D_*.
\end{split}
\end{align}
We do not impose constraints for the BPS states since their scaling dimension is completely fixed by the R-charge. When we find a linear functional satisfying the above conditions, we rule out the spectrum where all the charged non-BPS states have scaling dimension greater than $\Delta_\ast$. Therefore $\Delta_\ast$ provides an upper bound on the spectrum of non-BPS charged states.

Note that (\ref{N=2 CHFK}) has two continuous parameters $\D$ and $Q$. It is in fact challenging to convert
the modular bootstrap equation (\ref{N=2 CHFK}) with two variables to a positive semi-definite
programming problem. This conversion is related to the Hilbert's $17$th problem.
To circumvent this issue, we take the approach developed in \cite{Dyer2017} where
the search space is restricted to the functional expressed as a sum of squares of polynomials.
A brief review of the approach is given in appendix \ref{app:PMP}.

\begin{figure}[h!]
\begin{center}\includegraphics[width=.85\textwidth]{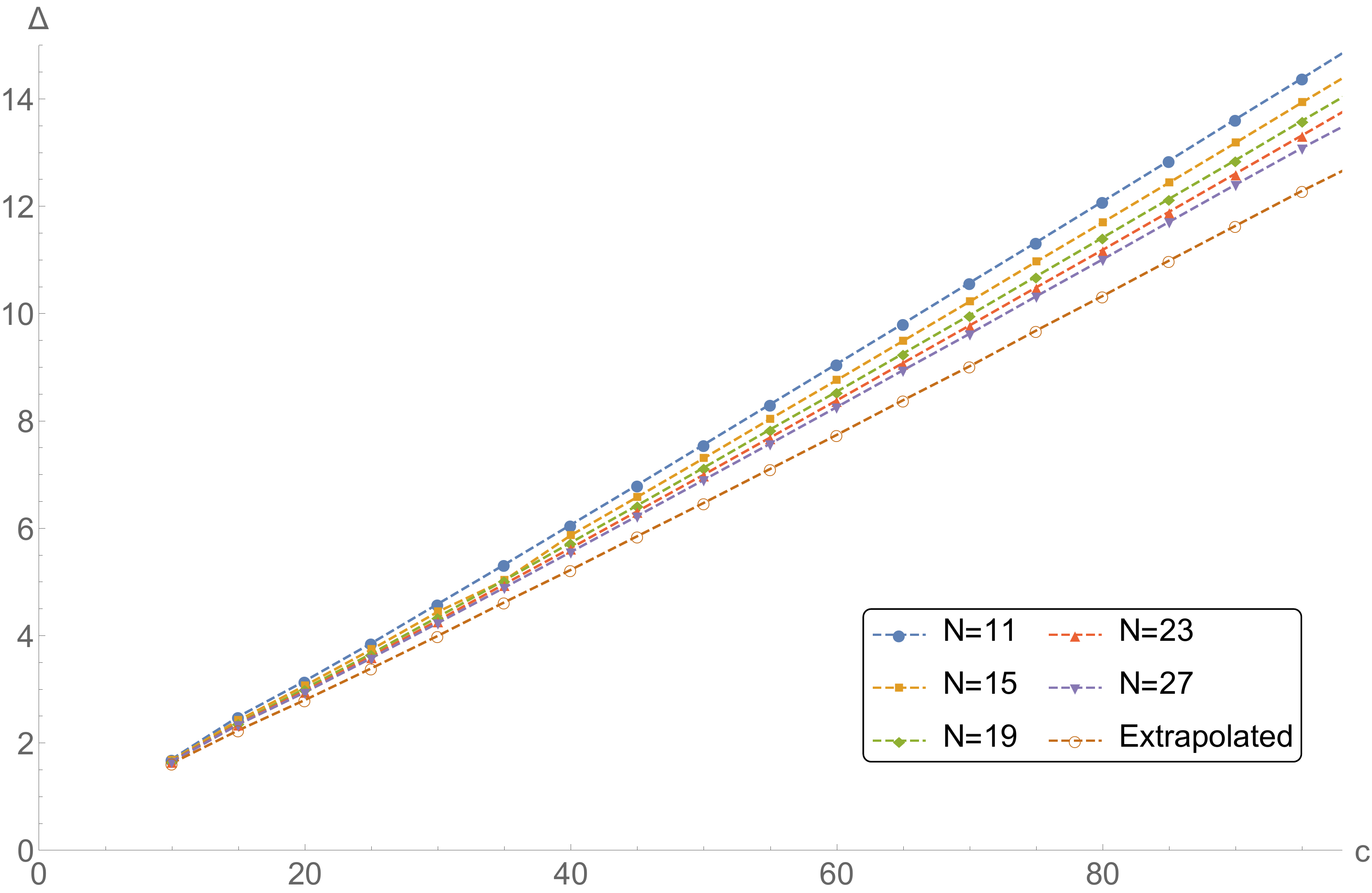}
\end{center}
\caption{The upper bound on the dimension of the lowest charged primary in $\CN=(2, 2)$ SCFT. The bound improves as we increase $N$. }
\label{N=2HFKQ}
\end{figure}

The numerical upper bound on the $\D_\ast$ we obtain is presented in figure \ref{N=2HFKQ}. In the large central charge limit, we find that the slope extrapolates to $0.12597$. Based on this observation, we conjecture that the upper bound on the $\D_\ast$ of charged (non-BPS) state is constrained as
\begin{align}
\D \le \frac{c}{8}  + \mathcal{O}(1) \ .
\end{align}
Our result implies that any $\CN=(2, 2)$ SCFTs must have at least one light charged (non-BPS) state since not all the charged states can have scaling dimension larger than $\frac{c}{8}$. Or to put in another way, the lightest charged (non-BPS) state cannot be heavier than $\frac{c}{8}$. We also investigate the lower bound on the charged states in the following subsection.

%%%%%%%%%%%%%%%%%%%%%%%%%%%%%%%%%%%%%%%%%%%%%%%%%%%%%%
\subsection{Charge Bounds}
%%%%%%%%%%%%%%%%%%%%%%%%%%%%%%%%%%%%%%%%%%%%%%%%%%%%%%

It is clear from (\ref{S01}) that a CFT with Kac-Moody symmetry
has to contain charged states in the spectrum.
It is then natural to ask what charges must be present.
In \cite{Benjamin2016}, the authors addressed this question and
derived a constraint on the $U(1)$ charge, with
a very simple linear functional of the form
\begin{align}
  \a\Big[f(\b,z)\Big] =  \left. \left( \frac{\partial}{\partial z} \right)^2  f(\b, z)
  \right|_{\b=1,z=0}.
\end{align}
Applying this to the modular constraint, they show that arbitrary CFT with a $U(1)$ global symmetry requires
\begin{align}
  \sum_{i} e^{-2\pi \D_i} \left( Q_i^2 - \frac{1}{4\pi} \right)  = 0  ,
  \label{Charge Constraint}
\end{align}
to be satisfied, when $U(1)$ current is normalized to have level one $k=1$.
Here the sum is taken over all states including primaries and descendants with scaling dimension $\Delta_i$ and $U(1)$-charge $Q_i$.
The identity (\ref{Charge Constraint}) implies that in an any two-dimensional CFTs with $U(1)$ symmetry there exists at least one charged state with $Q^2 > 1/(4\pi)$, otherwise it violates the modular constraint.

The above simple constraint on charge $Q^2 > 1/(4\pi)$ can be significantly improved for the case of $\CN=(1,1)$ and $\CN=(2,2)$ SCFTs. To see this, we search for a spin-dependent linear functional $\b_2$ of the form
\begin{align}
\label{LF2_type2}
  \b_2 = \sum_{p=0}^N \sum_{m+n+2l=2p+1} \tilde \b_{m,n,l}
  \left. \left(  \frac{\partial}{\partial t} \right)^m
  \left( \frac{\partial}{\partial \bar t} \right)^n \left( \frac{\partial}{\partial \omega} \right)^{2l}
  \right|_{t=\bar t=\omega=0},
\end{align}
with $\tau \equiv i e^t$ and $y \equiv e^{2\pi i z} \equiv \omega e^{\frac{t}{2}}$,
satisfying the relations below
\begin{align}
\begin{split}
  \b_2\Big[\CZ_\text{vac} (\t, \bar \t, z,0) \Big]&=1,
  \\
  \b_2 \Big[ \CZ_{h,\bar h}^{Q,\bar Q}
  (\t, \bar \t, z,0) \Big] & \geq 0 \quad \text{for} \quad |Q| < Q_* ,
\end{split} \label{eq:MChighestQN1}
\end{align}
for the $\CN=(1,1)$ theory, and
\begin{align}
\begin{split}
  \b_2\Big[\CZ_\text{vac}(\t, \bar \t, z,0)  \Big]&=1,
  \\
  \b_2 \Big[\CZ_{h,\bar h}^{Q,\bar Q}
  (\t, \bar \t, z,0)  \Big] & \geq 0 \quad \text{for} \quad |Q | < Q_*,
  \\
  \b_2 \Big[ \CZ_{h,\bar h}^\text{rest} (\t,\bar \t,y,0) \Big] & \geq 0,
\end{split} \label{eq:MChighestQN2}
\end{align}
for the $\CN=(2,2)$ theory. $\CZ_{h,\bar h}^\text{rest}$ collectively denote
half and quarter BPS state contributions, i.e., $\CZ_{h,\bar h}^{r,\bar r}$,
$\CZ_{h,\bar h}^{r,\bar Q}$ and $\CZ_{h,\bar h}^{Q,\bar r}$. As before, we only impose the charge condition on the non-BPS states. This is to find inconsistency in the spectrum under the assumption that all the states have charges lower than $Q_\ast$. If such a functional exists, there cannot be a modular-invariant theory
unless one state with charge greater than $Q_\ast$ exists.
In other words, any unitary SCFTs must have at least one charged primary state with $Q>\text{min}(Q_\ast(c))$.

\begin{figure}[h!]
\begin{center}\includegraphics[width=.85\textwidth]{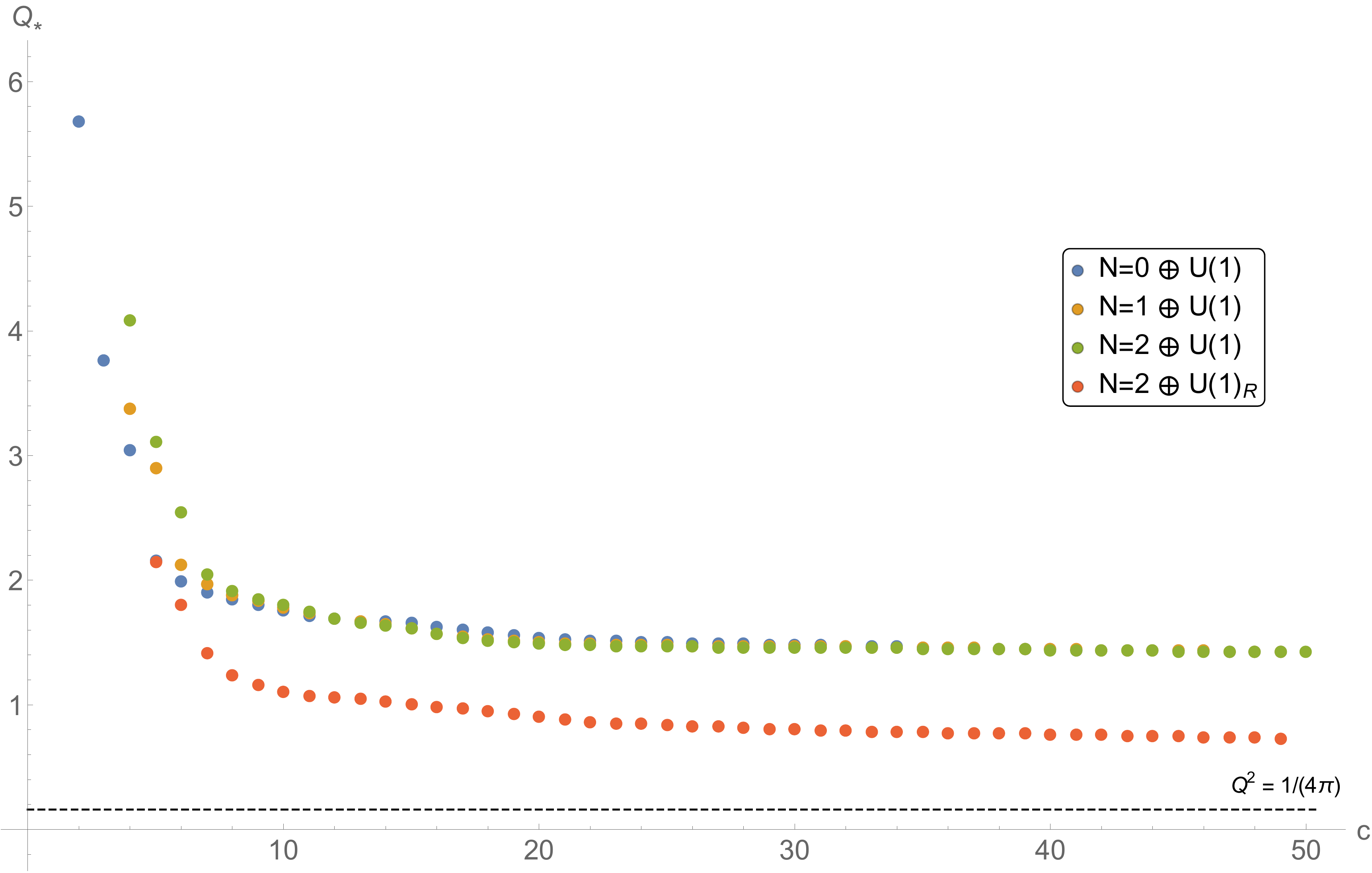}
\end{center}
\caption{Lower bounds on the highest charge. There must be at least one charged state with the charge greater than the bound. The level of $U(1)$ current was normalized by 1. Notice that the three cases for the non-R $U(1)$ charge bounds almost overlap.}
\label{Charge Upper Bound}
\end{figure}

Figure \ref{Charge Upper Bound} shows the numerical lower bound on $Q_\ast$ versus the central charge $c$ for various superconformal theories.
We find improved lower bounds on $Q_\ast$ as a function of $c$ compared to the previous bound $\frac{1}{\sqrt{4\pi}}$.  For the non-R symmetry cases,
we find that the numerical bounds overlap as the central charge becomes large.
Notice that the bounds become nearly flat and become lower as $c \to \infty$.
There is a sharp distinction between R and non-R symmetries, namely the constraint is weaker for the R-charges. Note here that
$U(1)$ R-current is now normalized to have level one rather than canonical level $c/3$ to compare
various results coherently. Therefore, we get a universal constraint for the highest charged state as follows:
\begin{align}
 Q_{\textrm{highest}} \ge \left\{ \begin{array}{rl} 1.42391 & \textrm{for non-R symmetry} \\
  0.73089  & \textrm{for R-symmetry}
  \end{array} \right.
\end{align}

To find the upper bounds on the lowest $U(1)$ charge, we look for the spin-dependent linear functional $\b_2$ satisfying
\begin{align}
\begin{split}
  \b_2\Big[\CZ_\text{vac} (\t, \bar \t, z,0)  \Big] &= 1,
  \\
  \b_2  \Big[ \CZ_{h,\bar h}^{Q=0,\bar Q} (\t, \bar \t, z,0) \Big] & \geq 0,
  \\
  \b_2 \Big[ \CZ_{h,\bar h}^{Q,\bar Q} (\t, \bar \t, z,0)  \Big] & \geq 0 \quad \text{for} \quad |Q| > Q_*
\end{split}
\end{align}
for the $\CN=(1,1)$ theory, and
\begin{align}
\begin{split}
  \b_2\Big[\CZ_\text{vac} (\tau, \bar{\tau}, z, 0) \Big] & =1,
  \\
  \b_2  \Big[ \CZ_{h,\bar h}^{Q=0,\bar Q} (\t, \bar \t, z,0) \Big] & \geq 0,
  \\
  \b_2 \Big[ \CZ_{h,\bar h}^{Q,\bar Q} (\t, \bar \t, z,0)  \Big] & \geq 0 \quad \text{for} \quad |Q| > Q_*,
  \\
  \b_2 \Big[ \CZ_{h,\bar h}^\text{rest} (\t,\bar \t,y,0) \Big] & \geq 0
\end{split}
\end{align}
for the $\CN=(2,2)$ theory.  %

\begin{figure}[h!]
\begin{center}\includegraphics[width=.85\textwidth]{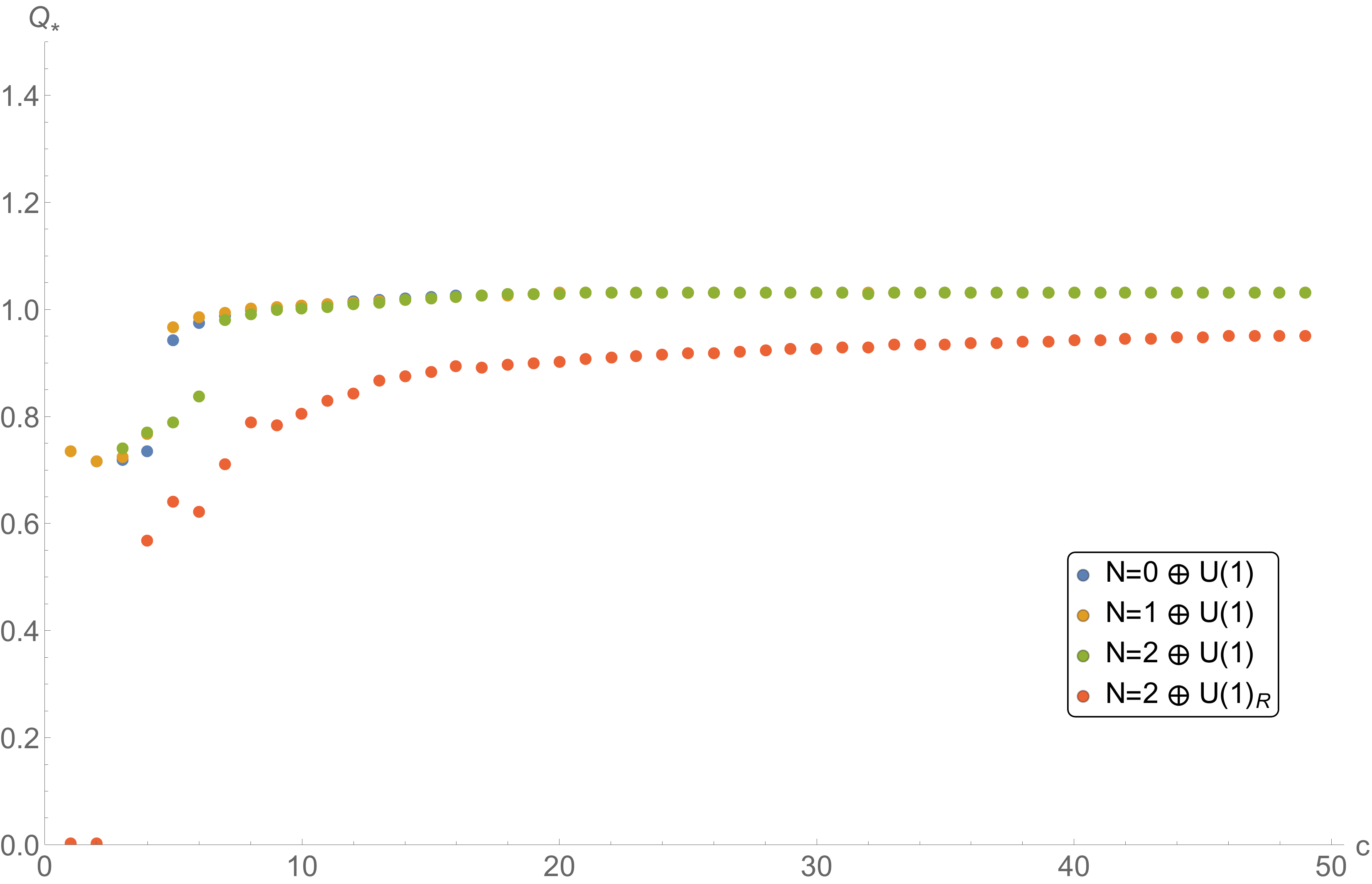}
\end{center}
\caption{Upper bounds on the lowest charge. There must be at least one charged state with the charge lower than the bound. The level of $U(1)$ current was normalized by 1. Notice that the three cases for the non-R $U(1)$ charge bounds almost overlap. }
\label{Charge Lower Bound}
\end{figure}

The numerical upper bounds on the lowest charge we obtain for various theories are depicted in figure \ref{Charge Lower Bound}. For a non-R $U(1)$ flavor symmetry, it turns out that the lowest charge should be constrained by $Q_{\ast} \le 1$ regardless of the amount of supersymmetry. We get a stronger numerical bound for the $U(1)$ R-symmetry of the $\CN=(2,2)$ theory. We again normalize
the $U(1)$ R-current to have non-canonical level $k=1$. For the bounds on the canonically normaized $U(1)_R$ charge, one should multiply $\sqrt{c/3}$ to the result in figure \ref{Charge Lower Bound}.
We find that the upper bounds on the lowest charge become constant as we increase the central charge so that we find the universal constraint
\begin{align}
 Q_{\textrm{lowest}} \le \left\{ \begin{array}{rl} 1.03125 & \textrm{for non-R symmetry} \\
  0.94978  & \textrm{for R-symmetry}
  \end{array} \right.
\end{align}
The bounds for the for small central charges for the non-R is lower than $1$. Also, the bounds for the R charge, canonically normalized, is lower than $\sqrt{c/3}$ for every value of $c$.

%%%%%%%%%%%%%%%%%%%%%%%%%%%%%%%%%%%%%%%%%%%%%%%
\subsection{On Weak Gravity Conjecture}
%%%%%%%%%%%%%%%%%%%%%%%%%%%%%%%%%%%%%%%%%%%%%%%

The Weak Gravity Conjecture (WGC) \cite{Arkani-Hamed2006} asserts that there must be a light charged particle in any $U(1)$ gauge theory consistently coupled to gravity. It is part of the more general program trying to explore constraints on the low-energy effective theory that can be UV completed consistently while coupling to quantum gravity \cite{Vafa:2005ui,Ooguri:2006in}. 

This conjecture predicts the existence of light charged particle of mass $m$ and $U(1)$ charge $q$ satisfying
\begin{align}
  m < q \left( g M_\text{pl} \right) \ ,
\end{align}
in four or higher dimensional space-time\cite{Arkani-Hamed2006}. A heuristic argument for the WGC is based on several assumptions such as Cosmic Censorship Hypothesis, the legitimate $g\to 0$ limit, and the validity of semi-classical description of black holes to the Planck scale, and has its own subtleties and loopholes.

Therefore, it would be interesting to verify the WGC in the context of AdS$_3$/CFT$_2$ correspondence, where the conjecture can be translated into a concrete statement in CFTs. More precisely, we would like to check if the modular invariance of two-dimensional CFT can guarantee the existence of the light charged states.

There are qualitative differences between three and higher dimensional WGC. First, the arguments that led to WGC in higher dimensions cannot be directly translated into AdS$_3$, since the gravitons do not have local degrees of freedom in AdS$_3$. Second, at least classically, the charged BTZ black hole has no obvious notion of extremality: the $U(1)$ gauge charge can be supported only by the flat connection along the non-contractible circle at infinity. In other words, having a finite electric charge does not seem to increase the energy and thus back-react to the geometry. Therefore, it is not clear whether the WGC has to be true in AdS$_3$ or not.

A couple of different proposals for the WGC have been proposed. One of such proposals is a straight-forward generalization of \cite{Nakayama2015}. The authors of \cite{Nakayama2015} have proposed that the WGC in AdS$_5$ can be written in terms of CFT$_4$ data as
\begin{align}
  \frac{\D^2}{Q^2} \lesssim  \frac{c_T}{c_V} \ ,
  \label{proposal1}
\end{align}
where $c_T$ and $c_V$ are the central charges of the stress-energy tensor and the conserved current, defined as
\begin{align}
\begin{split}
   \langle J_\mu(x) J_\nu(0) \rangle & = \frac{c_V}{x^6} I_{\mu\nu}(x) \ ,
    \\
   \langle T_{\mu\nu}(x) T_{\rho\sigma}(0) \rangle & = \frac{c_T}{x^8} I_{\m\n,\r\s}(x) \ ,
\end{split}
\end{align}
with
\begin{align}
  I_{\m\n} = \d_{\m\n} - \frac{x_\m x_\n}{x^2}\ , \qquad
  I_{\m\n,\r\s} = \frac{ I_{\m\r} I_{\n\s} + I_{\m\s} I_{\n\r} }{2}
  - \frac14 \d_{\m\n} \d_{\r\s} \ .
\end{align}
One can imagine that (\ref{proposal1}) may continue to hold even in two-dimensional CFTs. In other words, any unitary two-dimensional CFTs having holographic dual description contain a charged state of scaling dimension $\D$ and charge $Q$
satisfying
\begin{align}
  \frac{\D^2}{Q^2} \lesssim \frac{c}{k}\ .
  \label{proposal001}
\end{align}

Another proposal was made in \cite{Montero2016}. It is known that, in the bulk side, one can define the Brown-Henneaux stress-energy tensor $T^\text{BH}$ \cite{Brown:1986nw}
\begin{align}
  T^\text{BH}_{ij} = \frac{1}{8\pi G_\text{N} \ell_\text{AdS}} \Big[ g^{(2)}_{ij} - \text{Tr}\big[ g^{(2)}\big] g_{ij}^{(0)} \Big] \ ,
  \label{sum01}
\end{align}
where $g_{ij}^{(0)}$ is the conformal boundary metric and $g_{ij}^{(2)}$ is the the leading correction to an asymptotic expansion of the metric. In addition to $T^\text{BH}$, the authors of \cite{Montero2016} have shown that there is an additional contribution to the stress energy tensor from the boundary term $T^{U(1)}$. It has the form of
\begin{align}
  T^{U(1)}_{ij} = \frac{k}{4\pi} \Big[ A_i A_j - \frac12 A^k A_k g_{ij} \Big]\ ,
  \label{sum02}
\end{align}
and it supplements the Chern-Simons coupling in the bulk AdS$_3$. Then, the total stress-energy tensor $T^\text{tot}$ is a sum of two terms (\ref{sum01}) and (\ref{sum02}), i.e.,
\begin{align}
  T^\text{tot}_{ij} = T^\text{BH}_{ij} + T^{U(1)}_{ij}\ .
  \label{energy}
\end{align}

From the above description, one can easily show that the mass of a charged state is constrained to satisfy
\begin{align}
  M \ell_\text{AdS} \geq \frac{Q^2}{2k}\ .
  \label{bound}
\end{align}
It implies that piling up an electric charge in fact costs energy, and thus the charged black holes in AdS$_3$ can indeed have a notion of extremality. As discussed in \cite{Montero2016}, one can read off the CFT notion of the extremality bound from (\ref{energy}). This means that a state in CFT$_2$ with $\D \geq \frac{c}{12} + \frac{Q^2}{2k}$ can have a semiclassical description as a BTZ black hole with charge $Q$. Therefore, we can state the WGC in AdS$_3$ in terms of CFT$_2$: any holographic two-dimensional CFTs should contain a light charged state satisfying
\begin{align}
  \D \leq \frac{c}{12} + \frac{Q^2}{2k}\ .
  \label{proposal2}
\end{align}

In order to support or falsify the two proposals (\ref{proposal1}) and (\ref{proposal2}), we investigate the bound on the scaling dimension of the lowest lying primary state in a sector of charge $Q$. To this end, we consider the modular constraints with the spin-independent linear functional $\b_1$ and appropriate assumptions in the spectrum.

\paragraph{Testing the first proposal}
\begin{figure}[h!]
\begin{center}\includegraphics[width=.85\textwidth]{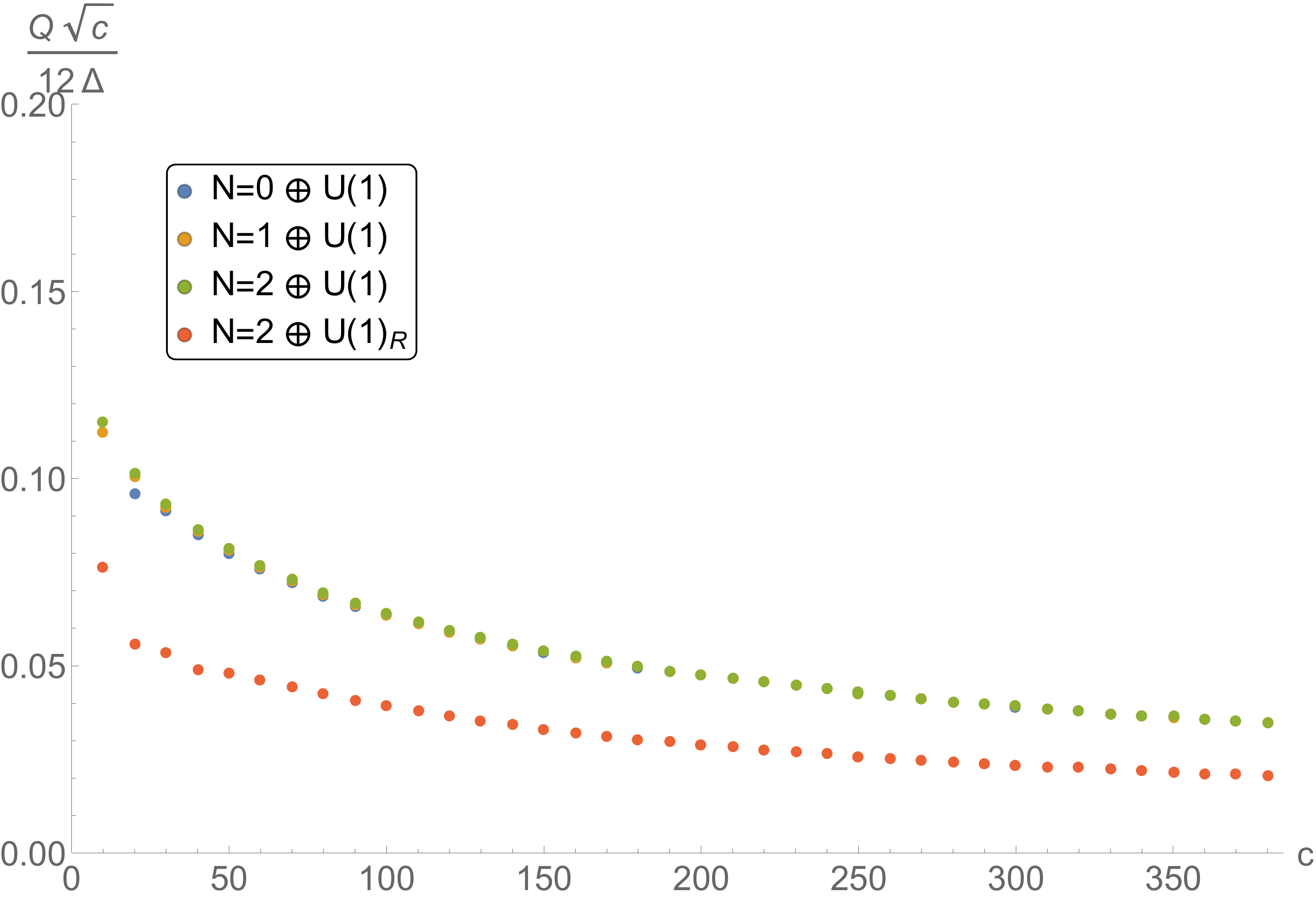}
\end{center}
\caption{Lower bounds on the charge to mass ratio. Non R-charge bounds almost overlap. Again, the level of $U(1)$ current was normalized by 1.}
\label{QMratio}
\end{figure}

In order to test the proposal \eqref{proposal1}, we impose a gap in the ratio of conformal dimension $\D$ to $U(1)$ charge $Q$. Assume that all the charged states have the mass to charge ratio greater than a certain value and look for an inconsistency. More precisely, we search for a spin-independent linear functional $\b_1$ satisfying\footnote{The parameter $\r$ was firstly introduced in \cite{Dyer2017} to
place bounds on the ratio of charge to dimension for bosonic CFTs with $U(1)$ flavor symmetry.}
\begin{align}
\begin{split}
  \b_1 \Big[ \CZ_\text{vac}(i\b,-i\b,z,0) \Big] & = 1,
  \\
  \b_1 \Big[ \CZ_\D^{Q} (i\b,-i\b,z,0) \Big] & \geq 0
  \quad \text{for} \quad |Q| \le \frac{12 \Delta}{\sqrt{c}} \r ,
\end{split}
\end{align}
where $\CZ_\D^Q(\t,\bar \t,z,\bar z)$ denotes the contribution from each primary
of conformal dimension $\D$ and (left-moving) $U(1)$ charge $Q$ collectively.
Here we normalize the $U(1)$ current so that the level is given by one, i.e., $k=1$. Figure \ref{QMratio} shows the lower bounds on $\r$ as a function of $c$. We observe that the numerical lower bound on $\frac{Q}{\D}$ becomes linear in $c$ in the large central charge limit. This implies that any bosonic or superconformal CFTs with a non-R $U(1)$ symmetry should contain a light charged state satisfying $\frac{\D}{Q} \lesssim c$. We thus conclude that our numerical results $\frac{\D}{Q} \lesssim c$ parametrically disagree with (\ref{proposal001}), which predicts the behavior of $\frac{\D}{Q} \lesssim \sqrt{c}$.

\begin{figure}[h!]
\begin{center}\includegraphics[width=.85\textwidth]{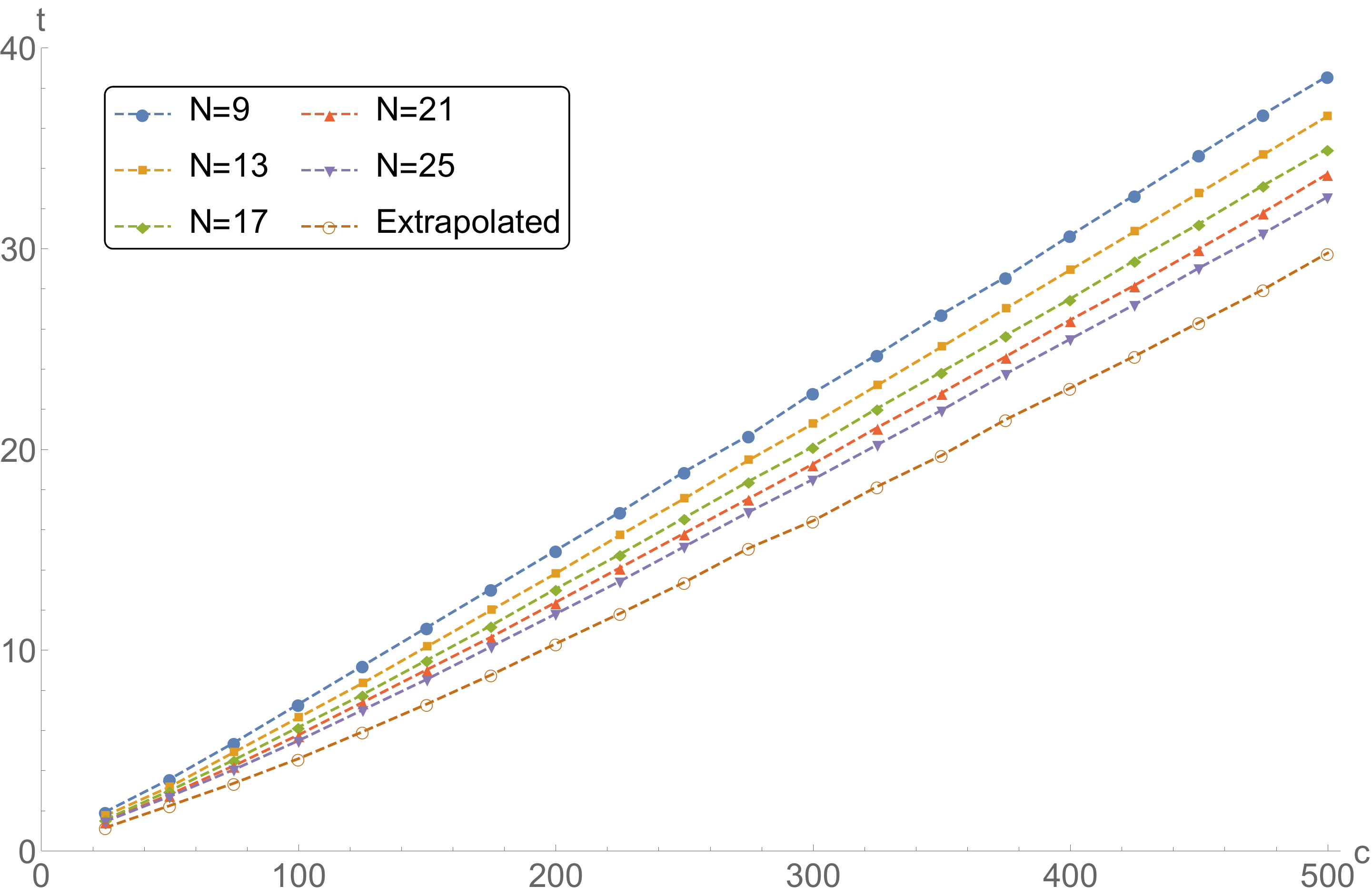}
\end{center}
\caption{Testing WGC: We find that $\Delta - \frac{3Q^2}{2c}$ is bounded above.}
\label{WGC}
\end{figure}

\paragraph{Testing second proposal}

Let us now we examine the second proposal \eqref{proposal2} using the modular bootstrap.
To be more specific, we test the proposal for the WGC in the context of $\CN=2$
supergravity in AdS$_3$.
For this purpose, we seek a spin-independent linear functional $\b_1$ that act positively on the contribution from the vacuum state and non-negatively on the contributions from other $\CN = 2$ primaries with bounded $U(1)$ R-charges. More precisely, restrict the charges to satisfy
\begin{align}
  Q^2 \le \frac{2c}{3} \left(\D-\frac{c}{12}-t \right)
  ~\text{ with }~ t \in \mathbb{R}_{\geq0} \ .
\end{align}
Figure \ref{WGC} shows the numerical upper bounds on $t$ we obtained. We plot bounds for some fixed values of $N$, the dimension of the linear functional $\a$, and then extrapolate to $N \to \infty$. We obtain the extrapolated slope $dt/dc$ to be approximately $0.061137$, or roughly $\frac{1}{16}$. Based on the numerical results, we conjecture that any $\CN=(2,2)$ SCFT contains a light charged state satisfying
\begin{align}
\D \le \frac{7 c}{48} + \frac{3Q^2}{2c}
\label{WGC conjecture}
\end{align}
which is compatible with \eqref{proposal2} proposed by \cite{Montero2016}. It is not clear if the numerical bound (\ref{WGC conjecture}) weaker than \eqref{proposal2} is a short-coming of the numerical method or if there exists a physical theory saturating the weaker bound. We however suspect that a more numerically sophisticated and improved machinery would provide the ultimate bound \eqref{proposal2}.

%%%%%%%%%%%%%%%%%%%%%%%%%%%%%%%%%%%%%%%%%%%%
%%%%%%%%%%%%%%%%%%%%%%%%%%%%%%%%%%%%%%%%%%%%
\section*{Acknowledgments}
We would like to thank Liam Fitzpatrick, Jeffrey Harvey, Emil Martinec, Sunil Mukhi, Kimyeong Lee, Gary Shiu and Yuan Xin for useful discussions.
We thank KIAS Center for Advanced Computation for providing computing resources.
The research of S.L. is supported in part by the National Research Foundation of Korea (NRF) Grant NRF-$2017$R$1$C$1$B$1011440$.
S.L. and J.S. thank the Aspen Center for Physics (supported by National Science Foundation grant PHY-1607611), Asian Pacific Center for Theoretical Physics and Galileo Galilei Institute for Theoretical Physics for the hospitality. 
J.S. thanks the Simons Center for Geometry and Physics during the Simons Summer Workshop 2018 for hospitality. J.B. also thanks the Bootstrap 2018 workshop at Caltech for hospitality.

%%%%%%%%%%%%%%%%%%%%%%%%%%%%%%%%%%%%%%%%%%%%%%%%%%%%%%%%%%%%%%%%%%%%%
\appendix
\section{Modular Differential Equation with $\Gamma_\theta$} \label{app:MDE}
%%%%%%%%%%%%%%%%%%%%%%%%%%%%%%%%%%%%%%%%%%%%%%%%%%%%%%%%%%%%%%%%%%%%%

It is known that the characters of rational conformal field theories (RCFT) are given by the solutions of $n$-th order modular differential equation\cite{Mathur1988}. Such modular differential equation (MDE) takes the form of
\begin{align}
\label{MDE}
 \left[ \CD_\tau^n + \sum_{k=0}^{n-1} \phi_{2(n-k)}(\tau) \CD_\tau^k \right] f(\tau)= 0,
\end{align}
where $\CD_\tau^n$ is an abbreviation of $n$ successive action of the covariant Serre derivative
\begin{align}
 \CD_\tau^n \equiv \CD_{\tau, 2n} \circ \CD_{\tau, 2n-2} \circ \ldots \circ \CD_{\tau, 2} \circ \CD_{\tau, 0} \ .
\end{align}
Here the Serre derivative on a modular form of weight $r$ is defined as $\CD_{\tau, r} \equiv \partial_\tau - \frac{i \pi r}{6} E_2(\tau)$ with $E_2(\tau)$ being the second Eisenstein series of weight two.
The coefficient $\phi_{2k}(\t)$ is a modular form of weight $2k$, which has singularities at the zeros of the Wr\"onskian $W=\det W^i_{\ j}$ with $W^i_{\ j} = \CD^i_\t f_j(\t)$ and $i,j=1,\ldots, n$.
Since $\phi_{2k}(\t)$ is a modular form, it can be expressed as a linear combination of the monomials generated by the Eisenstein series $E_4 (\tau)$ and $E_6 (\tau)$. It was found that many of the interesting rational theories that are given by the solutions to the MDE (such as WZW models for the Deligne exceptional series with level 1) are also realized at the numerical boundaries of the modular bootstrap program \cite{BLS}.

In this section, we would like to apply the same idea of using MDE to the SCFT. Our central object is the NS sector partition function, which is invariant only under the congruence subgroup $\Gamma_\theta$, not the full modular group $SL(2,\mathbb{Z})$ \footnote{The congruence subgroup $\Gamma_\theta$ is not the same as $\Gamma_0(2)$ or $\Gamma^0(2)$. The latter acts on the R-NS, NS-R sector partition functions. The modular differential equation for this case is the one studied in \cite{Beem:2017ooy}. On the other hand, we focus on the NS-NS partition function.}. Therefore, $\phi_{2k}(\t)$ is now given by a modular form of $\Gamma_\theta$, whose transformation law is given by
\begin{align}
\label{modular transformation}
M_{2k}(\tau+2) = M_{2k}(\tau), \qquad M_{2k}(-1/\tau) = \tau^{2k} M_{2k}(\tau).
\end{align}
A modular form of weight $2k$ can be constructed using the Jacobi-theta functions. For example, the function
\begin{align}
\label{Modular Forms2}
M_2(q) &\equiv \vartheta_{01}(\tau)^4 - \vartheta_{10}(\tau)^4  = 1 -24 q^{\frac{1}{2}} + 24 q -96 q^{\frac{3}{2}} + 24 q^2  + \cdots.
\end{align}
is a $\Gamma_\theta$-modular form of weight two. Note that in contrast to the $SL(2,\mathbb{Z})$ case, there exists a weight-two modular form of $\Gamma_\theta$.
Other $\Gamma_\theta$ modular forms can be given in terms of Jacobi theta function as follows:
\begin{align}
\label{Modular Forms46}
\begin{split}
M_4^{(1)}(\tau) &= \vartheta_{00}(q)^8 = 1+16 q^{\frac{1}{2}} + 112 q + 448 q^{\frac{3}{2}} + 1136 q^2 + 2016 q^{\frac{5}{2}}  + \cdots  \\
M_4^{(2)}(\tau) &= \vartheta_{10}(q)^8 + \vartheta_{01}(q)^8 = 1-16 q^{\frac{1}{2}} + 368 q - 448 q^{\frac{3}{2}} + \cdots \\
M_6^{(1)}(\tau) &= \vartheta_{01}(q)^{12} - \vartheta_{10}(q)^{12} + \vartheta_{10}(q)^{8} \vartheta_{01}(q)^{4} - \vartheta_{10}(q)^{4} \vartheta_{01}(q)^{8}  \\
           &= 1-40 q^{\frac{1}{2}} + 776 q -9760 q^{\frac{3}{2}} + 24328 q^2 -125040 q^{\frac{5}{2}}  + \cdots  \\
M_6^{(2)}(\tau) &= \vartheta_{01}(q)^{12} -  \vartheta_{10}(q)^{12}+ \vartheta_{10}(q)^{4} \vartheta_{01}(q)^{8} -  \vartheta_{10}(q)^{8} \vartheta_{01}(q)^{4}  \\
           &= 1-8 q^{\frac{1}{2}} -248 q -1952 q^{\frac{3}{2}} -8440 q^2 -25008 q^{\frac{5}{2}}  + \cdots,
\end{split}
\end{align}
Let us note that the above modular forms are related to the Eisenstein series as
\begin{align}
\begin{split}
M_4^{(1)}(\tau)  + M_4^{(2)}(\tau)&=E_4(\tau)  \ , \\
5M_6^{(2)}(q)  - M_6^{(1)}(q)&= 4E_6(\tau) \ .
\end{split}
\end{align}
This is expected since the $\Gamma_\theta$ is a subgroup of the $SL(2,\mathbb{Z})$.

The modular differential equation for the $\Gamma_\theta$ can be constructed using the modular forms \eqref{Modular Forms2} and \eqref{Modular Forms46}.
The second order modular differential operator should have the modular weight four, therefore it has the following form:
\begin{equation}
\label{2ndSMDE}
\left[ \mathcal{D}_\tau^2 + \mu_1 M_2(\tau) \mathcal{D}_{\tau}  + \mu_2 M_4^{(1)}(\tau)  + \mu_3  M_4^{(2)}(\tau) \right] f(\tau) = 0,
\end{equation}
with undetermined constants $\mu_1,\mu_2$ and $\mu_3$.
Similarly, the third order modular differential equation can be written in the form of
\begin{align}
\label{3rdSMDE}
\begin{split}
\left[ \mathcal{D}_\tau^3  + \mu_1 M_2 \mathcal{D}_\tau^2 + \mu_2 M_4^{(1)}(\tau) \mathcal{D}_\tau  + \mu_3 M_4^{(2)}(\tau) \mathcal{D}_\tau  + \mu_4 M_6^{(1)} + \mu_5 M_6^{(2)}   \right] f(\tau) = 0 \ ,
\end{split}
\end{align}
with undetermined constants $\mu_i$.

Let us look for the solutions of the second order modular differential equation \eqref{2ndSMDE} to have either of the following forms:
\begin{align}
\label{2ndMDEansatz}
\begin{split}
f_1(\tau) &= q^{-\frac{c}{24}} \left( 1 + a_1 q^{1/2} + a_2 q + a_3 q^{3/2} + a_4 q^2 + a_5 q^{5/2} + \cdots \right) \ ,  \\
f_2(\tau) &= q^{-\frac{c}{24}} \left( 1 + b_2 q + b_3 q^{3/2} + b_4 q^2 + b_5 q^{5/2}+ \cdots \right) \ .
\end{split}
\end{align}
They can be identified as vacuum characters of certain rational SCFTs (RSCFT). By plugging in the ansatz to the MDE \eqref{2ndMDEansatz}, it is possible to determine the constants $\mu_i$ in the MDE and also the coefficients $a_i$ and $b_i$.
For example, when $c=4$, the vacuum character $f_1 (\tau) = q^{-1/6}(1+8 q^{1/2} + 28 q + 64 q^{3/2} + 134 q^2 + 288 q^{5/2} + \cdots)$ satisfy the MDE
\begin{align}
 \left[ \mathcal{D}_\tau^2 + \left( \frac{1}{2} - h \right) M_2(\tau) \mathcal{D}_{\tau} + \frac{6h-4}{36}   M_4^{(1)}(\tau) + \frac{5-12h}{36}  M_4^{(2)}(\tau) \right] f_1(\tau) = 0 ,
\end{align}
regardless of the value $h$. Some of the solutions are listed in table \ref{2ndSMDETable}. Our solutions in table \ref{2ndSMDETable} exactly agree with the characters of the vertex operator superalgebra (VOSA) studied in \cite{Hoehn2007}.

\begin{table}[t]
\centering
{%
\begin{tabular}{|c |c || c| c| }
\hline
  \rule{0in}{2.5ex}$c$ & $f_1(\t)$ &  $c$ &$f_2(\t)$
  \\ \hline \hline
  \rule{0in}{5ex}$\frac{3}{2}$ & $q^{-\frac{1}{16}} \left[ 1 + 3 q^{\frac{1}{2}} + 3 q + 4 q^{\frac{3}{2}} + \cdots \right]$ & $9$ &
  $q^{-\frac{3}{8}} \left[ 1 + 261 q + 456 q^{\frac{3}{2}} +\cdots \right]$
  \\ [0.5ex]
  \rule{0in}{5ex}$2$ & $q^{-\frac{1}{12}} \left[ 1 + 4 q^{\frac{1}{2}} + 6 q + 8 q^{\frac{3}{2}} + \cdots \right]$ & $\frac{19}{2}$ &
   $q^{-\frac{19}{48}} \left[ 1 + 266 q + 703 q^{\frac{3}{2}} +\cdots \right]$
  \\ [0.5ex]
  \rule{0in}{5ex}$\frac{5}{2}$ & $q^{-\frac{5}{48}} \left[ 1 + 5 q^{\frac{1}{2}} + 10 q + 15 q^{\frac{3}{2}} +\cdots \right]$ & $11$ &
   $q^{-\frac{11}{24}} \left[ 1 + 275 q + 1496 q^{\frac{3}{2}} +\cdots \right]$
  \\ [0.5ex]
  \rule{0in}{5ex}$3$ & $q^{-\frac{1}{8}} \left[ 1 + 6 q^{\frac{1}{2}} + 15 q + 26 q^{\frac{3}{2}} +\cdots \right]$ & $12$ &
   $q^{-\frac{1}{2}} \left[ 1 + 276 q + 2048 q^{\frac{3}{2}} +\cdots \right]$
  \\ [1.5ex]
  \rule{0in}{2.5ex}$4$ & $q^{-\frac{1}{6}} \left[ 1 + 8 q^{\frac{1}{2}} + 28 q + 64 q^{\frac{3}{2}} +\cdots \right]$ & $13$ &
  $q^{-\frac{13}{24}} \left[ 1 + 273 q + 2600 q^{\frac{3}{2}} +\cdots \right]$
  \\ [1.5ex]
  \rule{0in}{2.5ex}$9$ & $q^{-\frac{3}{8}} \left[ 1 + 18 q^{\frac{1}{2}} + 153 q + 834 q^{\frac{3}{2}} +\cdots \right]$ & $14$ &
  $q^{-\frac{7}{12}} \left[ 1 + 266 q + 3136 q^{\frac{3}{2}} +\cdots \right]$
  \\ [1.5ex]
  \rule{0in}{2.5ex}$12$ & $q^{-\frac{1}{2}} \left[ 1 + 24 q^{\frac{1}{2}} + 276 q + 2048 q^{\frac{3}{2}} +\cdots \right]$ & $15$ &
   $q^{-\frac{5}{8}} \left[ 1 + 255 q + 3640 q^{\frac{3}{2}} +\cdots \right]$\\ [1.5ex]
   \hline
\end{tabular}%
\caption{Solutions for the second order modular differential equation.}
\label{2ndSMDETable}
}
\end{table}

The solutions $f_1(\t)$ are identified with the partition function of free fermions,
\begin{equation}
f_{1}(\tau) = \left[ \prod_{i=1}^{\infty} (1+ q^{i-\frac{1}{2}}) \right]^{2c}.
\end{equation}
For any number of free-fermions, the partition function of above solves the second order modular differential equation.
On the other hand, the solution $f_2(\t)$ at $c=12$ turns out to be the same as the $K$-function,
\begin{align}
K(\t) = \left(\frac{\eta(q)^2}{\eta(q^2)\eta(q^{\frac{1}{2}})}\right)^{24} - 24,
\end{align}
which was suggested as the partition function of $\mathcal{N}=1$ super extremal CFT\cite{Witten2007}.

We look for the solutions of the third order differential equation \eqref{3rdSMDE} with the following ansatzs
\begin{align}
\label{3rdSMDEAnsatz}
\begin{split}
g_1(\tau) &= q^{-\frac{c}{24}} \left( 1 + c_3 q^{3/2} + c_4 q^2 + c_5 q^{5/2}+ \cdots \right)  \ , \\
g_2(\tau) &= q^{-\frac{c}{24}} \left( 1 + d_4 q^2 + d_5 q^{5/2}+ \cdots \right) \ .
\end{split}
\end{align}
for vacuum character of certain rational CFTs. We found solutions that have the positive integer coefficients $c_i$ and $d_i$, some of which are listed in table \ref{3rdSMDETable}. Among them, the solutions $g_1(\tau)$ match with the extremal characters of the VOSAs found in \cite{Hoehn2007}.
On the other hand, there are no solutions of the type $g_2(\tau)$. This is because it cannot be decomposed into $\CN=1$ or $\CN=2$ super-Virasoro characters. Therefore, $g_2(\tau)$ cannot be the vacuum character of $\CN=1$ or $\CN=2$ superconformal field theories.

\begin{table}[t]
\centering
{%
\begin{tabular}{|c |c ||c| c| }
 \hline
  \rule{0in}{2.5ex}$c$ & $g_1(\t)$ & $c$ &  $g_2(\t)$
  \\ \hline \hline
  \rule{0in}{5ex}$16$ & $q^{-\frac{8}{24}} \left[ 1 + 7936 q^{\frac{3}{2}} + 2296 q^2 + \cdots \right]$ & $24$ &
  $q^{-1} \left[ 1 + 196884 q^2 +\cdots \right]$
  \\ [0.5ex]
  \rule{0in}{5ex}$\frac{33}{2}$ & $q^{-\frac{33}{48}} \left[  1 + 7766 q^{\frac{3}{2}} + 11220 q^2  + \cdots \right]$ & $\frac{49}{2}$ &
  $q^{-\frac{49}{48}} \left[ 1 + 188258 q^2 +\cdots \right]$
  \\ [0.5ex]
  \rule{0in}{5ex}$17$ & $q^{-\frac{17}{24}} \left[  1 + 7582 q^{\frac{3}{2}} + 19907 q^2  +\cdots \right]$ & $25$ &
   $q^{-\frac{25}{24}} \left[ 1 + 179675 q^2 +\cdots \right]$\\ [1.5ex]
   \hline
\end{tabular}%
\caption{The list of solutions for the third order modular differential equation.}
\label{3rdSMDETable}
}
\end{table}
%

%%%%%%%%%%%%%%%%%%%%%%%%%%%%%%%%%%%%%%%%%%%%%%%%%%%%%%%%%%%
%%%%%%%%%%%%%%%%%%%%%%%%%%%%%%%%%%%%%%%%%%%%%%%%%%%%%%%%%%%
\section{Polynomial matrix problem} \label{app:PMP}
%%%%%%%%%%%%%%%%%%%%%%%%%%%%%%%%%%%%%%%%%%%%%%%%%%%%%%%%%%%
%%%%%%%%%%%%%%%%%%%%%%%%%%%%%%%%%%%%%%%%%%%%%%%%%%%%%%%%%%%

The presence of $U(1)$ global symmetry entails additional chemical potential, thus one should translate two variable polynomial matrix problem to the semi-definite programming. Namely, we seek polynomial $f(x,y)$ of degree $d_x, 2d_y$ such that
\begin{align}
\begin{split}
f(x,y) \ge 0 \quad \mbox{for} \quad x \in \mathbb{R}_{\ge 0}, \ y \in \mathbb{R}.
\end{split}
\end{align}
The polynomial $f(x,y)$ is nonnegative if one can find semi-definite matrix $\mathbb{A}$ and $\mathbb{B}$ such that
\begin{align}
\label{Two sdp}
\begin{split}
f(x,y) &= \left( g_{e}(x,y) \right)^2 + x \left( g_{o}(x,y) \right)^2 \\
       &= \left( \sum_{m,n} \sum_{a,b} A_{mn,ab} x^{m} x^{b} y^a y^b \right) + x \left( \sum_{m,n} \sum_{a,b} B_{mn,ab} x^{m} x^{b} y^a y^b \right)\\
       &= \mbox{Tr} \left[ \mathbb{A} \cdot Q^e(x) \otimes Q(y) \right] + x \mbox{Tr} \left[ \mathbb{B} \cdot Q^o(x) \otimes Q(y) \right].
\end{split}
\end{align}
Here, we denote
\begin{align}
\begin{split}
Q^{e}_{nm}(x) =
\begin{pmatrix}
1 & x & x^2 & \cdots & x^{d_e}\\
x & x^2 &  &  & \\
x^2 &  & \ddots &  & \vdots \\
\vdots & & & &  \\
x^{d_e} & & \cdots & & x^{2d_e}
\end{pmatrix}
, \quad
Q^{e}_{nm}(x) =
\begin{pmatrix}
1 & y & y^2 & \cdots & y^{d_y}\\
y & y^2 &  &  & \\
y^2 &  & \ddots &  & \vdots \\
\vdots & & & &  \\
y^{d_y} & & \cdots & & y^{2d_y}
\end{pmatrix}
. \nonumber
\end{split}
\end{align}
Once we know the value of polynomial at different random $2d_y+1$ points, we can fix the $y$-dependence of polynomial. Namely, we have
\begin{align}
\label{Two sdp yk}
\begin{split}
f(x,y_k)  \equiv f_{k}(x) &= \mbox{Tr} \left[ \mathbb{A} \cdot Q^e(x) \otimes Q(y_k) \right] + x \mbox{Tr} \left[ \mathbb{B} \cdot Q^o(x) \otimes Q(y_k) \right] \\
         &= A_{mn,ab} Q_{mn}^{e}(x) \sum_{i} C_{ki} e^{i}_{ab} + x B_{mn,ab} Q_{mn}^{o}(x) \sum_{i} C_{ki} e^{i}_{ab}.
\end{split}
\end{align}
In the second line of \eqref{Two sdp yk}, we used the expansion $Q(y_k)= \sum_i C_{ki} e^{i}$ where the basis $e^{i}$ are defined as
\begin{scriptsize}
\begin{align}
\begin{split}
e^{1} =
\begin{pmatrix}
1 & 0 & 0 & \cdots & 0\\
0 & 0 &  &  & \\
0 &  & \ddots &  & \vdots \\
\vdots & & & &  \\
0 & & \cdots & & 0
\end{pmatrix}, \quad
e^{2} = \frac{1}{\sqrt{2}}
\begin{pmatrix}
0 & 1 & 0 & \cdots & 0\\
1 & 0 &  &  & \\
0 &  & \ddots &  & \vdots \\
\vdots & & & &  \\
0 & & \cdots & & 0
\end{pmatrix}, \quad
e^{3} = \frac{1}{\sqrt{3}}
\begin{pmatrix}
0 & 0 & 1 & \cdots & 0\\
0 & 1 &  &  & \\
1 &  & \ddots &  & \vdots \\
\vdots & & & &  \\
0 & & \cdots & & 0
\end{pmatrix}
\cdots.
\nonumber
\end{split}
\end{align}
\end{scriptsize}
Note that the basis $e^{i}$ are orthonormal to each other in that Tr$[e^{i} e^{j}] =\delta^{ij}$.

Now, \eqref{Two sdp yk} can be expressed as a standard form of
the semi-definite programming problem,
\begin{align}
\begin{split}
\sum_{i,k=1}^{2d_y+1} \left( C^{-1} \right)_{ik} f_k(x) e^{i}_{ab} = A_{mn,ab} Q^{e}_{mn}(x) + x B_{mn,ab}Q^{o}_{mn}(x).
\end{split}
\end{align}

\newpage
\bibliographystyle{jhep}
\bibliography{Reference}

\end{document}